%% file: paper_main.tex
\documentclass[11pt]{article}

\usepackage{amssymb,amsmath,amsthm,amsfonts}
\usepackage{thmtools} 

\usepackage[T1]{fontenc}
\usepackage[tt=false, type1=true]{libertine}
\usepackage[libertine]{newtxmath}

\usepackage[margin=1in]{geometry}

\usepackage{enumitem}
\setlist{nosep,topsep=0pt,leftmargin=*}

\usepackage{hyperref}

\usepackage[sortcites,backend=bibtex,sorting=nyt,style=alphabetic,natbib=true]{biblatex}
\usepackage{booktabs} 

\usepackage{float}

\newtheorem{theorem}{Theorem}[section]
\newtheorem{corollary}[theorem]{Corollary}
\newtheorem{lemma}[theorem]{Lemma}

\newtheorem{proposition}[theorem]{Proposition}
\theoremstyle{definition}

\newtheorem{definition}[theorem]{Definition}

\usepackage{titling}
\usepackage{etoolbox}
\pretitle{\begin{center}\Large\bfseries} 

\input{commands}
 
\newcommand{\acknowledgements}{\thanks{
Chido Onyeze, David X. Lin, and Siddhartha Banerjee were supported by NSF grant ECCS-1847393 and through a SPROUT Award from Cornell Engineering.
Éva Tardos was supported in part by ONR MURI grant N000142412742.
}}

\newcommand{\email}[1]{\protect\href{mailto:#1}{#1}}

\title{Dynamic Allocation of Public Goods with Approximate Core Equilibria\acknowledgements}

\usepackage{authblk} 

\author[1]{Chido Onyeze}
\author[1]{David X. Lin}
\author[1]{Siddhartha Banerjee}
\author[1]{\'Eva Tardos}

\affil[1]{Cornell University (
\email{chidoonyeze@cs.cornell.edu}
\email{dxl2@cornell.edu}, \email{sbanerjee@cornell.edu}, \email{eva.tardos@cornell.edu})}

\date{}

\begin{document}

\begin{titlepage}

\pagenumbering{gobble}  
\maketitle

\begin{abstract}

    \input{Paper_Sections/abstract}

\end{abstract}


\end{titlepage}

\clearpage
\pagenumbering{arabic}  
\setcounter{page}{1}    

\input{Paper_Sections/Introduction}
\input{Paper_Sections/Prelimiaries}

\input{Paper_Sections/Monetary_To_Non_Monetary_Framework}

\input{Paper_Sections/Characterization_Of_Pareto_Optimality}

\input{Paper_Sections/Equilibrium_Outcomes}

\input{Paper_Sections/Lower_Bounds}
\input{Paper_Sections/Extensions}
\input{Paper_Sections/Conclusions}

\newpage
\printbibliography

\newpage
\appendix
\input{Paper_Sections/Appendix/assumption_on_mechanism}

\newpage
\input{Paper_Sections/Appendix/equilibrium_existence}

\newpage
\input{Paper_Sections/Appendix/ex-ante_core_relaxation_proofs}
\newpage
\input{Paper_Sections/Appendix/equilibrium_proof_IC}
\input{Paper_Sections/Appendix/equilibrium_proof_Non_IC}

\newpage
\input{Paper_Sections/Appendix/multi-item-dwl-and-core}

\newpage
\input{Paper_Sections/Appendix/lower_bounds_proofs}
\newpage

\input{Paper_Sections/Appendix/proof_of_regularity}

\end{document}

%% file: commands.tex
\usepackage{comment}
\usepackage{enumitem,xspace,nicefrac,tcolorbox,xifthen,tikz,subcaption,physics,suffix,wrapfig,algorithm,algpseudocode,bbm}
\usepackage{pgfplots}
\pgfplotsset{compat=1.18}

\usetikzlibrary{shapes, arrows, positioning, pgfplots.fillbetween, patterns}

\definecolor{darkgreen}{rgb}{0.0, 0.5, 0.0}

\usepackage{thmtools,cancel,dsfont}
\usepackage[bb=dsserif]{mathalpha}
\usepackage[capitalize]{cleveref}

\usepackage{xcolor}
    \definecolor{myred}{HTML}{ea4335}
    \definecolor{mygreen}{HTML}{41a756}
    \definecolor{myblue}{HTML}{4285f4}

\IfPackageLoadedTF{hyperref}{}
{
\usepackage[colorlinks=true,allcolors=myblue]{hyperref}
}

\usepackage{comment} 
\usepackage{thm-restate}

\usepackage[suppress]{color-edits}

\addauthor[Eva]{et}{mygreen}
\addauthor[Sid]{sb}{cyan}
\addauthor[Chido]{co}{purple}
\addauthor[David]{dl}{blue}
\definecolor{@gray}{HTML}{edc3c5}

\DeclareMathOperator*{\argmax}{arg\,max}

\newcommand{\inner}[2]{\left \langle #1, #2 \right \rangle}

\newcommand{\inN}{\in \mathbb{N}}

\newcommand{\inR}{\in \mathbb{R}}

\newcommand{\N}{\mathbb{N}}

\newcommand{\R}{\mathbb{R}}

\DeclareMathOperator*{\E}{\mathbb{E}}
\DeclareMathOperator*{\pr}{\mathbb{P}}

\newcommand{\ind}[1]{\mathbbm{1}\left[#1\right]}

\newcommand{\graph}{\text{Gr}}

\newcommand{\utility}{\mathcal{U}}

\newcommand{\alloc}{\mathbf{A}}
\newcommand{\allocSet}{\mathcal{A}}
\newcommand{\val}{V}

\newcommand{\repTotalVal}{\reportedVal_{\text{tot}}}
\newcommand{\valvec}{V}
\newcommand{\reportedVal}{\widetilde{V}}
\newcommand{\costFunctionSet}{\mathcal{P}}
\newcommand{\reportedValAlt}{\widetilde{W}}
\newcommand{\valDist}{\mathcal{D}}
\newcommand{\mech}{\mathcal{M}}

\newcommand{\mechProp}{\mech_{\text{prop}}}
\newcommand{\mechMou}{\mech_{\text{mou}}}
\newcommand{\mechPot}{\mech_{\text{pot}}}

\newcommand{\nonMonMech}{\widetilde{\mech}\xspace}

\newcommand{\strat}{\mathcal{S}}

\newcommand{\harm}{\mathcal{H}}

\newcommand{\costFunc}{\mathcal{C}}

\newcommand{\newObjective}{Dead-Weight Loss\xspace}

\newcommand{\payments}{\mathbf{p}}

\newcommand{\socialCost}{SC}

\newcommand{\dwl}{\textsc{DWL}\xspace}

\renewcommand{\Pr}{\pr}
\renewcommand{\setminus}{-}
\renewcommand{\subset}{\subseteq}

%% file: Paper_Sections/abstract.tex
We consider the problem of repeatedly allocating multiple shareable public goods that have limited availability in an online setting without the use of money. In our setting, agents have additive values, and the value each agent receives from getting access to the goods in each period is drawn i.i.d. from some joint distribution $\mathcal{D}$ (that can be arbitrarily correlated between agents). The principal also has global constraints on the set of goods they can select over the horizon, which is represented via a submodular allocation-cost function. Our goal is to select the periods to allocate the good to ensure high value for each group of agents. 

We develop mechanisms for this problem using an artificial currency, where we give each agent a budget proportional to their (exogenous) fair share. The correlated value distribution makes this an especially challenging problem, as agents may attempt to free-ride by declaring low valuations for the good when they know other agents have high values—hoping those agents will bear a larger share of the cost of the resource. We offer a black-box reduction from monetary mechanisms for the allocation of a costly excludable public good.
We focus on pacing strategies, the natural strategies when using AI agents, where agents report a scaled version of their value to the mechanism. Our main results show that when using a truthful monetary mechanism as our building block, the resulting online mechanism has a focal equilibrium in which each agent plays a pacing strategy whose outcome results in an allocation that is a $(\mathcal{H}_n-1)$-\emph{approximation of the core}, where $\mathcal{H}_n$ is the Harmonic number, and $n$ is the number of players. Remarkably, we are able to achieve an approximate core solution as a Nash outcome without explicit collaboration or coordination between the agents.  With the agents' values in incomparable units, comparing the Nash outcome to the core is a reasonable measure of both efficiency and fairness.

%% file: Paper_Sections/Introduction.tex
\section{Introduction}

Consider a set of resources shared over multiple periods between a set of agents -- for example, a scientific instrument like a telescope shared by several researchers. There is a large recent literature on mechanisms allocating such a resource to one user at-a-time, in a way that is both fair and efficient, and without using money (as is desirable for shared resources like scientific instruments). These works crucially assume that the resource in each period can benefit only one user. However many research groups may be interested in having access to telescope data. This brings us into the domain of \emph{public goods allocation}, a cornerstone of microeconomics and the source of some of the most famous impossibility theorems in the field. 

When monetary transfers are allowed, allocating even a single public good raises challenges not present with private goods: users may be tempted to \emph{free ride}, declaring low valuations in hope that others bear all costs. Given a resource of cost $c$ and users with private values $\val_i$ for accessing it, the `first-best' decision (i.e., with full information) is to allocate the resource if $\sum_i\val_i\ge c$, and make users pay enough to cover the cost. However, it is known that due to free-riding, any mechanism can have arbitrarily bad welfare compared to this solution.
The situation is improved when one can \emph{exclude} some users (for example, by delaying the sharing of data to some set of agents); in this case, we know truthful monetary mechanisms (due to \citet{moulin2001strategyproof} for a single good, and \citet{dobzinski2017combinatorial} for multiple) that get the best possible approximation of the first-best welfare.

In this work, we consider the problem of designing \emph{non-monetary} mechanisms for the repeated allocation of excludable public goods. \cref{sec:prelims} provides a formal model, but at high level, and focusing first on the simpler setting of a single public good per round, our setting is as follows:
Over a horizon of $T$ periods, a principal is tasked with allocating (or not) a public good in each period among $n$ agents. If the principal chooses to allocate the good in any period, they can further specify which subset of agents get access to the good. The principal is constrained to allocate in at most $\alpha T$ periods (for some $\alpha \leq 1$, for example, due to limited finance, or energy consumption constraints). 
Each agent $i$ has private valuation $\val_i[t]$ for the good in period $t$.
The values are \emph{drawn i.i.d.} across periods $t\in[T]$ from some joint distribution $\valDist$, that may be arbitrarily correlated between agents. An agent's ex-post utility is the average of the total value she receives from periods in which she is given access to the good. More generally, we allow for multiple items to be allocated in each period, with the agent utilities and global allocation constraints depending both on the items allocated and the set of users who get access to the items in each period.

To model heterogeneous influence of each agent in allocation decisions, we assume each agent $i$ is associated with an exogenous share $\alpha_i$; setting all $\alpha_i$ equal means agents have equal influence. The goal of the principal is to select an allocation that satisfies global allocation constraints, and is both \emph{efficient} (resulting in high ex-post utility for agents) and \emph{fair} with respect to agents' shares; however, it is not clear how one can quantify these notions in a public goods setting. Each individual agent is of course self-serving, and so some mechanism is required to incentivize them to reveal their preferences. With values coming from a possibly highly correlated distribution $\valDist$, free riding is a major concern. Finally, one would like the mechanism to be simple and practical, and detail-free (in particular, agnostic of~$\valDist$).


Our work tackles these by combining two ideas: (approximate) \emph{core outcomes} and \emph{pseudomarkets}.
The former gives a benchmark for quantifying allocation quality in terms of how well they satisfy any potential coalition of blocking agents.
The latter lets us take a monetary mechanism $\mech$ for one-shot public goods allocation, and use it in a black-box way for repeated non-monetary allocation: We give each agent~$i$ a budget of $\alpha_i T$ artificial credits and ask them to use this currency to participate in $\mech$ in each round to determine allocations.
In this context, we study a class of simple \emph{value-scaling} policies, where each agent $i$ reports value $\val_i[t]/\beta_i$ in period $t$ for some chosen $\beta_i$; these have attracted particular recent interest, as one can think of them as the outcome of using auto-bidders, with $\beta_i$ being optimized to help agents optimally use their budgets. Our main result shows that \emph{value-scaling policies admit a Nash equilibrium which results in a strong core approximation}: when using monetary mechanism $\mech$, we show the resulting core approximation factor equals the \emph{\newObjective} (i.e., the difference between the first-best and realized welfare of the agents) of $\mech$. This is a widely-used performance metric in economics for evaluating monetary mechanisms, and our work thus provides a clean connection between inefficiency in one-shot monetary settings and core approximation in our context.

\subsection{Overview of Results and Techniques}

To build intuition, we focus first on a single good per period, before describing how our results extend to the general setting.

\emph{Black-Box Monetary Mechanism Reduction.} We introduce a black-box reduction (\cref{alg:Mon2NonMon}) from any one-shot monetary mechanism $\mech$ for allocating a single excludable good with cost $c$, to a non-monetary mechanism $\nonMonMech$ for our setting by giving each agent $i$ an initial budget of artificial credits proportional to their share $\alpha_i$. The resulting $\nonMonMech$  always selects at most an $\alpha$ fraction of the goods, provided the mechanism $\mech$ is \emph{cost covering} (i.e., the sum of the payments is at least the cost $c$ if the good is allocated). Moreover, it also gets an approximate equilibrium with strong performance guarantees if $\mech$ satisfies some additional requirements (discussed below).
Our reduction crucially is agnostic of agents' joint value distribution $\valDist$.

An example of a simple monetary mechanism in case of a single item is the proportional mechanism, denoted $\mechProp$: given reported values $\reportedVal_i$ from agents, we allocate to all agents if $\sum_i \reportedVal_i\ge 1$, and charge agent $i$ proportional to their bid $\payments_i=\reportedVal_i/\sum_j \reportedVal_j$. While this mechanism is useful for illustrating our approach, it is specialized for a single item, and also is highly prone to free-riding -- even with money, it is not incentive compatible for players to report their true values. This is remedied by more complex monetary mechanisms, which are incentive compatible, and also generalize to multiple items~\cite{moulin2001strategyproof,dobzinski2017combinatorial}.


\medskip
\noindent\emph{Agent Behavior under $\nonMonMech$.} When players use AI agents or auto-bidders to adapt to the environment, a natural outcome of this process is for the agent to use a simple \emph{value-scaling strategy} -- they report $\val_i[t]/\beta_i$ for some fixed $\beta_i \in \R_+$ when their true value is $\val_i[t]$ -- and select $\beta_i$ in a way to optimize their utility.  
We evaluate the outcome of our mechanism assuming all agents use value scaling, and focus on settings where agents scale their values so as to run out of credits at the end of the horizon. 
When using $\mechProp$ in our reduction, we show an approximate Nash equilibrium when agents can \emph{only} use value-scaling strategies. 
We also show that when the monetary mechanism $\mech$ is incentive compatible, then the same value scaling outcome is a true approximate equilibrium, where no agent can gain more than $o(T)$ value from deviating to \emph{any} adaptive strategy. 

\medskip
\noindent\emph{Evaluating Outcomes.} Next, we want to evaluate the quality of the outcomes realized under these mechanisms. With a non-monetary mechanism, agents' values are not on the same scale and hence, using the social welfare (i.e., sum of agent utilities) is not a reasonable measure. 
Instead, we focus on a benchmark of approximate \emph{core outcomes} with non-transferable utilities introduced in~\cite{fain2016core}, which has been growing in popularity in the computational social choice literature. 
The core of a game is a solution where no subgroup of agents can realize 
an alternate solution that is better for each agent in the group. 
In our setting, this corresponds to verifying that no group $S\subset[n]$ can select an allocation of size $(\sum_{i\in S} \alpha_i)T$ that makes each member of $S$ better off.  Such a guarantee is known to imply various properties about the allocation such as Pareto optimality (see~\cite{fain2016core,fain2018fair} for details).

\medskip
\noindent\emph{Approximating the Core with Online Mechanisms.} To connect outcomes in our online mechanism to the core, we need the monetary mechanism $\mech$ we use as in our reduction to satisfy some mild structural properties (see~\cref{app:assumptions_on_mech}).
In more detail: a monetary mechanism $\mech$ allocating a good with cost $1$ among agents $[n]$ takes as input reported values $\{\reportedVal_i\}_{i\in[n]}$ and outputs an allocation $\alloc^*\subset[n]$, and payments $\payments_i$ for each $i\in[n]$. As mentioned, $\mech$ is cost-covering if $\sum_i \payments_i \ge 1$ when the item is allocated. To understand how using $\mech$ affects outcome quality, we need to first define the \emph{\newObjective} ($\dwl$) of $\mech$. In general, the $\dwl$ of a mechanism is the difference between the first-best welfare and the realized welfare for the agents (for example, see~\cite{cummings2020algorithmic} for its use in price discrimination, or ~\cite{roughgarden2009quantifying} for monetary public good mechanisms). For our particular setting, this simplifies to the following (see~\cref{def:dwl} for details)\footnote{Here we assume the cost of the good is $1$; for general $c$, we normalize by the cost to make $\dwl$ scale invariant (see~\cref{def:dwl_multiitem} for the more general form, as well as for the generalization to multiple items).}:

$$
\dwl(\mech,\reportedVal)=\left(\sum_{i \in [n]} \reportedVal_i - 1\right)^+ - \left(\sum_{i \in \alloc^*\left(\reportedVal\right)} \left(\reportedVal_i - \payments_i\left(\reportedVal\right)\right)\right)$$
The proportional mechanism $\mechProp$ satisfies $\dwl(\mechProp,\reportedVal)=0$; it has 0 budget surplus, and either all players get the item, or the total reported value is below $1$. Getting $\dwl=0$ is known to be impossible for incentive compatible mechanisms in our setting, however, the IC mechanisms we consider (in particular, those of~\citet{moulin2001strategyproof} and \citet{dobzinski2017combinatorial}) satisfy 
$\dwl(\mech,\reportedVal)\le \harm_n - 1$.

Given the above, we can paraphrase our main result (\cref{thm:mainresult}): 
\begin{center}
\begin{quote}
\emph{If monetary mechanism $\mech$ satisfies $\dwl(\mech,\reportedVal)\le \gamma$ for some $\gamma\ge 0$, and if agents can only use value-scaling strategies in $\nonMonMech$, then there exists a profile of strategies we refer to as the {focal pacing equilibrium}, such that agents are in approximate Nash equilibrium and the resulting allocation is in the $\left(\gamma, o(1)\right)$-approximate core w.h.p}. 
\end{quote}
\end{center}

To see the significance of this result, note that the core is a collaborative solution concept, and existing approximation results typically require cooperation and full knowledge of agent values. However, in our result, agents realize an approximate core allocation at an approximate \emph{Nash equilibrium}, and while making online decisions. Moreover, we directly translate an inefficiency guarantee ($\dwl$) into a core approximation. 
We note that the $o(1)$ additive factor is due to random fluctuation, and is unavoidable (see~\cref{subsec:coreApproxToEq}).

\medskip
\noindent\emph{Using Incentive Compatible (IC) Mechanisms in our Reduction.} Using an IC mechanism as a building block makes our focal pacing equilibrium a stronger notion: the value scaling profile remains a best response even considering \emph{all} adaptive strategies. The cost is a higher core approximation -- however, we can now leverage known truthful mechanisms which have $\dwl(\mech,\reportedVal)\le \harm_n - 1$.

\medskip
\noindent\emph{Core Approximation at Equilibrium.} 
Putting these results together, we show that, using mechanism $\mechProp$, if agents could only use value-scaling strategies, $\nonMonMech$ has a class of approximate equilibrium strategies such that the resulting allocation is, w.h.p., in the $(0, o(1))$-approximate core (\cref{thm: auto-bidding equilibrium}). Using instead the mechanisms of $\mechMou$ \cite{moulin2001strategyproof} or $\mechPot$ \cite{dobzinski2017combinatorial} that are truthful and incentive compatible as monetary mechanisms then there exists a class of approximate equilibrium strategies (w.r.t. all adaptive strategies) such that the resulting allocation in $\nonMonMech$ is, w.h.p., in the $\left(\harm_n-1, o(1)\right)$-approximate core (\cref{cor: H_n approx result}). 

\medskip 
\noindent\emph{Focal Equilibrium and Tightness of our Analysis.} Our results above are focusing on what we called a \emph{focal pacing equilibrium}, an equilibrium where each agent is using value-scaling strategies, and agents don't run out of money (in expectation). Such pacing equilibria were studied in the context of auctions for allocating items \cite{DBLP:journals/ior/ConitzerKSM22}. In their context, when goods are not shareable, scaling values to avoid running out of money early is a best response, as it optimizes the agents ``bang-for-buck''. In the case of shareable items this may not be the case: if other agents bid too aggressively and run out of money early, the best response of an agent may be to do the same, as others having no money severely limits sharing opportunities. We think of the pacing equilibrium when players don't run out of money as \emph{focal} as when all players increase their scalars to have their money last longer, this is better for the group involved.
We show that the connection of such focal equilibria to the $\dwl$ of the monetary mechanism is tight in our analysis. If $\mech$ is incentive compatible and $\sup_{\reportedVal}\dwl(\mech, \reportedVal) \geq \gamma$ for some reported valuations $\reportedVal$, then there is a measure $\mu$ of value distributions such that there is an approximate equilibrium of value-scaling strategies in $\nonMonMech$ where agents do not run out of money early, and yet it induces an allocation that is not in the $(\gamma-\epsilon, o(1))$-approximate core with high probability for any $\epsilon > 0$ (\cref{thm: lower bound result}).
In particular, previous results on one-shot cost-sharing mechanisms show that $\sup_{\reportedVal}\dwl(\mech, \reportedVal) \geq \harm_n-1$  for any incentive-compatible $\mech$ satisfying some additional mild conditions \cite{dobzinski2018shapley}, so that the induced allocation in $\nonMonMech$ is not in the $(\harm_n-1-\epsilon, o(1))$-approximate core (\cref{cor: lower bound result for ic mechanisms}). 



\medskip
\noindent\emph{Generalization to the settings with multiple available items.} 
Finally, our results generalize to the following more general setting: in each of $T$ periods, $m$ public goods are available for allocation. The principal's global constraints on the allocation are modeled as before by assuming an overall budget of at most $\alpha T$, but now the overall cost of an allocation is given via a submodular function that depends on the set of agents given access to each good. Each agent $i$ in period $t$ has a private valuation $\val_i[t]\in [0,1]^m$, with $\val_{i,k}[t]$ denoting the value agent $i$ gets for access to good $k$ in period $t$; valuations are additive across periods. The valuation vectors and cost functions are \emph{drawn i.i.d.} across periods $t\in[T]$ from some joint distribution $\valDist$, that may be arbitrarily correlated between agents and cost functions. We show that we can appropriately generalize the concepts previously described and achieve the same main result even in this more general setting (\cref{thm:mainresult_multiitem}).

\paragraph{Organization.} 

The proof outline of our main theorem spans the next few sections. In~\cref{sec:exantecore}, we define the ex-ante approximate core, allowing us to focus on time-independent strategies with a small additive loss. In~\cref{sec: eq outcomes}, we show how value-scaling strategies yield a Nash equilibrium and how the mechanism’s $\dwl$ translates into an ex-ante core approximation. In~\cref{sec:lower_bounds}, we construct a distribution $\valDist$ showing the tightness of our guarantee. In~\cref{sec:extensions}, we show how our results immediately apply to the general setting described in \cref{ssec:model}.

\input{Paper_Sections/Introduction/Related_Works}

%% file: Paper_Sections/Introduction/Related_Works.tex
\paragraph*{Related Work.}
Our work intersects the literature on public goods allocation, repeated resource allocation without money, and fairness in computational social choice. 
The classical public goods problem suffers from fundamental inefficiencies due to the free-rider problem. For non-excludable public goods, classic impossibility theorems imply that no mechanism can achieve full efficiency and strategyproofness 
while exactly balancing the budget \cite{green1979incentives, mascolell1995}.
For excludable public goods (club goods), the theory of clubs \cite{buchanan1965economic,sandler1980economic} allows for pricing mechanisms that partially mitigate inefficiencies, though the most efficient of these mechanisms (in particular, the Moulin mechanisms \cite{moulin2001strategyproof}) still incur provable lower bounds on social cost \cite{roughgarden2009quantifying,dobzinski2018shapley,deb1999voluntary}.

Our use of the approximate core as a benchmark for efficiency and fairness is inspired by a recent line of work in computational social choice. The first to propose this benchmark was the work of~\citet{fain2016core,fain2018fair}, based on earlier works in the economics literature~\cite{foley1970lindahl,scarf1967core}. More recently, the idea has been used in a variety of settings, in particular, in multiwinner and constrained allocation settings. Recent work has applied this framework to problems such as proportionally fair clustering \cite{chen2019proportionally}, group fairness in committee selection \cite{cheng2020group}, and stability under various constraints \cite{jiang2020approximately,munagala2022approximate,mavrov2023fair}. Crucially, however, most of these works concentrate on computational questions regarding integral allocations and approximations in \emph{full-information} settings. In contrast, the focus of our work is on designing strategyproof mechanisms for achieving good core outcomes online.

A large body of work studies repeated allocation without money, motivated by applications such as course allocation \cite{budish2017course}, food banks \cite{prendergast2022allocation}, etc. The single-item-per-round model with random valuations was introduced by \citet{guo2009competitive}. The work of \citet{gorokh2021monetary} provided a unified black-box approach for emulating monetary mechanisms via artificial currencies, building on the idea of linking repeated allocations to incentivize truthful behavior \cite{jackson}. While our work takes inspiration from this literature, we note that unlike \cite{jackson,gorokh2021monetary} which require extensive knowledge of the value distribution, our mechanism is \emph{distribution agnostic}.
We note that requiring our mechanism be distribution agnostic makes the problem considerably more difficult. In our setting, with extensive value of agents' value distribution, it may be possible to enforce truthfulness and cover the cost of the allocations without much welfare loss by ``promising utility'' to an agent in the future based on the value distribution as long as they report truthfully and decreasing future utility under non-truthful reports, as done by \cite{balseiro2021non} in a monetary online public good allocation setting. Instead, we focus on simple distribution-agnostic mechanisms, incurring a welfare loss to enforce truthfulness similar to that of one-shot monetary mechanisms for our setting.
This is more in the spirit of more recent work for private goods that focuses on robust individual-level guarantees. \citet{gorokh2019remarkable} showed that the repeated first-price pseudo-market admits a policy via which an agent is guaranteed to receive at least a $1/2$ fraction of their ideal utility (i.e., the maximum utility they can get from a set of goods proportional to a pre-specified share) regardless of others' behavior. Subsequent work has recovered these robustness guarantees and studied equilibria for an alternate mechanism called Dynamic Max-Min Fairness which is not based on a pseudo-market~\cite{fikioris2023online,onyeze2025allocating}. While ideal utility guarantees imply core outcomes, its use is highly specific to the private goods setting, and does not extend to public goods in any natural way.

Finally, we note our use of value scaling has parallels to budget scaling in pacing equilibria, which have been studied in the context of repeated monetary auctions with budgets~\cite{DBLP:journals/ior/ConitzerKSM22,conitzer2022pacing,gaitonde2023budget}. Our naming of the focal pacing equilibrium acknowledges this resemblance. A critical difference however is that unlike with monetary budgets which have the same units as that of utility, it is much less clear how to determine the scaling factors in our setting, and how these translate to core outcomes.

%% file: Paper_Sections/Prelimiaries.tex
\section{Preliminaries}
\label{sec:prelims}

We first introduce our setting of dynamic allocation of excludable public goods, as well as our benchmark concept of the core. Throughout this work, for $n \inN$, we denote $[n] =\{1, \dots, n\}$ and $\R_+^n = \{x \in \R: x_i \geq 0 \text{ for all } i \in [n]\}$. We denote $\harm_k = \sum_{j=1}^k\frac{1}{j}$ as the $k$-th harmonic number, and for $x,y \in \R^n$ and $S \subset [n]$, denote $\inner{x}{y}_S = \sum_{i\in S} x_iy_i$. Finally, we say random events happen `with high probability' (henceforth abbreviated w.h.p.) over an interval of length $T$ if they occur with probability $1-o(1)$ in $T$.

\input{Paper_Sections/Preliminaries/Model}

\input{Paper_Sections/Preliminaries/Core}

\input{Paper_Sections/Preliminaries/Auction}

%% file: Paper_Sections/Preliminaries/Model.tex
\subsection{Dynamic Allocation of Excludable Public Goods}
\label{ssec:model}

We now formalize the allocation problem that we consider in this paper in full generality.
Fix $n, m \in \N$. Our setting consists of a principal, $n$ agents and $T$ rounds. In each round $t$, $m$ indivisible goods arrive as well as a cost function for the items $c[t]: 2^{[n]\times [m]} \rightarrow [0, 1]$. In each round, the principal must make an \emph{irrevocable} choice of an allocation of the items in the round to the agents $\alloc \subset [n]\times [m] $. An allocation specifies the set of goods to which each agent is granted access where $(i, k) \in \alloc$ implies that agent $i$ is granted access to good $k$. Furthermore, let $\alloc_i = \{k \in [m]: (i,k) \in \alloc\}$.  The cost of the allocation in round $t$ is given by $c[t](\alloc)$.

Let $\allocSet = (\alloc[t])_{t\in [T]}$ be the final allocation of the principal over $T$ rounds where $\alloc[t]$ is the allocation chosen by the principal in round $t$. Let $\costFunc(\allocSet)$ be the total cost of the final allocation ie. $\costFunc(\allocSet) = \sum_{t = 1}^T c[t](\alloc[t])$. The principal is constrained on the total cost of the final allocation they selects. Specifically, it must hold that $\costFunc(\allocSet) \leq \alpha T$ for some constant $\alpha \leq 1$.

Each agent $i \in [n]$ has associated private value $\val_i[t] \in [0, 1]^m$ for the goods in round $t$, 
which becomes known to the agent at the beginning of round $t$ (and is unknown to all other agents/the principal). The (ex-post) utility that agent $i$ gets from a final allocation $\allocSet$ is given by 
\begin{align*}
\utility_i\left(\allocSet, T\right) = \frac{1}{T}\sum_{t\in [T]} \sum_{k \in \alloc_i[t]}\val_{i,k}[t].
\end{align*}

Let $\costFunctionSet$ be a set of cost functions $c: 2^{[n] \times [m]} \rightarrow [0,1]$ that are:
\begin{enumerate}
    \item \emph{submodular}: For all $\mathbf A, \mathbf B \subset [n] \times [m]$, $c(\mathbf A \cup \mathbf B) + c(\mathbf A \cap \mathbf B) \leq c(\mathbf A) + c(\mathbf B)$.
     \item \emph{monotone}: For all $\mathbf A, \mathbf B \subset [n] \times [m]$, $c(\mathbf A) \leq c(\mathbf A \cup \mathbf B)$.
     \item \emph{normalized}: $c(\mathbf A) = 0$ if and only if $\mathbf A = \emptyset$. 
\end{enumerate}

We consider the setting where the valuations and cost function in each round $(\val[t], c[t])$ is drawn from some joint distribution $\valDist$ over $[0, 1]^{n \times m} \times \costFunctionSet$ such that conditional on $c[t] = c$, the distribution of $\val[t]$ is absolutely continuous on $[0, 1]^{n \times m}$ for all $c \in \costFunctionSet$. Further, we assume $(\val[t], c[t])$ are drawn i.i.d. between rounds. Note that this allows different agents to have very different value distributions, for values to be arbitrarily correlated between agents, for both values and costs to be arbitrarily correlated across goods, and for costs to depend on both which goods are allocated and to whom they are allocated. 
The joint distribution $\valDist$ is unknown to the principal, and may or may not be known to the agents\footnote{In particular, we assume $\valDist$ is common knowledge when reasoning about equilibria, but when arguing that certain simple mechanisms lead to approximate equilibria, we want mechanisms to be distribution agnostic.}.

\paragraph{Single Good, 0-1 Cost Setting}
For ease of exposition, we present our results and techniques first for the simpler single-good setting (i.e., with $m=1$, cost function $c(\alloc) = \ind{\alloc \neq \emptyset}$). 
Here, the principal must select up to $\alpha T$ rounds to allocate the good, and also select the set of agents who get access to the good in each period. An allocation $\alloc$ represents the set of agents who get access to the good. Ideally, the principal would give every agent access to the good when the good is allocated. However, this may lead to perverse incentives on how agents report as we discuss in the following sections. That said, we compare our resulting allocation with the full-information first-best allocation.

We state our main \cref{thm:mainresult} for this setting, and for the more general setting in \cref{thm:mainresult_extension}.

%% file: Paper_Sections/Preliminaries/Core.tex
\subsection{Non-Monetary Mechanisms and Core Allocations}

Our focus in this work is on \emph{non-monetary} mechanisms for dynamic public goods allocation, where the principal allocates resources based on history and agents’ bids, but without monetary transfers. Without money, comparing agents’ utilities becomes meaningless, making standard welfare objectives like total utility inapplicable. Instead, we adopt a fairness benchmark based on the \emph{core} from cooperative game theory, adapted to public goods settings by \citet{fain2016core,fain2018fair}, building on earlier work in economics \cite{foley1970lindahl,scarf1967core}.

To model an agent's influence on the allocation, we assume each agent $i$ is assigned a vote share $\alpha_i \in (0,1)$ with $\sum_{i \in [n]} \alpha_i = \alpha$. These shares reflect the relative weight of agents’ preferences—equal shares ($\alpha_i = \alpha/n$) imply equal treatment. The core then captures the idea that no group of agents should be able to reallocate their collective share to strictly benefit all its members. We now formalize a generalized, approximate version of this concept.

\begin{definition}[\textbf{$(\gamma, \delta)$-Approximate Core}]\label[definition]{def: approximate core}
For $\gamma, \delta \geq 0$, a feasible allocation $\allocSet$ (i.e., with $\costFunc(\allocSet) \leq \alpha T$) is said to have a $(\gamma, \delta)$-blocking coalition $S \subset [n]$ if there is some other allocation $\widetilde\allocSet$ with $\costFunc\left(\widetilde\allocSet\right) \leq (\sum_{i \in S}\alpha_i)T$ such that 
for all $i \in S$,
\begin{align*}
\utility_i\left(\widetilde\allocSet, T \right) > (1+\gamma)\cdot \utility_i\left(\allocSet, T\right) + \delta.
\end{align*}    
We say a feasible allocation $\allocSet$ is in the $(\gamma, \delta)$-\emph{approximate core}
if there are no $(\gamma, \delta)$-blocking coalitions.  
\end{definition}
A $(0,0)$-approximate core allocation corresponds to a \emph{true core} allocation. The principal wants to choose an allocation in the $(\gamma, \delta)$-approximate core for as small a value of $\gamma$ and $\delta$ as possible; in particular, our guarantees have $\gamma=O(\log n)$ and $\delta=o(1)$ (i.e., vanishing with $T$). Note also that this additive $\delta=o(1)$ loss is primarily due to stochastic fluctuations as we are comparing an online allocation against an ex post benchmark, and is unavoidable even when using adaptive strategies~\footnote{For example, suppose $\alpha=1/2$, and there is a \emph{single} agent with value $\val[t]=\{0,1,2\}$ with probabilities $(1/2-1/\sqrt{T},1/\sqrt{T},1/2)$ respectively. There is no loss due to incentives (since there is a single agent), but nevertheless any policy realizes an allocation that is at best in the $(0,\Theta(1/\sqrt{T}))$-core (following from standard lower bounds for prophet inequalities; see for example~\cite{alaei2014bayesian}).}. In~\cref{sec:exantecore}, we discuss how this is eliminated when considering an ex-ante definition of the core.

%% file: Paper_Sections/Preliminaries/Auction.tex
\subsection{Monetary Mechanisms for One-Shot Public Good Allocation} 
\label{subsec: One Shot Monetary Mechanisms Description}

Our main results in this work involve obtaining good approximate core allocations for the dynamic non-monetary allocation via \emph{black-box reductions} from monetary mechanisms for one-shot allocation. We now briefly describe salient features of such mechanisms that we need for our results.

As before, we consider a setting with $n$ agents, but now with a \emph{single} good with cost $1$. 
Each agent has a value $\val_i$ for receiving access to the good. A mechanism $\mech$ in this setting is a pair of maps from agents' reported values $\reportedVal$ to a set of allocated agents $\alloc^*\left(\reportedVal\right)$ and payments $\payments_i\left(\reportedVal\right)$ for each agent $i$; in addition, we define the cost $C(\alloc)$ of an allocation $\alloc$ as $1$ if $\alloc\neq \emptyset$, else $0$.
Note that in this setting, the `first-best' welfare (i.e., that achievable under full information) is $W^{\star}=(\sum_{i\in[n]}\val_i - 1)^+$, achieved by giving access to the good to all agents, and charging payments $\payments_i\leq \val_i$ so as to cover the required cost.

Our aim is to harness existing mechanisms for this setting in a black-box fashion to get good approximate core outcomes. To this end, we need some requirements on the mechanisms. In particular, two critical requirements for mechanisms admitted to our reduction are:
\begin{itemize}
\item  \emph{Individually Rationality (IR)}: agents are charged at most their reported values, and only if given access to the good; formally, for all $\reportedVal \in \R^n_+$  and $i \in [n]$, $\payments_i\left(\reportedVal\right) \leq \reportedVal_i\cdot \ind{i \in \alloc^*\left(\reportedVal\right)}$.
\item \emph{Cost-Covering (CC)} (or {budget balanced}): The mechanism $\mech$ must have a non-negative {budget surplus}, i.e., raise the cost of the good if it is selected; formally
$\sum_{i\in[n]}\payments_i\left(\reportedVal\right)\geq \ind{\alloc^*\left(\reportedVal\right)\neq\emptyset}$.
\end{itemize} 
Besides these, we impose some additional assumptions on the form of $\mech$ (given in \cref{app:assumptions_on_mech}). While these are somewhat more technical, 
they are satisfied by all known \emph{incentive compatible (IC)} mechanisms in this setting. 
However, our conditions also admit other reasonable (though not IC) mechanisms such as the Proportional Mechanism $\mechProp$ that we introduced earlier.

Finally, the relevant metric of any given mechanism that controls our core guarantees is the \emph{\newObjective} (or Efficiency Loss). This is defined for any general setting and mechanism as the difference between the `first-best' welfare for the agents (i.e., with full information assuming agents have to cover the cost) and that realized by the mechanism. In our specific setting (one-shot excludable public good allocation with fixed cost), this is defined as follows:
\begin{restatable}{definition}{defDwl}[\newObjective (DWL)]
\label{def:dwl}
For any monetary mechanism $\mech$ which is IR and CC, if agents report values $\reportedVal \in \R^n_+$, then the \newObjective (DWL) is given by 
\begin{align*}
\dwl\left(\mech, \reportedVal\right) &
= \left(\sum_{i \in [n]} \reportedVal_i - 1\right)^+ - \left(\sum_{i \in \alloc^*\left(\reportedVal\right)} \left(\reportedVal_i - \payments_i\left(\reportedVal\right)\right)\right)
= \left(\sum_{i \notin \alloc^*\left(\reportedVal\right)} \reportedVal_i + \sum_{i \in [n]} \payments_i\left(\reportedVal\right) - 1\right)^+.
\end{align*}
\end{restatable}
It is possible for non-incentive compatible mechanisms to always achieve $0$ \newObjective, e.g., $\mechProp$. However, for any incentive-compatible mechanism $\mathcal M$ satisfying (IR) and (CC), the worst-case \newObjective must be at least $\Omega(\log n)$ \cite{dobzinski2018shapley}.

%% file: Paper_Sections/Monetary_To_Non_Monetary_Framework.tex
\section{Blackbox Monetary to Non-Monetary Reduction}
\label{sec: reduction}

In this section, we introduce our reduction from one-shot monetary mechanism $\mech$ to a non-monetary mechanism $\nonMonMech$, and state our main result in~\cref{thm:mainresult}. We also discuss some example monetary mechanisms and the guarantees we get when using these.

\subsection{A General Template for Artificial Currency Mechanisms} 
\label{ssec:auctions}

We  recall our definition of the one-shot monetary allocation mechanism we introduced in~\cref{subsec: One Shot Monetary Mechanisms Description}.

Given such a monetary mechanism, our reduction (see~\cref{alg:Mon2NonMon}) is now as follows: 
\begin{itemize}
    \item At the start of the mechanism, agent $i$ is given a budget of $\alpha_i T$ units of artificial currency ($B_i[0] = \alpha_i T$).
    \item In each round $t$, we set the cost of the good to be 1 and run an instance of the monetary mechanism $\mech$ to determine the allocation $\alloc[t]=\alloc^*\left(\reportedVal[t]\right)$. 
    \item Agents can report any value until their budget goes below $\sup_{\val\in\R_+^n}\payments_i(\val)$ (which is finite by assumption), at which point all future reports are set to $0$. (Alternately, we can take the minimum of their report and budget; we choose the former for simplicity.)
    \item If the allocation from the mechanism is non-empty, we charge each agent an amount of artificial currency equal to the payment of the agent from $\mech$ (setting $B_i[t] = B_i[t-1] - \payments_i\left(\reportedVal[t]\right)$).
\end{itemize}
\medskip
An immediate consequence of assuming $\mech$ is cost covering is that we immediately get ex-post feasibility:
\begin{proposition}
    The final allocation produced by the induced non-monetary mechanism $\nonMonMech$ is feasible.
\end{proposition}
\begin{proof}
Since $\mech$ is cost covering, on any round on which the good is allocated to any agents at least one unit of artificial currency is removed from the mechanism. Since, there is a total of $\alpha T$ units of artificial currency in the mechanism, in the final allocation, at most $\alpha T$ goods are allocated.
\end{proof}
\input{Paper_Sections/algorithm_1}

\subsection{Agent Strategies and Equilibrium}
\label{ssec:scalingstrategies}

We next consider how an agent might behave in the non-monetary mechanism $\nonMonMech$ in~\cref{alg:Mon2NonMon}. We define an adaptive strategy $\phi_i$ for agent $i$ to be sequence of maps $(\phi_i[t])_{t \in [T]}$ where $\phi_i[t]: (\R^n)^{t-1} \times [0, 1] \rightarrow \R$. Under the strategy $\phi_i$, on round $t$, assuming agents report in previous rounds were $\reportedVal[1], \cdots, \reportedVal[t-1] \in \R^n$ and agent $i$'s realized value is $\val_i[t]$, the report of agent $i$ in the round is given by $$\reportedVal_i[t] = \phi_i[t]\left(\reportedVal[1], \cdots, \reportedVal[t-1], \val_i[t]\right).$$

We also define the class of \emph{time-independent strategies}, in which the report of the agent does not dependent on the reports on previous rounds. These strategies can be represented as maps $\phi_i: [0, 1] \rightarrow \R$ where for realized value $\val_i[t]$ on round $t$, the agent reports $\reportedVal_i[t] = \phi_i(\val_i[t])$. 

For ease of notion, we drop the dependence on previous reports when discussing adaptive strategies: $$\phi_i[t]\left(\reportedVal[1], \cdots, \reportedVal[t-1], \val[t]\right) = \phi_i[t](\val[t]).$$ 

\begin{definition}[Average Expected Utility] \label[definition]{def:expect utility}
    Given a mechanism $\mech$ and profile of strategies for the agents $\phi = (\phi_1, \cdots, \phi_n)$, let $\utility_i(\phi, T)$ be the expected per round utility for agent $i$ when agents use the strategy profile under $\nonMonMech$ over $T$ rounds.  Formally, given stopping times $\tau_1, \cdots, \tau_n$ where $\tau_i$ is the minimum of $T$ and the round in which agent $i$'s remaining budget is depleted (i.e., goes below $\payments_{\max}$), we have: 
    $$\utility_i(\phi,T) = \frac{1}{T}\sum_{t=1}^T \E_{\val[t] \sim \valDist}\bigg[\val_i[t]\cdot \mathbbm1\Big[i \in \alloc^*\left(\reportedVal_1[t], \cdots, \reportedVal_n[t]\right)\Big]\bigg]$$
    where $\reportedVal_i[t] = \phi_i[t](\val_i[t])\cdot \ind{t \leq \tau_i}$ for all $i$.
    
\end{definition}

\begin{definition}[Expected Cost Before Depletion] \label[definition]{def:expect cost}
Given a profile of time-independent strategies $\phi = (\phi_1, \cdots, \phi_n)$, let $\costFunc_i(\phi)$ be the expected payment for agent $i$ on a round before any agent has depleted their budget when agents use the strategy profile under $\nonMonMech$:
$$\costFunc_i(\phi) = \E_{\val \sim \valDist}\left[ \payments_i\left(\reportedVal_1, \cdots, \reportedVal_n\right)\right]$$ where $\reportedVal_i = \phi_i(\val_i)$ for all $i$.
\end{definition}

Using this, we can now define when a profile of strategies is an approximate Nash equilibrium. 
\begin{definition}[Approximate Best Response and Approximate Nash Equilibrium]
    Given a class of strategies $\strat$, a strategy $\phi_i$ is an approximate best response for agent $i$ with respect to $\strat$ given the strategy profile of the other agents $\phi_{-i}$ if 
    $$\utility_i(\phi_i,\phi_{-i}, T) \geq \utility_i(\phi'_i,\phi_{-i}, T) - o(1)$$
    for all $\phi'_i \in \strat$. We say a strategy profile $\phi = (\phi_1, \cdots,\phi_{n})$ is an approximate Nash equilibrium with respect to $\strat$ if, for all $i \in [n]$, $\phi_i \in \strat$ is an approximate best response for agent $i$ wrt. $\strat$ given the strategy profile of the other agents $\phi_{-i}$.
\end{definition}
One might hope that under a suitably nice monetary mechanism $\mech$, the simple strategy of truthful reporting (i.e., $\phi(v) = v$) is reasonable. Unfortunately, the following example illustrates this is not the case. 

\medskip
\noindent\emph{Example.} Consider an instance with a single agent with value drawn uniformly from $[0, \frac{1}{2}]$ for each good. For any IR and CC mechanism $\mech$, if the agent reports their value truthfully, then they receive none of the goods since, to receive a good, they need to be charged 1 unit of artificial currency, which violates individual rationality. 
However, the agent has no intrinsic value for the artificial currency, and so prefers to misreport their value to be allocated the good as many times as their budget allows. 
\medskip

We note that this issue can be resolved by using an appropriate \emph{value-scaling strategy}: $\phi(v) = v/\beta$ for some $\beta \in \R_+$. The remainder of this work will consider what outcome we can achieve when agents use strategies of this type.

\input{Paper_Sections/Main_Result}

%% file: Paper_Sections/algorithm_1.tex
\begin{algorithm}[t]
\caption{Monetary To Non-Monetary Reduction}\label{alg:Mon2NonMon}
\begin{algorithmic}
\Require Monetary Mechanism $\mech = (\alloc^*, \payments)$, Agent Shares $(\alpha_i)_{i \in [n]}$
\State Assign each agent a budget $B_i[0] = \alpha_i T$
\For{$t = 1, \cdots, T$}
    \State Obtain agent reports $\reportedVal_i[t] \in \R_+$
    \State To ensure budget feasibility, set $\reportedVal_i[t] \gets \reportedVal_i[t] \cdot \ind{B_i[t-1] \geq \payments_{i,\max}}$ where $\payments_{i,\max} = \sup_{\val \in \R^n_+} \payments_i(\val)$ 
    \State Run mechanism $\mech$ to get allocation $\alloc^*(\reportedVal[t])$ and artificial payments $\payments_i(\reportedVal[t])$ 
    \State Enact $\alloc^*(\reportedVal[t])$ and update budgets $B_i[t] = B_i[t-1] - \payments_i(\reportedVal[t])$
\EndFor
\end{algorithmic}
\end{algorithm}

%% file: Paper_Sections/Main_Result.tex
\subsection{Value-Scaling Strategies and Approximate Core Outcomes}
\label{subsec:coreApproxToEqOutcomes}

We now state our main theorem, which shows that if we use~\cref{alg:Mon2NonMon} with any suitable monetary mechanism $\mech$ (i.e., one which is IR, CC, and satisfying the technical conditions in~\cref{app:assumptions_on_mech}), then there is an approximate equilibrium of $\nonMonMech$ in the class of \emph{value-scaling strategies}, i.e., using $\phi^\beta(v) = \frac{v}{\beta}$ for some $\beta \in \R_+$,  such that the maximum \newObjective of $\mech$ translates into a core approximation of the allocation in $\nonMonMech$ at equilibrium. Moreover, if the mechanism $\mech$ is also incentive compatible, this strategy profile is an approximate equilibrium in the class of all strategies.

\begin{theorem}
\label{thm:mainresult}
Let $\gamma = \sup_{\reportedVal}\dwl(\mech, \reportedVal)$. Assume that one of the following hold.
\begin{enumerate}
\item Agents are restricted to only using value-scaling strategies.
\item Agents can use any adaptive strategies and $\mech$ is incentive compatible.
\end{enumerate}
Then, there exists an approximate Nash equilibrium $\phi = (\phi^{\beta_1}, \dots, \phi^{\beta_n})$ where each player plays a value-scaling strategy in $\nonMonMech$ such that the resulting allocation $\allocSet$ is in the $(\gamma,o(1))$-approximate core with probability at least $1-o(1)$.
\end{theorem}






Before proceeding, we first offer a high level outline of the proof of this main theorem; this is then further fleshed out in \cref{sec:exantecore,sec: eq outcomes}, with some details, as well as the extension to multi-item settings, deferred to the appendix. 
The main steps of the proof of~\cref{thm:mainresult} can be summarized as follows:\\
\begin{enumerate}
    \item We first want to relax the online problem into a one-shot allocation problem. To do so, we define an \emph{ex-ante relaxation} of our allocation problem, and a corresponding ex-ante definition of the core, and give a sufficient condition for an allocation the be in the approximate ex-ante core (\cref{lem: core-characterization}).
    
    \item Next, we analyze the performance of value-scaling policies with respect to this ex-ante core. Recall we define $\costFunc_i(\phi)$ to be the expected `cost-until-budget-depletion' for agent $i$ under some strategy profile $\phi = (\phi_1, \cdots, \phi_n)$. We show that any profile of value-scaling strategies such that $\costFunc_i(\phi^{\beta_1}, \dots, \phi^{\beta_n}) = \alpha_i$ for all $i$ induces an ex-ante allocation that is in the approximate ex-ante core, where the approximation depends on $\sup_{\reportedVal}\dwl(\mech, \reportedVal)$ (\cref{lem: social cost to core approx}).
    
    \item We then pass from ex-ante allocations back to ex-post allocations, by showing that an ex-ante allocation induces a ex-post allocation that is w.h.p. in the approximate $(\gamma, o(1))$-approximate core if the ex-ante allocation is in the $\gamma$-approximate core (\cref{cor: ex ante core implies expost core}).

    \item Next, we show that if all other agents use time-independent strategies $\phi_{-i}$, and a specific agent $i$ uses the value-scaling strategy $\phi^{\beta_i}$ such that $\costFunc_j(\phi^{\beta_i}, \phi_{-i}) \leq \alpha_j$ for all $j$ with equality for $j = i$, then w.h.p. agent $i$ can gain at most $o(T)$ more value from any deviation to another value-scaling strategy. Furthermore, when the mechanism is (IC), any deviation of any sequence reports in hindsight achieves at most $o(T)$ more value (\cref{lem: best value scale,thm:best_response_with_ic}).

    \item Finally, we show that there exists $\beta \in \R^n_+$ such that  $\costFunc_i(\phi^{\beta_1}, \dots, \phi^{\beta_n}) = \alpha_i$ using a fixed point argument in \cref{app:existence of eq}. This leverages our assumption that $V[t]$ is drawn from an absolutely continuous distribution, but can be extended to discrete distributions by adding small perturbations to the values. \\  
\end{enumerate}

Before explaining these proof steps in detail, we first provide some concrete examples of how this result helps translate existing guarantees for one-shot mechanisms into approximate core guarantees.

\subsubsection*{Pacing Equilibria}

We first consider a setting where agents are made to use value-scaling strategies: given the mechanism $\mech$, each agent $i$ can choose a $\beta_i$ upfront, but then in each round $t$, the reported value is constrained to be $\phi_i(\val_i[t]) = \val_i[t]/\beta_i$. One way to think about this is that bids are made by an `auto-bidder' which is programmed initially with a $\beta$, and then reports the scaled true value in each round.

Let $\repTotalVal = \sum_{i \in [n]}\reportedVal_{i}$ denote the total reported value in any round.
Consider the simple \emph{Proportional Mechanism} $\mechProp$: the good is allocated to all agents if the total reported value exceeds $1$ and each agent pays an amount that is proportional to the value they report, i.e., if $\repTotalVal \geq 1$, then $\alloc^*\left(\reportedVal\right) = [n]$ and  $\payments_i\left(\reportedVal\right) = \reportedVal_{i} /\repTotalVal$ for all $i \in [n]$, else $\alloc^*\left(\reportedVal\right) = \emptyset$.
It is clear $\mechProp$ is IR and CC, and, in \cref{sec:regularity}, we check that it satisfies the regularity conditions defined in \cref{app:assumptions_on_mech}. Furthermore, for all $\reportedVal \in \R^n_+$, $\dwl\left(\mechProp,\reportedVal\right)=0$ because $\mechProp$ always allocates to everyone as long as $\repTotalVal\geq 1$ and is exactly budget-balanced~\footnote{Note this does not contradict any impossibility as $\mechProp$ is not (IC); rather, we are constraining agents to report $\val_i[t]/\beta_i$.}. Hence, the non-monetary mechanism induced by $\mechProp$ satisfies the following:

\begin{theorem}[Pacing Equilibrium] \label{thm: auto-bidding equilibrium}
Consider~\cref{alg:Mon2NonMon} using $\mechProp$ as input: then there exists a profile of value-scaling strategies $(\phi^{\beta_1},\cdots,\phi^{\beta_n})$ such that $\costFunc_i(\phi^{\beta_1},\cdots,\phi^{\beta_n}) = \alpha_i$ for all $i$. Furthermore, any such strategy profile is an approximate Nash equilibrium in the class of value-scaling strategies, and the resulting ex-ante allocation is in the $(0,o(1))$-approximate core with probability at least $1-o(1)$.
\end{theorem}

\subsubsection*{Incentive Compatible Auctions}

We next consider mechanisms $\mech$ which are \emph{incentive compatible} (IC): for each agent, independent of the reports of all other agents, reporting their value truthfully maximizes their quasi-linear utility. Formally, for all $\val_i$, $\reportedVal_i$ and $\reportedVal_{-i}$, 
$$\val_i\cdot \ind{ i \in \alloc^*(\val_i, \reportedVal_{-i})} - \payments_i(\val_i, \reportedVal_{-i}) \geq \val_i\cdot \ind{ i \in \alloc^*(\reportedVal_i, \reportedVal_{-i})} - \payments_i(\reportedVal_i, \reportedVal_{-i}).$$ 
In this case, we show that the equilibrium of value-scaling strategies is still supported as equilibrium if agents could deviate to \emph{any} adaptive strategy:

\begin{theorem}\label{thm: IC mechanism equilibrium}
Assume $\mech$ is (IC). Let $\gamma = \sup_{\reportedVal}\dwl(\mech,\reportedVal)$. Then, any profile of value-scaling strategies $(\phi^{\beta_1},\cdots,\phi^{\beta_n})$ such that $\costFunc_i(\phi^{\beta_1},\cdots,\phi^{\beta_n}) = \alpha_i$ for all $i \in [n]$ is an approximate Nash equilibrium within the class of all adaptive strategies. Furthermore, the allocation resulting from agents using this profile is in the $(\gamma,o(1))$-approximate core with probability at least $1-o(1)$.
\end{theorem}

To instantiate this theorem, consider the \emph{Moulin Mechanism} \cite{moulin2001strategyproof}, $\mechMou$, in which we allocate the good to the largest collection of agents who are willing to pay the same amount for the good (summing to the cost) and have them pay that amount: 

\begin{itemize}
    \item For $\reportedVal_{i} \inR^n_+$, $\alloc^*(\reportedVal) = \argmax_{S \subset [n]} \left\{|S|: \text{For all }i \in S, \reportedVal_i \geq 1/|S| \right\}$
    \item For $\reportedVal_{i} \inR^n_+$ and $i \in \alloc^*(\reportedVal)$, $\payments_i(\reportedVal) = 1/|\alloc^*(\reportedVal)|$.
\end{itemize}
\medskip

\cite{dobzinski2017combinatorial} also introduce a VCG-style mechanism they call the \emph{Potential Mechanism} $\mechPot$ that achieves the same \newObjective bound. We make note of this mechanism because, unlike the Moulin Mechanism, the Potential Mechanism cleanly generalizes to the setting with multiple goods and submodular cost. We discuss this further in \cref{sec:extensions}.

Prior work has shown that both mechanisms are individually rational, cost-covering, and incentive compatible. In \cref{sec:regularity}, we show that they both satisfy the conditions in~\cref{app:assumptions_on_mech}. Furthermore, for all $\reportedVal \in \R^n_+$, $\dwl(\mech,\reportedVal) \leq \harm_n - 1$ for either mechanism, which we check in \cref{ssec: mechanisms for multiple items,sec:regularity} (which is the best possible, up to a constant factor, for individually rational, cost covering and incentive compatible mechanisms; this follows from \cite{dobzinski2018shapley} and \cref{lem:high_social_cost_implies_high_dwl}). 
Using either of these mechanisms in our reduction, we achieve a mechanism with an approximate Nash equilibrium in which the resulting allocation is, w.h.p., in the $(\harm_n-1, o(1))$ approximate core (\cref{cor: H_n approx result}).

%% file: Paper_Sections/Characterization_Of_Pareto_Optimality.tex
\section{The Ex-Ante Core Relaxation}
\label{sec:exantecore}

To help simplify our analysis of our mechanism and its outcomes, we first define an \emph{ex-ante} (i.e., in expectation) version of the core, and show in \cref{cor: ex ante core implies expost core} how its deviation from the ex-post allocation can be controlled. This approach parallels the use of ex-ante relaxations for item pricing~\citep{alaei2014bayesian}, which defines fractional allocations over the space of valuation profiles $\underline{\val} = \{\val_i\}_{i \in [n]} \subset [0,1]^n$. Our main result here in~\cref{lem: core-characterization} is a succinct geometric representation of ex-ante core outcomes, which then helps us translate between the core and the $\dwl$ of monetary mechanisms.

Recall agents' valuations are drawn i.i.d. from joint distribution $\valDist$ over $[0,1]^n$. As shorthand, for any set $\alloc \subset [0,1]^n$, we use $\mu(\alloc)=\Pr_{\valvec\sim\valDist}[\valvec\in \alloc]$ to be the measure of the set under $\valDist$. Now we define:
\begin{definition}[Approximate Ex-Ante Core] \label[definition]{def: ex-ante approximate core}
An ex-ante allocation policy is a collection $\alloc = \{\alloc_i\}_{i \in [n]}$, where for each agent $i\in[n]$, $\alloc_i$ is a measurable set in $[0,1]^n$ which represents the set of valuation vectors (of all agents) under which agent $i$ is included in the allocation set. 

\begin{itemize}
\item An ex-ante allocation policy $\alloc=\{\alloc_i\}_{i\in[n]}$ is said to be \emph{ex-ante feasible} if $\mu(\alloc)\triangleq \mu\left(\bigcup \alloc_i\right) \leq \alpha$. 
\item Finally, a feasible $\alloc$ is said to be in the \emph{$\gamma$-approximate ex-ante core} if it has \emph{no $\gamma$-blocking coalition}: for all $S \subset [n]$ and any measurable set $\alloc' \subset[0,1]^n$ satisfying $\mu(\alloc') \leq \sum_{i \in S}\alpha_i$:
\begin{align*}
(1+\gamma)\cdot\E_{\valvec \sim \valDist}\left[\val_i \cdot \ind{\valvec \in \alloc_i}\right] \geq \E_{\valvec \sim \valDist}\left[\val_i \cdot \ind{\valvec\in \alloc'}\right]\,\qquad\mbox{for some $i\in S$}
\end{align*}
\end{itemize}
\end{definition}
Note that to define the ex-ante blocking allocation policy for set $S$, we set the allocation set $\alloc_i'$ to be the same for all $i\in S$; this is w.l.o.g. as all valuations are non-negative. Also, unlike~\cref{def: approximate core}, we no longer need an additive term as that was needed to accommodate the deviations from expectations.

To see why the ex-ante core is useful, suppose the principal commits to allocating to agent $i$ in all rounds where the realized values are in $\alloc_i$, for some allocation policy $\{\alloc_i\}_{i \in [n]}$ in the $\gamma$-approximate ex-ante core. 
By standard concentration bounds, we expect the resulting ex-post allocation to be \textit{almost} feasible (w.h.p., over allocating by no more than $o(T)$ goods). Indeed, in \cref{thm: ex-ante core to real core}, we show that a collection of sets $\{\alloc_i\}_{i \in [n]}$ in the $\gamma$-approximate ex-ante core induces an allocation on the realized goods that is ex-post feasible and, w.h.p., in the $(\gamma, o(1))$-approximate core. 
Furthermore, any mechanism $\mech$ and strategy profile $(\phi_1, \cdots, \phi_n)$ consisting of time-independent strategies can be associated with an \emph{induced ex-ante allocation policy} $\alloc_i = \{\valvec\in [0, 1]^n: i \in \alloc^*(\phi_1(\val_1),\cdots, \phi_n(\val_n))\}$ for each $i\in[n]$; in \cref{cor: ex ante core implies expost core}, we show that if this induced allocation is in the $\gamma$-approximate ex-ante core and $\E[\payments_i(\phi_1(V_1), \dots, \phi_n(V_n)] \leq \alpha_i$, then the ex-post allocation under $\nonMonMech$ is, w.h.p., in the $(\gamma, o(1))$-approximate core. This allows us to henceforth focus on the ex-ante core, with the implicit assumption that a $\gamma$-approximate ex-ante core bound translates to a $(\gamma, o(1))$-approximate ex-post core guarantee w.h.p.

Our current ex-ante core definition, though useful in converting our online allocation problem to a static one, is not yet of much use as it is unclear what form such allocation policies take.
The main result of this section is the following characterization of which ex-ante allocation policies $\{\alloc_i\}_{i \in [n]}$ are in the $\gamma$-approximate ex-ante core:

\begin{lemma}
\label[lemma]{lem: core-characterization}
An ex-ante allocation policy $\alloc=\{\alloc_i\}_{i\in[n]}$ is in the $\gamma$-approximate ex-ante core if $\mu\left(\alloc\right) \leq \alpha$ (i.e., it is feasible) and, for all $S \subset [n]$, there exist $\beta \inR_+^n$ such that the half-space $H = \left\{x \in [0, 1]^n: \sum_{i\in S}\frac{x_i}{\beta_i} \geq 1\right\}$ satisfies $\mu(H) \geq \sum_{i \in S}\alpha_i$ and:
\begin{align}
\label{eq:exanteblock}
(1+\gamma)\cdot\sum_{i \in S}\E_{\valvec \sim \valDist}\left[ \frac{\val_i}{\beta_i} \cdot \ind{\valvec\in \alloc_i}\right] \geq \sum_{i\in S}\E_{\valvec \sim \valDist}\left[\frac{\val_i}{\beta_i} \cdot \ind{\valvec\in H}\right].
\end{align}
\end{lemma}

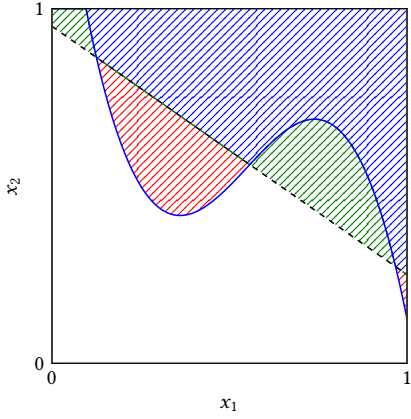
\begin{figure}[t]
\centering
    \begin{minipage}{0.35\textwidth}
    \raggedleft
        \resizebox{\textwidth}{!}{
            \input{Paper_Sections/plots/hyperplanes_are_optimal_plot}
        }
    \end{minipage}
    \hfill
    \begin{minipage}{0.55\textwidth}
        \caption{\em \small This diagram illustrates the fact that half-spaces are Pareto optimal, which is the key idea to the proof of \cref{lem: core-characterization}. Assume the region $H$ above the dashed black line (corresponding to $\{x \in [0, 1]^2: \sum_{i\in S}\frac{x_i}{\beta_i}  = 1\}$ for some $\beta$), and the region $R$ above the solid blue curve (corresponding to some ex-ante allocation policy), have equal measure. Then, the measure of the green region is equal to the measure of the red region. Furthermore, for all points in the green region $\sum_{i\in S}\frac{x_i}{\beta_i}  \geq 1$ and for all points in the red region $\sum_{i\in S}\frac{x_i}{\beta_i}  < 1$. Hence, $\E_{\valvec \sim \valDist}\left[\sum_{i\in S}\frac{x_i}{\beta_i} \cdot \ind{\valvec \in H}\right] \geq \E_{\valvec \sim \valDist}\left[\sum_{i\in S}\frac{\val_i}{\beta_i} \cdot \ind{x \in R}\right]$. Thus, $R$ cannot Pareto dominate $H$. This argument generalizes to any number of dimensions.}
        \label{fig: lines are optimal plot}
    \end{minipage}
\end{figure}

The critical insight of the above lemma is that we can reduce the problem of determining if an allocation is in the approximate ex-ante core into finding a single large half-spaces for all $S \subset [n]$ that satisfy the inequality in~\eqref{eq:exanteblock}. The intuition behind this is the fact that among all ex-ante allocation policies of measure at most $\sum_{i\in S} \alpha_i$, those corresponding to half-spaces are Pareto optimal, and in fact, just one such half-space guarantees that no set $\alloc'$ can offer at least $(1+\gamma)$ additional value to \emph{each} agent. See~\cref{fig: lines are optimal plot} for illustration of the optimality of half-spaces.
Of technical interest is the fact that the converse of this also holds true: any Pareto optimal ex-ante allocation policy must differ from some half space in a set of measure 0; we formally define and argue this in~\cref{thm: pareto optimal necessary condition characterization} using a minimax argument.

\begin{proof}[Proof of \cref{lem: core-characterization}]
By assumption $\{\alloc_i\}_{i \in [n]}$ is feasible. Consider potential blocking set $S \subset [n]$ and any set of valuations $\alloc' \subset [0, 1]^n$ with $\mu(\alloc') \leq \sum_{i \in S}\alpha_i$. Assume, for contradiction, that $$(1+\gamma)\E_{\valvec \sim \valDist}\left[\val_i \cdot \ind{\valvec \in \alloc_i}\right] < \E_{\valvec \sim \valDist}\left[\val_i \cdot \ind{\valvec\in \alloc'}\right]$$ for all $i \in S$. Then, combining with~\eqref{eq:exanteblock}, we have 
\begin{align*}
\E_{\valvec \sim \valDist}\left[\sum_{i\in S}\frac{\val_i}{\beta_i} \cdot \ind{\valvec \in H}\right] \leq (1+\gamma) \E_{\valvec \sim \valDist}\left[\sum_{i \in S}\frac{\val_i}{\beta_i}  \cdot \ind{\valvec \in \alloc_i}\right] < \E_{\valvec \sim \valDist}\left[\sum_{i\in S}\frac{\val_i}{\beta_i} \cdot \ind{\valvec \in \alloc'}\right].
\end{align*}
Recall we define the hyperplane $H$ as the set of all $\valvec \in[0,1]^n$ such that $\sum_{i\in S}\frac{\val_i}{\beta_i} \geq 1$ (and hence, for any $\valvec \not\in H$, $\sum_{i\in S}\frac{\val_i}{\beta_i} < 1$).
We now see that 
\begin{align*}
\E_{\valvec \sim \valDist}\left[\sum_{i\in S}\frac{\val_i}{\beta_i} \cdot \left( \ind{\valvec \in H} -  \ind{\valvec \in \alloc'}\right)\right] 
&= \E_{\valvec \sim \valDist}\left[\sum_{i\in S}\frac{\val_i}{\beta_i} \cdot \left(\ind{\valvec \in H \setminus \alloc'} -  \ind{\valvec \in \alloc' \setminus H}\right)\right]\\
&> \E_{\valvec \sim \valDist}\left[\ind{\valvec \in H \setminus \alloc'}\right] - \E_{\valvec \sim \valDist}\left[\ind{\valvec \in \alloc' \setminus H}\right]\\
&= \mu(H \setminus \alloc') + \mu(H \cap \alloc') - \mu(\alloc'\setminus H)-\mu(H \cap \alloc')\\
&= \mu(H) - \mu(\alloc') \geq 0
\end{align*} 
Hence, $E_{\valvec \sim \valDist}\left[\sum_{i\in S}\frac{\val_i}{\beta_i} \cdot \ind{\valvec \in H}\right] > \E_{\valvec \sim \valDist}\left[\sum_{i\in S}\frac{\val_i}{\beta_i} \cdot \ind{\valvec \in \alloc'}\right]$, a contradiction. 
\end{proof}

%% file: Paper_Sections/plots/hyperplanes_are_optimal_plot.tex
\begin{tikzpicture}
  \begin{axis}[
    axis lines=box,
    xmin=0, xmax=1,
    ymin=0, ymax=1,
    xtick={0,1}, ytick={0,1},
    width=8cm, height=8cm,
    xlabel={$x_1$}, ylabel={$x_2$},
    domain=0:1,
    samples=300,
    clip=true,
    axis on top,
    grid=both,
    legend style={at={(0.5,-0.15)}, anchor=north, cells={anchor=west}}
  ]

    \addplot [
      name path=unchanged_region,
      darkgreen,
      thick,
      dashed,
      domain=0:1,
    ] {max(min(1, max(0, 10*(1-x-.1)*(1-x-.5)*(1-x-.75) +0.5)), min(1, max(0, 0.95 - 0.7*x))};

    \addplot [
      name path=curve,
      blue,
      thick,
      solid,
      domain=0:1,
    ] {min(1, max(0, 10*(1-x-.1)*(1-x-.5)*(1-x-.75) +0.5))};
    
    \addplot [
      name path=line_shift,
      black,
      thick,
      dashed,
      domain=0:1,
    ] {min(1, max(0, 0.95 - 0.7*x)};

    \path[name path=top] (axis cs:0,1) -- (axis cs:1,1);

    \addplot [
      pattern=north east lines,
      pattern color=red,
    ] fill between [
        of=line_shift and curve,
        soft clip={domain= 0.12:0.55},
    ];

    \addplot [
      pattern=north east lines,
      pattern color=darkgreen,
    ] fill between [
        of=line_shift and curve,
        soft clip={domain= 0:0.12},
    ];

    \addplot [
      pattern=north east lines,
      pattern color=darkgreen,
    ] fill between [
        of=line_shift and curve,
        soft clip={domain= 0.55:0.96},
    ];

    \addplot [
      pattern=north east lines,
      pattern color=red,
    ] fill between [
        of=line_shift and curve,
        soft clip={domain= 0.96:1},
    ];
    
    \addplot [
      pattern=north east lines,
      pattern color=blue,
      forget plot
    ] fill between [of=top and unchanged_region];

    \draw[black, thick] (axis cs:0,0) rectangle (axis cs:1,1);

  \end{axis}
\end{tikzpicture}

%% file: Paper_Sections/Equilibrium_Outcomes.tex
\section{Equilibrium Outcomes under Value-Scaling Strategies} 
\label{sec: eq outcomes}

Having defined the ex-ante core, we are now ready to describe how we can translate between the $\dwl$ of a given monetary mechanism and the ex-ante core.
Specifically, we show that, given a suitable monetary mechanism $\mech$ and the induced non-monetary mechanism $\nonMonMech$, if agents can only use value-scaling strategies, then at any equilibrium in which agents do not spend their budget too quickly, the resulting ex-ante allocation is in the approximate core. 

\subsection{Core Approximation at Equilibrium} \label{subsec:coreApproxToEq}

Recall the definition of $\costFunc_i(\phi_1, \cdots, \phi_n)$ in \cref{def:expect cost}. We now have the following:

\begin{restatable}{theorem}{ThmConsumerCostToCore}[\newObjective to Core Approximation] \label{lem: social cost to core approx}
    Fix a monetary mechanism $\mech$ and the induced non-monetary mechanism $\nonMonMech$. 
        Set $\gamma = \sup_{\reportedVal}\dwl(\mech,\reportedVal)$.
    If agents use strategy profile $(\phi^{\beta_1}, \cdots, \phi^{\beta_n})$ such that $\costFunc_i(\phi^{\beta_1}, \cdots, \phi^{\beta_n}) = \alpha_i$ for all $i \in [n]$, the ex-ante allocation policy induced by this strategy profile is in the ex-ante $\gamma$-approximate core.
\end{restatable}

To prove this result, we show that for $\beta$ satisfying he assumptions of the theorem, \cref{lem: core-characterization} holds for the half-space $H_{z^*} = \left\{x \in [0, 1]^n: \sum_{i\in S}\frac{x_i}{\beta_i} \geq z^*\right\}$ where $z^*$ is the value such that $\mu(H_{z^*}) = \sum_{i\in S}\alpha_i$. The proof can be found in \cref{ssec:dwl_to_core}.

In the following lemma, we see that when agents use value-scaling strategies that satisfy the conditions of \cref{lem: social cost to core approx}, agents are, in fact, best responding within the class of value-scaling strategies:

\begin{restatable}{lemma}{ThmBestResponseInVS}
\label[lemma]{lem: best value scale}
    Fix a strategy profile of time-independent strategies for agents other than $i$, $\phi_{-i}$, and  $\beta \in \R_+$ such that $\costFunc_j(\phi^\beta, \phi_{-i})\leq \alpha_j$ for all $j \neq i$ and $\costFunc_i(\phi^\beta, \phi_{-i}) = \alpha_i$. Then, for any $\beta' \in \R_+$, $$\utility_i(\phi^{\beta}, \phi_{-i}, T) \geq \utility_i(\phi^{\beta'}, \phi_{-i}, T) - o(1).$$
\end{restatable}

\begin{proof}[Proof Sketch]
    By the assumptions on $\costFunc_i(\phi^\beta, \phi_{-i})$, w.h.p., no agent depletes their budget until at least round $T - o(T)$ and by round $T$ agent $i$ has spent at least $\alpha_i T - o(T)$ of their budget. We now see that increasing $\beta$ means the agent reports a smaller value in each round; hence, may lose rounds that they had previously won. Assuming they never run out budget after this deviation, we see that they gain value from at most $o(T)$ more rounds. On the other hand, decreasing $\beta$ means that the agent will pay at least as much as they paid previously for each round they win. They may also win new rounds in which they have a value per unit payment ratio of less than $\beta$ (ie. $\frac{\val_i}{\payments_i} < \beta$). Hence, the overall value per unit of payment decreases. Hence, the total value achieved when the budget of the agent depletes is reduced by this deviation (up to $o(T)$ factors). The details of this proof are given in \cref{app: proof of pace eq}.
\end{proof}

When the mechanism $\mech$ is also assumed to be incentive compatible, we can strengthen this result to show that no deviation to \emph{any adaptive strategy} is beneficial:

\begin{restatable}{lemma}{ThmBestResponsewithIC} \label[lemma]{thm:best_response_with_ic}
    Assume the mechanism $\mech$ is (IC). Fix a strategy profile of time-independent strategies for agents other than $i$, $\phi_{-i}$, and  $\beta \in \R_+$ such that $\costFunc_j(\phi^\beta, \phi_{-i}) \leq \alpha_j$ for all $j \neq i$ and $\costFunc_i(\phi^\beta, \phi_{-i}) = \alpha_i$. Then, for any adaptive strategy $\phi'_i$, $$\utility_i(\phi^{\beta}, \phi_{-i}, T) \geq \utility_i(\phi'_i, \phi_{-i}, T) - o(1).$$
\end{restatable}

\begin{proof}[Proof Sketch]
    We show that given a sample path of realized values in which no agent depletes their budget until at least round $T - o(T)$ and in which agent $i$ spends at least $\alpha_i T - o(T)$ of their budget, no alternative sequence of reported values by agent $i$ could increase their utility by more than $o(T)$. We see this as follows. Let $\reportedVal_{-i}[t]$ be the value reported by other agents in round $t$. Since other agents use time-independent strategies, these reports are not affected by agent $i$'s strategy. Let $\reportedValAlt_i[t]$ be an arbitrary sequence of alternative reports by agent $i$ (induced by some strategy). Since the mechanism is (IC), it holds that 
    $$\reportedVal_i[t]\cdot \ind{ i \in \alloc^*\left(\reportedVal_i[t], \reportedVal_{-i}[t]\right)}- \payments_i\left(\reportedVal_i[t], \reportedVal_{-i}[t]\right) \geq \reportedVal_i[t] \cdot \ind{ i \in \alloc^*\left(\reportedValAlt_i[t], \reportedVal_{-i}[t]\right)} - \payments_i\left(\reportedValAlt_i[t], \reportedVal_{-i}[t]\right)$$ where $\reportedVal_i[t] = \frac{\val_i[t]}{\beta}$. We now sum over all rounds from 1 to $T - o(T)$ and multiply through by $\frac{\beta}{T}$ to achieve
    \begin{align*}
        \frac{1}{T}\sum_{t=1}^{T - o(T)}\val_i[t] &\cdot \ind{ i \in \alloc^*\left(\reportedVal_i[t], \reportedVal_{-i}[t]\right)}- \frac{\beta}{T}\sum_{t=1}^{T - o(T)}\payments_i\left(\reportedVal_i[t], \reportedVal_{-i}[t]\right)\\
        &\geq \frac{1}{T}\sum_{t=1}^{T}\val_i[t] \cdot \ind{ i \in \alloc^*\left(\reportedValAlt_i[t], \reportedVal_{-i}[t]\right)} - \frac{\beta}{T}\sum_{t=1}^{T}\payments_i\left(\reportedValAlt_i[t], \reportedVal_{-i}[t]\right) - o(1)
    \end{align*} We now note that since we assume agent $i$ spends $\alpha_i T - o(T)$ of their budget, $\sum_{t=1}^{T - o(T)}\payments_i\left(\reportedVal_i[t], \reportedVal_{-i}[t]\right) \geq \alpha_i T - o(T)$. Assume $\sum_{t=1}^{T}\payments_i\left(\reportedValAlt_i[t], \reportedVal_{-i}[t]\right) \leq \alpha_i T$ since this is the most currency agent $i$ can spend. Hence, 
     \begin{align*}
        \utility_i(\phi^{\beta}, \phi_{-i}, T) & \geq \frac{1}{T}\sum_{t=1}^{T - o(T)}\val_i[t] \cdot \ind{ i \in \alloc^*\left(\reportedVal_i[t], \reportedVal_{-i}[t]\right)}\\
        &\geq \utility_i(\phi'_i, \phi_{-i}, T)  - \frac{\beta}{T}\sum_{t=1}^{T}\payments_i\left(\reportedValAlt_i[t], \reportedVal_{-i}[t]\right) - o(1) + \frac{\beta}{T}\sum_{t=1}^{T - o(T)}\payments_i\left(\reportedVal_i[t], \reportedVal_{-i}[t]\right)\\
        & \geq \utility_i(\phi'_i, \phi_{-i}, T)  - o(1)
    \end{align*}
    We note that this sketch is correct up to the fact that we assume that $\sum_{t=1}^{T}\payments_i\left(\reportedValAlt_i[t], \reportedVal_{-i}[t]\right) \leq \alpha_i T$ and we don't take into account other agents depleting their budget. We note that this  only helps the argument since other agents depleting their budget reduces the quasi-linear utility of the agent by the monotone assumption (See \cref{app:assumptions_on_mech}). We make this intuition concrete in the full proof of this result in \cref{thm: nash equilibrium}.
\end{proof}

Using fixed point arguments, we show in \cref{thm: equilibrium exist} that there must exist a collection $\beta_1, \cdots, \beta_n \in \R_+$ such that  $\costFunc_i(\phi^{\beta_1}, \cdots, \phi^{\beta_n}) = \alpha_i$ for all $i \in [n]$ (making use of the assumption that the value distribution is absolutely continuous); hence, a strategy profile satisfying the assumptions of \cref{lem: social cost to core approx} must exist.

Combining these results, we achieve the following guarantee on the equilibrium outcome for the non-monetary mechanism induced by either the Moulin mechanism $\mechMou$ or the Potential mechanism $\mechPot$:
\begin{corollary}[Incentive Compatible Equilibrium] \label[corollary]{cor: H_n approx result}
    Let $\nonMonMech$ be the non-monetary mechanism induced by either $\mechPot$ or $\mechMou$.  Any profile of value-scaling strategies $(\phi^{\beta_1},\cdots,\phi^{\beta_n})$ such that $\costFunc_i(\phi^{\beta_1},\cdots,\phi^{\beta_n}) = \alpha_i$ for all $i \in [n]$ (which must exist) is an approximate Nash equilibrium of adaptive strategies and the ex-ante allocation resulting from agents using this strategy profile is in the $(\harm_n-1)-$approximate ex-ante core. 
\end{corollary}

%% file: Paper_Sections/Lower_Bounds.tex
\section{Lower Bounds} \label{sec:lower_bounds}

In this section, we prove that our result in \cref{cor: H_n approx result} is tight. More generally, we prove a lower bound on the core-approximation of the induced allocation in $\nonMonMech$ for any incentive-compatible mechanism $\mech$ that additionally satisfies the following:
\begin{itemize}
    \item Equal Treatment (ET): If $\reportedVal_i = \reportedVal_j$, then $\payments_i(\reportedVal) = \payments_j(\reportedVal)$ and $i\in \alloc^*(\reportedVal)$ if and only if $ j\in\alloc^*(\reportedVal)$.
    \item Stand Alone (SA): If $\reportedVal_i \geq 1$, then $i\in\alloc^*(\reportedVal)$.
\end{itemize}

In particular, these conditions are satisfied by $\mechPot$ and $\mechMou$, the mechanisms in \cref{cor: H_n approx result}.
\begin{restatable}{theorem}{ThmGeneralLowerBound}
\label{thm: lower bound result}
Suppose the monetary mechanism $\mech$ satisfies (IC), (ET), and (SA) (in addition to the assumptions in \cref{app:assumptions_on_mech}). 
Set $\gamma = \sup_{\reportedVal}\dwl(\mech, \reportedVal)$.
Then, for every $\epsilon > 0$, there exists shares $(\alpha_1, \dots, \alpha_n)$, an absolutely continuous value distribution $\mu$, and value scaling parameters $\beta_1,\dots,\beta_n$ such that in $\nonMonMech$, when the agents use the strategy profile $(\phi^{\beta_1}, \dots, \phi^{\beta_n})$, we have   $\mathcal{C}_i(\phi^{\beta_1},\dots,\phi^{\beta_n}) = \alpha_i$, and the induced ex-ante allocation is not in the ex-ante $(\gamma-\epsilon)$-approximate core.
\end{restatable}

We prove the full theorem formally in \cref{app:lower_bounds_proofs}. To illustrate the theorem, let us prove it in the special case where $\mathcal M = \mechMou$ where we also allow $\mu$ not to be absolutely continuous for simplicity. Let each $\alpha_i = 1/(2n)$, so $\alpha = \sum_i \alpha_i = 1/2$. Let $\sigma$ be a random permutation of $[n]$, and let
\[
    \mu = \begin{cases}(1/\sigma(1) - \epsilon, 1/\sigma(2) - \epsilon, 1/\sigma(3)  - \epsilon, \dots, 1/\sigma(n) - \epsilon) & \text{with probability $\alpha$}\\e_i & \text{with probability $\alpha_i$}\end{cases}.
\]
Intuitively, each agent $i$ has a ``selfish round'' that comes up with probability $\alpha_i$ in which she has value $1$ and everyone else has value $0$. With probability $\alpha$, there will be a ``shared round'' in which every agent $i$ has value $1/\sigma(i)-\epsilon$. Suppose each agent uses $\beta_i=1$ so their reported values are just their actual values. Then, in $\nonMonMech$, none of the shared rounds will be selected since there is no value of $k$ in which there are at least $k$ agents with reported value at least $1/k$. On agent $i$'s selfish round, the mechanism will allocate the good to them and charge them $1$. Hence, each agent's expected spending and expected utility are both exactly $\alpha_i$. However, if we select the shared rounds, which have probability $\alpha$ equal to the total spending $\sum_i\alpha_i$, the total utility would be $\alpha \sum_i (1/\sigma(i) - \epsilon) = \alpha(\harm_n - n\epsilon)$. By symmetry, every agent has the same expected utility, so the expected utility of any particular agent $i$ when we take the shared rounds is $\frac1n\cdot\alpha(\harm_n - n\epsilon) = \alpha_i(\harm_n - n\epsilon)$. Therefore, each agent $i$ is doing a multiplicative factor of $\harm_n-n\epsilon$ better, so, the induced allocation in $\nonMonMech$ is not in the $(\harm_n - n\epsilon - 1)$-core, which tends to $\harm_n-1$ as $\epsilon\to0$.

Previous work has given lower bounds on how well one-shot monetary mechanisms can approximate the \textit{social cost}, where the social cost of an allocation $\alloc$ is
\begin{equation*}
    \socialCost(\alloc) = c(\alloc) + \sum_{i\notin\alloc}\val_i
\end{equation*}
In \cref{lem:high_social_cost_implies_high_dwl}, we show that a lower bound $\eta$ on the multiplicative approximation ratios of the social cost implies a lower-bound of $\eta-1$ on the \newObjective.
\citet{dobzinski2018shapley} proves that mechanism $\mech$ satisfying (IR), (CC), and (IC) does not provide a better social cost approximation than $\harm_n/16$, and additionally, if $\mech$ satisfies (ET) and has $0$ budget surplus (i.e. $\sum_{i\in[n]}\payments_i(\reportedVal) = c(\alloc^*(\reportedVal))$ for all $\reportedVal$), then $\mathcal M$ does not provide a better social cost approximation than $\harm_n$. By combining these results with \cref{thm: lower bound result}, we obtain the following.
\begin{corollary} \label{cor: lower bound result for ic mechanisms}
    Suppose the monetary mechanism $\mech$ satisfies (IC), (ET), and (SA). Then, for every $\epsilon > 0$, there exists shares $(\alpha_1,\dots,\alpha_n)$, an absolutely continuous value distribution $\mu$, and value scaling parameters $\beta_1,\dots,\beta_n$ such that in $\nonMonMech$, when the agents use the strategy profile $(\phi^{\beta_1},\dots,\phi^{\beta_n})$, we have $\mathcal C_i(\phi^{\beta_1},\dots,\phi^{\beta_n}) = \alpha_i$, and the induced ex-ante allocation is not in the ex-ante $(\harm_n/16-1-\epsilon)$-approximate core. 
    If we also assume that $\mech$ always has $0$ budget surplus, then the induced ex-ante allocation is not in the ex-ante $(\harm_n-1-\epsilon)$-approximate core.
\end{corollary}

%% file: Paper_Sections/Extensions.tex
\section{General Setting} \label{sec:extensions}
In the general setting, we show that \cref{thm:mainresult,cor: H_n approx result} still hold after appropriately generalizing the conditions.

First, we generalize \cref{def:dwl} to the general setting. Recall that given reported values $\reportedVal\in\R_n^+$ and cost function $c:2^{[n]\times [m]}$, a monetary mechanism $\mech$ outputs an allocation $\alloc^*(\reportedVal,c) \subset [n] \times [m]$ and payments $\payments(\reportedVal,c)$.
\begin{restatable}{definition}{defDwlMultiItem}[\newObjective (\dwl)] \label{def:dwl_multiitem}
    For any monetary mechanism $\mech$ which is IR and CC, given reported values $\reportedVal$ and cost $c$, the \newObjective (\dwl) is given by
\begin{equation*}
    \dwl\left(\mech, \reportedVal,c\right) = \max_{\alloc'\neq\emptyset} \frac{\left(\sum_{i\in [n]}\sum_{k\in \alloc_i'}\reportedVal_{i,k}-c(\alloc')\right)^+ - \left(\sum_{i\in[n]}\sum_{k\in\alloc_i^*(\reportedVal, c)}\reportedVal_{i,k} - \sum_{i\in[n]}\payments_i(\reportedVal,c)\right)}{c(\alloc')}.
\end{equation*}
\end{restatable}
Intuitively, this is the maximum additive difference in the social welfare between an allocation $\alloc'$ with cost $c(\alloc')$ and the mechanism's allocation $\alloc^*(\reportedVal,c)$ and payments $\payments(\reportedVal,c)$, normalized by the cost of the allocation $\alloc'$. It is easy to check that if $m=1$ and $c(\alloc)=\ind{\alloc\neq\emptyset}$, this definition is the same as \cref{def:dwl}. We detail this check in \cref{ssec: dwl in general setting}.

We now state the main result in full generality (noting that \cref{def: approximate core} of $(\gamma,\delta)$-approximate core  is the same in this setting).
\begin{theorem}
\label{thm:mainresult_extension}
\label{thm:mainresult_multiitem}
Fix a monetary mechanism $\mech$, and $\nonMonMech$ be the non-monetary mechanism induced by $\mech$. Let $\gamma = \sup_{(\reportedVal,c)}\dwl(\mech, \reportedVal,c)$. Assume that one of the following hold.
\begin{enumerate}
    \item Agents are restricted to only using value-scaling strategies.
    \item Agents can use any adaptive strategies and $\mech$ is incentive compatible.
\end{enumerate}
Then, there exists an approximate Nash equilibrium $\phi = (\phi^{\beta_1}, \dots, \phi^{\beta_n})$ where each player plays a value-scaling strategy in $\nonMonMech$ such that the resulting allocation $\allocSet$ is in the $(\gamma,o(1))$-approximate core with probability at least $1-o(1)$.
\end{theorem}

This result follows the same strategy used to prove \cref{thm:mainresult} (and the appendix proofs are written in the generality required for this result).

For an incentive-compatible mechanism, we use the Potential mechanism as given by \cite{dobzinski2017combinatorial}, which goes as follows. Given a cost function $c: 2^{[n]\times [m]} \rightarrow [0, 1]$, each agent reports a value for each good $\reportedVal_i \in \R_+^m$. Let 
$$P_{c}(\alloc) = \sum_{\emptyset\subsetneq I \subset [n]} \frac{c\left(\bigcup_{i \in I}\alloc_i \right)}{|I| \cdot \binom{n}{|I|}}.$$ 
Then, the allocation chosen by the Potential mechanism is given by
$$\alloc^*\left(\reportedVal,c\right) = \argmax_{\alloc \subset [n] \times [m]} \left\{\left(\sum_{j\in [n]}\sum_{k \in \alloc_j} \reportedVal_{j, k}\right) - P_{c}(\alloc)\right\}$$ where we choose a maximizer of greatest cardinality (allocates the most items to the most agents).  Furthermore, the agents are charges their VCG payment (externality): $$\payments_i\left(\reportedVal,c\right) = \max_{\alloc \subset ([n]-\{i\}) \times [m]}\left\{\left(\sum_{j\in [n]\setminus\{i\}}\sum_{k \in \alloc_j} \reportedVal_{j, k}\right) - P_{c}(\alloc)\right\} - \left[\left(\sum_{j\in [n]\setminus\{i\}}\sum_{k \in \alloc^*_j(\reportedVal,c)} \reportedVal_{j, k}\right) - P_{c}\left(\alloc^*\left(\reportedVal,c\right)\right)\right].$$

Being a VCG mechanism, $\mechPot$ is (IR) and (IC) and \cite{dobzinski2017combinatorial} shows that it satisfies (CC). We show that $\mechPot$ has \newObjective bounded by $\harm_n-1$ in \cref{ssec: mechanisms for multiple items} (and check the other required regularity conditions in \cref{thm: equilibrium exist}), yielding the following corollary.

\begin{corollary}
Let $\nonMonMech$ be the non-monetary mechanism induced by $\mechPot$. There exists an approximate Nash equilibrium $\phi = (\phi^{\beta_1}, \dots, \phi^{\beta_n})$ where each player plays a value-scaling strategy in $\nonMonMech$ such that the resulting allocation $\allocSet$ is in the $(\harm_n-1,o(1))$-approximate core with probability at least $1-o(1)$.
\end{corollary}

%% file: Paper_Sections/Conclusions.tex
\section{Conclusion}
Our results offer a black-box reduction from monetary mechanisms for excludable public goods to online allocation of such public goods without the use of money. We show that bounds on the \newObjective of such mechanisms directly translate into approximation bounds to the core for the online allocation problem for the focal value-scaling equilibria we study. Further, we also show that if the monetary mechanism is truthful, then this value scaling strategy remains a best response even when considering all time-independent strategies.

A possible future direction we hope to consider variants of our problem where, in each round, there are multiple goods but at most one good can be allocated to agents. The techniques in this work do not seem to extend to this case. A naive attempt could be to set the cost function $c(\alloc)$ as $\infty$ if $\alloc$ corresponds to allocating more than one good, but this cost function would not be submodular. Hence, we believe that different ideas will be needed to attain results in this proposed setting similar to the ones achieved in this paper.


%% file: Paper_Sections/Appendix/assumption_on_mechanism.tex
\section{Assumptions on One-Shot Mechanism \texorpdfstring{$\mech$}{}}
\label{app:assumptions_on_mech}

We discuss the properties we require of the monetary mechanism used in our reduction. We describe these properties in the generality of the multi-item per round and submodular cost setting. For notational convenience, throughout this appendix, we omit dependence on $c$ where clear and write $\alloc^*(\reportedVal)$ and $\payments(\reportedVal)$ for $\alloc(\reportedVal)$ and $\payments(\reportedVal)$. As we discuss in~\cref{ssec:model}, 
in this paper, we assume the mechanism $\mech$ satisfies two critical conditions for ensuring feasibility:\\

\begin{itemize}
    \item \emph{Individual Rationality (IR)}: The payment collected from an agent is no more than their reported value for the chosen allocation. Formally, for all $\reportedVal \in \R^{n \times m}_+$  and $i \in [n]$, $\payments_i\left(\reportedVal\right) \leq \sum_{k \in \alloc_i^*\left(\reportedVal\right)} \reportedVal_{i,k}$.

    \item \emph{Cost Covering (CC)}: The total payment collect from agents is at least the cost of the allocation chosen $c$. Formally, for all $\reportedVal \in \R^n_+$, $\sum_{i\in [n]}\payments_i\left(\reportedVal\right) \geq c\left(\alloc^*\left(\reportedVal\right)\right)$.\\
\end{itemize}

In addition, for our equilibrium and core-approximation guarantees, we require the following technical conditions in $\mech$, in order to rule out pathological mechanisms:

\begin{itemize}
    \item \emph{Monotone (MT)}: For any report by agents $\reportedVal \geq \reportedVal'$ and all $i \in [n]$:
\begin{enumerate}
    \item $\alloc_i^*(\reportedVal') \subset \alloc_i^*(\reportedVal)$ 
    \item $\payments_i(\reportedVal) \geq \payments_i(\reportedVal'_i, \reportedVal_{-i})$ 
    \item $\sum_{k \in \alloc^*_i(\reportedVal)} \reportedVal_{i,k} - \payments_i(\reportedVal) \geq \sum_{k \in \alloc^*_i(\reportedVal')} \reportedVal'_{i,k} - \payments_i(\reportedVal')$ 
    \item $\payments_i(\reportedVal) = \payments_i(\reportedVal'_i, \reportedVal_{-i})$ implies that $\alloc^*(\reportedVal) = \alloc^*(\reportedVal'_i, \reportedVal_{-i}) $
    \item $\alloc_i^*(\reportedVal) = \alloc_i^*(\reportedVal'_i, \reportedVal_{-i})$ implies $\alloc^*(\reportedVal) = \alloc^*\left(\reportedVal'_i, \reportedVal_{-i}\right)$ 
\end{enumerate}
    
    \item \emph{Consumer Sovereignty (CS)}: For all $i \in [n]$ and $k \in [m]$, there exists a value $\widehat\val_{i,k}$ such that, for any report by agents $\reportedVal$, $k \in \alloc_i^*(\reportedVal)$ if $\reportedVal_{i,k} \geq {\widehat\val_{i,k}}$.

    
    \item \emph{Payment Stability (PS):} Fix an agent $i$ and reports of others $\reportedVal_{-i}$. For any two reports $\reportedVal_i \leq \reportedVal_i'$  by agent $i$,
    \begin{equation*}
        \sum_{k\in \alloc^*_i(\reportedVal_i, \reportedVal_{-i})}\reportedVal_{i,k} - \payments_i(\reportedVal_i, \reportedVal_{-i}) \geq \sum_{k\in \alloc^*_i(\reportedVal_i', \reportedVal_{-i})}\reportedVal_{i,k} - \payments_i(\reportedVal_i', \reportedVal_{-i})
    \end{equation*}
    We note that this is, in a sense, a one-sided incentive compatibility condition: an agent cannot improve their quasi-linear utility by over-reporting their values.

    \item \emph{Bounded Payments (BP):} The payment functions $\payments_i(\reportedVal)$ are bounded above by an absolute constant $\payments_{\max}$.

    \item \emph{Measurable (ME)}: The allocations $\alloc^*(\reportedVal)$ and payments $\payments(\reportedVal)$ are measurable functions of $\reportedVal$.
\end{itemize}

\bigskip
We say a mechanism is regular if it satisfies these conditions. Of particular note is the monotonicity property, an immediate consequence of which is the following:

\begin{corollary}\label{cor: others say more means you pay less}
    Consider a mechanism $\mech$ that is (MT). Fix $\reportedVal$ and $\reportedVal'$ such that $\reportedVal\geq \reportedVal'$. Then $\alloc_i^*\left(\reportedVal\right) = \alloc_i^*\left(\reportedVal'_i, \reportedVal_{-i}\right)$ implies $\payments_j\left(\reportedVal\right) \leq \payments_j\left(\reportedVal'_i, \reportedVal_{-i}\right)$ for all $j \neq i$.
\end{corollary}
\begin{proof} Since the mechanism is (MT),
$$\sum_{k \in \alloc^*_j\left(\reportedVal\right)} \reportedVal_{j,k} - \payments_j\left(\reportedVal\right) \geq \sum_{k \in \alloc^*_j\left(\reportedVal'_i, \reportedVal_{-i}\right)} \reportedVal'_{j,k} - \payments_j\left(\reportedVal'_i, \reportedVal_{-i}\right).$$ By condition (5) of (MT), $\alloc^*_j\left(\reportedVal\right) =  \alloc^*_j\left(\reportedVal'_i, \reportedVal_{-i}\right)$. Hence,
$\payments_j\left(\reportedVal\right) \leq \payments_j\left(\reportedVal'_i, \reportedVal_{-i}\right)$ as claimed. 

\end{proof}

%% file: Paper_Sections/Appendix/equilibrium_existence.tex
\section{Existence of Equilibria}\label{app:existence of eq}

\begin{lemma} \label{lem:continuity_in_denominator_integral}
Let $\mu$ be an absolutely continuous measure on $[0,\infty]^n$ and let $g:[0,\infty]^n\to\mathbb R_+$ be integrable against $\mu$. Then, the function
\[
\beta \mapsto \int_{[0,\infty]^n} g\left(\frac{v_1}{\beta_1}, \dots, \frac{v_n}{\beta_n}\right)\,d\mu(v)
\]
defined for $\beta \in \mathbb R_+^n$ is continuous.
\end{lemma}
\begin{proof}
We must prove that for all $\beta'$,
\[
    \lim_{\beta\to\beta'}\left[\int_{[0,\infty]^n} g\left(\frac{v_1}{\beta_1}, \dots, \frac{v_n}{\beta_n}\right)\,d\mu(v) - \int_{[0,\infty]^n} g\left(\frac{v_1}{\beta_1'}, \dots, \frac{v_n}{\beta_n'}\right)\,d\mu(v)\right] = 0.
\]
Observe that the set of $g$ satisfying the above is closed under addition, scalar multiplication, and monotone limits. Thus, it suffices to consider the case where $g = \mathbbm1_{[a_1,b_1]\times\dots\times[a_n,b_n]}$ for some rectangle $[a_1,b_1]\times\dots\times[a_n,b_n] \subset [0,\infty]^n$. In this case,
\begin{equation} \label{eq:mu_measure_difference_indicator_g}
\begin{split}
    \int_{[0,\infty]^n}g&\left(\frac{v_1}{\beta_1}, \dots, \frac{v_n}{\beta_n}\right)\,d\mu(v) - \int_{[0,\infty]^n} g\left(\frac{v_1}{\beta_1'}, \dots, \frac{v_n}{\beta_n'}\right)\,d\mu(v)\\
    & = \mu([\beta_1a_1, \beta_1b_1]\times\dots\times[\beta_na_n,\beta_nb_n]) - \mu([\beta_1'a_1, \beta_1'b_1]\times\dots\times[\beta_n'a_n,\beta_n'b_n]).
\end{split}
\end{equation}
Denoting Lebesgue measure by $\lambda$, notice that 
\[
    \lambda([\beta_1a_1, \beta_1b_1]\times\dots\times[\beta_na_n,\beta_nb_n]) \to \lambda([\beta_1'a_1, \beta_1'b_1]\times\dots\times[\beta_n'a_n,\beta_n'b_n])
\]
as $\beta\to\beta'$. By the hypothesis that $\mu$ is absolutely continuous with respect to $\lambda$, we have that
\[
    \mu([\beta_1a_1, \beta_1b_1]\times\dots\times[\beta_na_n,\beta_nb_n]) \to \mu([\beta_1'a_1, \beta_1'b_1]\times\dots\times[\beta_n'a_n,\beta_n'b_n]),
\]
so \eqref{eq:mu_measure_difference_indicator_g} tends to $0$.
\end{proof}

\begin{lemma}\label[lemma]{lem: fixed points in continous funcitons}
    Let $g(\beta):\R^n_+ \rightarrow \R^n_+$ be a continuous function
    such that $g_i(\beta)$ is non-increasing in $\beta_i$. Let $D = [0, M_1]\times\cdots \times[0, M_n]$ for  $M_1, \cdots, M_n \in \R_+$. Let $\eta_i \in \R_+$ such that $g_i(M_i, \beta_{-i}) < \eta_i$ for all $i$ and $\beta \in D$.
    Then, there exists $\beta^* \in D$ such that, for all $i$, $$g_i(\beta^*) = \min\{\eta_i, g_i(0, \beta^*_{-i})\}.$$
\end{lemma}
\begin{proof}
    Note that $D$ is compact and convex. We define a set-valued map $\psi: D \rightarrow 2^{D}$ defined as follows: Let
     $$\psi_i(\beta) = \left\{\lambda \in \left[0, M_i\right]: g_i(\lambda, \beta_{-i}) = \min\{\eta_i, g_i(0, \beta_{-i})\} \right\}.$$ Then $\psi(\beta) = \psi_1(\beta)\times \cdots \times \psi_n(\beta)$. We note that $\psi_i(\beta)$ is non-empty for all $\beta$ since either $ g_i(0, \beta_{-i}) \leq \eta_i$ and $0 \in \psi_i(\beta)$ or $ g_i(0, \beta_{-i}) > \eta_i$ and there exist $\lambda \in [0, M_i]$ such that $g_i(\lambda, \beta_{-i}) = \eta_i$ by the Intermediate Value Theorem (since $g$ is continuous). Furthermore, since $g_i(\beta_i, \beta_{-i})$ is non-increasing in $\beta_i$, $\psi_i(\beta)$ is a closed interval. Hence, $\psi(\beta)$ is convex and non-empty for all $\beta \in D$. We now show that $\psi$ has closed graph:\\

     Let $\graph(\psi) = \{(x, y) \in  D \times D: g_i(y_i, x_{-i}) = \min\{\eta_i, g_i(0, x_{-i})\} \;\text{ for all $i \in [n]$}\}$. We now show that $\graph(\psi)^c$ is open. Consider $(u, v) \not\in \graph(\psi)$. Then for some $i$,  $|g_i(v_i, u_{-i}) - \min\{\eta_i, g_i(0, u_{-i})\}| = \Delta > 0$. Since $g_i$ is continuous on the compact set $D$, it is uniformly continuous on that set. Hence, there exists $\delta > 0$ such that if $|\beta' - \beta''| < \delta$, $|g_i(\beta') - g_i(\beta'')| < \frac{\Delta}{3}$. Consider any $(u', v') \in D \times D$ such that $|(u', v') - (u, v)| < \delta$. Then $|(v'_i, u'_{-i})-(v_i, u_{-i})| < \delta$ and $|u'_{-i} - u_{-i}| < \delta$. Hence, 
    \begin{eqnarray*}
        \left|g_i(v'_i, u'_{-i}) - \min\{\eta_i, g_i(0, u'_{-i})\}\right| &\geq& \left|g_i(v_i, u_{-i}) - \min\{\eta_i, g_i(0, u_{-i})\}\right|  -  \left|g_i(v'_i, u'_{-i}) - g_i(v_i, u_{-i})\right|\\
        &&\;\;\;\;\;-\; \left| \min\{\eta_i, g_i(0, u'_{-i})\}- \min\{\eta_i, g_i(0, u_{-i})\}\right|\\
        &\geq& \Delta  -  \frac{\Delta}{3} - \frac{\Delta}{3} > 0
    \end{eqnarray*}  
    Hence, $(u', v') \not\in \graph(\psi)$. Hence, $\graph(\psi)^c$ is open. Thus, $\psi$ has closed graph. We now apply the \emph{Kakutani Fixed Point Theorem} on $\psi$ to conclude that there exist $\beta^*$ such that $\beta^* \in \psi(\beta^*)$. We now note that $\beta^*$ satisfies the conditions of the result by the definition of $\psi$.
\end{proof}

\begin{lemma}\label[lemma]{lem: integral of continuous func is continuous}
    Let $g:\R_+^n \times \costFunctionSet \rightarrow \R_+$ be a function that is continuous as a function of the first variable for each fixed value of the second variable. Furthermore, assume $g(\beta, c) < g_{\max}$ for all $\beta, c$. Let $\mu_{\costFunctionSet}$ be a probability measure on $\costFunctionSet$. Then, the map \[
\beta \mapsto \int_{\costFunctionSet} g(\beta, c) \;d\mu_{\costFunctionSet}(c)
\]
is continuous.
\end{lemma}
\begin{proof}
Since $g(\beta, c)$ is continuous as a function of $\beta$, $\lim_{\beta' \rightarrow \beta}g(\beta', c) = g(\beta, c)$. Since $|g(\beta, c)| \leq |g_{\max}|$ and $\int_{\costFunctionSet} g_{\max} \;d\mu_{\costFunctionSet}(c) = g_{\max}$, we apply the dominated convergence theorem to deduce that
$$\lim_{\beta' \rightarrow \beta}\int_{\costFunctionSet} g(\beta', c) \;d\mu_{\costFunctionSet}(c) = \int_{\costFunctionSet} \lim_{\beta' \rightarrow \beta}g(\beta', c) \;d\mu_{\costFunctionSet}(c) = \int_{\costFunctionSet} g(\beta, c) \;d\mu_{\costFunctionSet}(c)$$ proving the claim.
\end{proof}

\begin{theorem}\label{thm: equilibrium exist}
Fix a mechanism $\mech$ that satisfies the assumption of \cref{app:assumptions_on_mech} in the general setting. Let $\costFunc_i(\beta) = \E_{(\val,c) \sim \valDist}\left[\payments_i\left(\frac{\val_1}{\beta_1}, \cdots, \frac{\val_n}{\beta_n}\right)\right]$. There exists $\beta^* \in \R^n_+$ such that:
    \begin{enumerate}
        \item for all $i \in [n]$, $\costFunc_i(\beta^*) \leq \alpha_i$
        \item for all $i \in [n]$, either $\costFunc_i(\beta^*) = \alpha_i$ or $\beta_i^* = 0$. 
    \end{enumerate}
\end{theorem}
\begin{proof}
    Fix $c \in \costFunctionSet$. Let $\mu_{\costFunctionSet}$ be the distribution measure over the cost functions. By \cref{lem: potential mech is monotone on agent}, we see that $\costFunc_i( \beta|c) = \E_{(\val,c) \sim \valDist}\left[\left.\payments_i\left(\frac{\val_1}{\beta_1}, \cdots, \frac{\val_n}{\beta_n}\right)\right|c\right]$ is non-increasing in $\beta_i$. Hence 
    $\costFunc_i( \beta) = \int_{\costFunctionSet} \costFunc_i( \beta|c) \; d\mu_{\costFunctionSet}(c)$ is non-increasing in $\beta_i$. 
    
    We now note that $\payments_i\left(\frac{\val_1}{\beta_1}, \cdots, \frac{\val_n}{\beta_n}\right)$  is bounded and continuous almost everywhere; hence, $\payments_i$ is integrable. Thus, by \cref{lem:continuity_in_denominator_integral}, $\costFunc_i( \beta|c)$ is continuous in $\beta$. Since $c \leq 1$ for all $c \in \costFunctionSet$, $\costFunc_i( \beta|c) < 1$. Hence, by \cref{lem: integral of continuous func is continuous}, $\costFunc_i(\beta)$ is continuous.

    Observe that, by (IR), $$\payments_i\left(\left.\frac{\val_1}{\beta_1}, \cdots, \frac{\val_n}{\beta_n}\right|c\right) \leq \sum_{k \in [m]} \frac{\val_{i,k}}{\beta_i} \leq \frac{m}{\beta_i}$$ for all $c$. Hence, $\costFunc_i\left(\frac{m}{ \alpha_i}, \beta_{-i}\right) \leq \alpha_i$. We now apply \cref{lem: fixed points in continous funcitons} with $g_i(\beta) = \costFunc_i( \beta)$, $M_i = \frac{m}{ \alpha_i}$ and $\eta_i = \alpha$ to achieve that there exists $\beta^* \in D = [0, M_1]\times\cdots\times [0, M_n]$ such that, for all $i$, $\costFunc_i( \beta^*) = \min\{\alpha_i, \costFunc_i(0, \beta^*_{-i})\}.$ Then, $\costFunc_i( \beta^*) \leq \alpha_i$ for all $i$.

    Consider $i$ such that $\costFunc_i(\beta^*) = \costFunc_i(0, \beta^*_{-i})$. Note that  $\payments_i\left(\frac{\val_i}{\beta^*_i}, \reportedVal_{-i}\right) \leq \payments_i\left(\infty, \reportedVal_{-i}\right)$ on all sample paths by monotonicity. Then, it must be that $\payments_i\left(\frac{\val_i}{\beta^*_i}, \reportedVal_{-i}\right) = \payments_i\left(\infty, \reportedVal_{-i}\right)$ on almost all sample paths. Hence, by condition 4 of (MT), we deduce that $\alloc^*_i\left(\left(\frac{\val_i}{\beta^*_i}, \reportedVal_{-i}\right), c\right) = \alloc_i^*\left(\left(\infty, \reportedVal_{-i}\right), c\right) = [m]$ on all most all sample paths. 
    
    By \cref{cor: others say more means you pay less}, $\payments_j\left(\frac{\val_i}{\beta^*_i}, \reportedVal_{-i}\right) \geq \payments_j\left(\infty, \reportedVal_{-i}\right)$ on all sample paths for all $j \neq i$. Hence, 
    $\costFunc_j(0, \beta^*_{-i}) \leq \costFunc_j(\beta^*) = \min\{\alpha_j, \costFunc_j(0, \beta^*_{-j})\} \leq \alpha_i$.  Let $M'_j = \beta^*_j$ for all $j \neq i$ and $M'_i = 0$. Let $D' = [0, M_1] \times \cdots \times [0, M_n]$. Then, by monotonicity, for any $\beta \in D'$, $\costFunc_j(\beta^*_j, \beta_{-j})  \leq \costFunc_j(0, \beta^*_{-i}) \leq \alpha_j$. We can now apply \cref{lem: fixed points in continous funcitons} again to achieve a $\beta^*$ such that $\beta_i^* = 0$. We repeat this argument until we achieve $\beta^*$ such that either $\costFunc_j(\beta^*) = \alpha_j$ or $\beta^* = 0$ for all $i$ as claimed.
    \end{proof}

In the special case of the single item $0$-$1$ item case, we can prove that $\beta_i^*$ as above does not have $\beta_i^*=0$ for any $i$, which gives us existence of $\beta_i^*$ such that $\mathcal C_i(\beta^*) = \alpha_i$ for each $i$, as claimed in the main text.

\begin{corollary}
In the special case where the cost function is always $c(\alloc) = \ind{\alloc\neq\emptyset}$, letting $\costFunc_i(\beta) = \E_{(\val,c) \sim \valDist}\left[\payments_i\left(\frac{\val_1}{\beta_1}, \cdots, \frac{\val_n}{\beta_n}\right)\right]$ as before, there exists $\beta_i^*\in\R_+^n$ such that $\mathcal C_i(\beta^*)=\alpha_i$ for all $i$.
\end{corollary}
\begin{proof}
Let $\beta_i^*$ be as in \cref{thm: equilibrium exist}. By contradiction, assume $\mathcal C_i(\beta^*)<\alpha_i$ for some $i$, so $\beta_i^*=0$ by the statement of \cref{thm: equilibrium exist}. Then, $V_i/\beta_i^*=\infty$ regardless of $V_i$, so by (CS), $\alloc_i^*\neq\emptyset$ with probability $1$, i.e., the single good is always allocated. Since the mechanism is cost covering, the total payments $\sum_{j\in[n]}\mathcal C_j(\beta^*)$ must be at least $1$ with probability $1$. Since each agent $j$ has budget $\alpha_j$ with $\sum_{j\in[n]}\alpha_j=1$ and $\mathcal C_j(\beta^*)\leq \alpha_j$ for each $j\in[n]$, this implies that $\mathcal C_j(\beta^*)=\alpha_j$ for each $j$, contradicting the assumption that $\mathcal C_i(\beta^*)<\alpha_i$.
\end{proof}

%% file: Paper_Sections/Appendix/ex-ante_core_relaxation_proofs.tex
\section{Ex-Ante Approximate Core to Approximate  Core Reduction}\label{app:exantecore}

For completeness, we define the generalized version of an ex-ante allocation and generalize \cref{lem: core-characterization}.

\begin{definition} \label{def: ex-ante Core extension}
    In the generalized setting, an ex-ante allocation policy $\alloc$ is a (possibly randomized) map from $[0, 1]^{n \times m} \times \costFunctionSet$ to $2^{[n] \times [m]}$, representing the set of agents given access to each good. We define the cost of an allocation $\costFunc(\alloc) = \E_{(\val, c) \sim \valDist}\left[c(\alloc(\val, c))\right]$.

    \begin{itemize}
\item An ex-ante allocation policy $\alloc$ is said to be \emph{ex-ante feasible} if $\mathcal{C}(\alloc) \leq \alpha$. 
\item A feasible $\alloc$ is said to be in the \emph{$\gamma$-approximate ex-ante core} if it has \emph{no $\gamma$-blocking coalition}: for all $S \subset [n]$ and any allocation policy $\alloc'$ satisfying $\mathcal{C}(\alloc') \leq \sum_{i \in S}\alpha_i$:
\begin{align*}
(1+\gamma)\cdot\E_{(\val,c) \sim \valDist}\left[\sum_{k\in \alloc_i(\val,c)}\val_{i,k} \right] > \E_{(\val,c) \sim \valDist}\left[\sum_{k\in \alloc'_i(\val,c)}\val_{i,k} \right] \,\qquad\mbox{for some $i\in S$} 
\end{align*}
or we have equality for all $i \in S$.
\end{itemize}
\end{definition}

We now provide a criteria for an ex-ante allocation policy $\alloc$ to be in the $\gamma$-approximate ex-ante core (analogous to \cref{lem: core-characterization}).  To do so, we define the allocation policy $\alloc^{\beta, S,z}$:
$$\alloc^{\beta, S, z}(\val, c) = \arg\max_{\alloc \subset [n]\times [m]} \left\{\sum_{i \in S}\sum_{k \in \alloc_i} \beta_i \cdot \val_{i,k} - z \cdot c\left(\alloc\right)\right\}$$ where we choose the maximizer of maximum cardinality (allocates the most good to the most agents).
Then, we have the following result:
\begin{lemma} \label{lem: ex-ante core generalized}
    Consider an ex-ante allocation policy $\alloc$. If $\costFunc(\alloc) \leq \alpha$ and, for all $S \subset [n]$, there exists $\beta \in \R^n_{> 0}$ and $z \geq 0$ such that $\costFunc(\alloc^{\beta, S,z}) \geq \sum_{i \in S} \alpha_i$ and 
    $$(1+\gamma)\cdot\E_{(\val,c) \sim \valDist}\left[\sum_{i\in S}\sum_{k\in \alloc_i(\val,c)}\beta_i \cdot\val_{i,k} \right] \geq \E_{(\val,c) \sim \valDist}\left[\sum_{i\in S}\sum_{k\in \alloc_i^{\beta, S,z}(\val,c)}\beta_i \cdot \val_{i,k} \right]$$ then $\alloc$ is in the $\gamma$-approximate ex-ante core.
\end{lemma}
\begin{proof}
    We first show that for any allocation policy $\alloc'$ such that $\costFunc(\alloc') \leq \costFunc(\alloc^{\beta, S, z})$,  
    $$\E_{(\val,c) \sim \valDist}\left[\sum_{i\in S}\sum_{k\in \alloc_i^{\beta, S, z}(\val,c)}\beta_i \cdot \val_{i,k} \right] \geq \E_{(\val,c) \sim \valDist}\left[\sum_{i\in S}\sum_{k\in \alloc'_i(\val,c)}\beta_i \cdot \val_{i,k} \right].$$
    By definition of $\alloc^{\beta, S, z}$,
    $$\sum_{i\in S}\sum_{k\in \alloc_i^{\beta, S, z}(\val,c)}\beta_i \cdot \val_{i,k} - z\cdot c\left(\alloc_i^{\beta, S, z}(\val,c)\right) \geq \sum_{i\in S}\sum_{k\in \alloc'_i(\val,c)}\beta_i \cdot \val_{i,k} - z\cdot c\left(\alloc'_i(\val,c)\right).$$ We now take an expectation and observe that $$\E_{(\val,c) \sim \valDist}\left[c\left(\alloc_i^{\beta, S, z}(\val,c)\right) - c\left(\alloc'_i(\val,c)\right)\right] = \costFunc(\alloc^{\beta, S, z}) - \costFunc(\alloc') \geq 0$$ to achieve the claim.

    We now consider $\alloc'$, a possible blocking allocation policy. By the claim, $$(1+\gamma)\cdot\E_{(\val,c) \sim \valDist}\left[\sum_{i\in S}\sum_{k\in \alloc_i(\val,c)}\beta_i \cdot\val_{i,k} \right] \geq \E_{(\val,c) \sim \valDist}\left[\sum_{i\in S}\sum_{k\in \alloc'_i(\val,c)}\beta_i \cdot \val_{i,k} \right].$$ Hence, it must hold that for some $i \in S$,
    $$(1+\gamma)\cdot\E_{(\val,c) \sim \valDist}\left[\sum_{k\in \alloc_i(\val,c)}\val_{i,k} \right] > \E_{(\val,c) \sim \valDist}\left[\sum_{k\in \alloc'_i(\val,c)}\val_{i,k} \right]$$ or we have equality for all $i \in S$, proving the result.
\end{proof}

The essential fact we use to prove \cref{lem: ex-ante core generalized} is that allocations of the form $\alloc^{\beta, S, z}$ are Pareto-optimal. We now show that inverse, that, in fact, every Pareto-optimal allocation is of this form. We first formally define a Pareto optimal allocation:

\begin{definition}\label[definition]{def: pareto-opt}
    Fix $S \subset [n]$. Consider an allocation $\alloc$. We say $\alloc$ is Pareto-dominated wrt. $S$ by another allocation (or distribution over allocations) $\alloc'$ if $\costFunc(\alloc') \leq \costFunc(\alloc)$ and 
    $$\E_{(\val,c) \sim \valDist}\left[\sum_{k\in \alloc'_i(\val,c)}\val_{i,k} \right] \geq \E_{(\val,c) \sim \valDist}\left[\sum_{k\in \alloc_i(\val,c)}\val_{i,k} \right]$$
    for all $i \in S$ with strict inequality on some $i \in S$. We say $\alloc$ is Pareto-optimal wrt. $S$ if it is not Pareto-dominated wrt. $S$ by any $\alloc'$.
\end{definition}

We now show that all Pareto-optimal allocations are of the form $\alloc^{\beta, S, z}$ almost everywhere:

\begin{theorem} \label{thm: pareto optimal necessary condition characterization}
    For any allocation $\alloc$ that is Pareto-optimal wrt. $S \subset [n]$. there exists $\beta \in \R_+^n$ and $z \geq 0$ such that $\alloc = \alloc^{\beta, S, z}$ for $\mu$-almost all $(\val, c)$.
\end{theorem}
\begin{proof}
   For all allocation $\alloc'$, let $\utility_i(\alloc') = \E_{(\val,c) \sim \valDist}\left[\sum_{k\in \alloc'_i(\val,c)}\val_{i,k} \right]$. We now show that result via a duality argument. Since $\alloc$ is Pareto-optimal, for every $\alloc'$ with $\costFunc(\alloc') \leq \costFunc(\alloc)$, there exist $\beta \in \Delta(n) \subset \R^n_+$ such that $\inner{\beta}{\utility(\alloc)}_S \geq \inner{\beta}{\utility(\alloc')}_S$. Specifically, since there exists $i \in S$ such that $\utility_i(\alloc) \geq \utility_i(\alloc)$, $\beta = e_i$ works. Let $K$ be the set of distributions over allocations $\alloc'$ such that $\costFunc(\alloc') \leq \costFunc(\alloc)$. Observe that $K$ is convex. We deduce that:
    $$\max_{\alloc' \in K} \min_{\beta \in \Delta(n)}\inner{\beta}{\utility(\alloc')}_S -\inner{\beta}{\utility(\alloc)}_S \leq 0.$$ 
    Clearly, $\Delta(n)$ is convex and compact. Furthermore, $\inner{\beta}{\utility(\alloc')}_S -\inner{\beta}{\utility(\alloc)}_S$ is continuous, convex in $\beta$ and concave in $\alloc'$. Hence, we apply Sion's Minimax Theorem to conclude that strong duality holds. Thus, 
    $$\min_{\beta \in \Delta(n)}\max_{\alloc' \in K} \inner{\beta}{\utility(\alloc')}_S -\inner{\beta}{\utility(\alloc)}_S \leq 0.$$ By the compactness of $\Delta(n)$ and the continuity of $\max_{\alloc' \in K} \inner{\beta}{\utility(\alloc')}_S -\inner{\beta}{\utility(\alloc)}_S$, there exist $\beta^*$ such that $\inner{\beta^*}{\utility(\alloc)} \geq \inner{\beta^*}{\utility(\alloc')}$ for all $\alloc'$ such that $\costFunc(\alloc') \leq \costFunc(\alloc)$. 
    
    We now select $z$ such that $\costFunc(\alloc^{\beta, S, z}) = \costFunc(\alloc)$. To see that such a $z$ must exist, observe that if otherwise, we have that $\costFunc(\alloc^{\beta, S, 0}) < \costFunc(\alloc)$. By definition $\alloc^{\beta, S, 0}_i(\val, c) = [m]$ for all $i$. Hence, $\alloc$ has a greater cost than allocating every good to every agent, which cannot be the case by the monotonicity of $c$. 
    
    We now apply the claim proven in \cref{lem: ex-ante core generalized}, we see that $\inner{\beta^*}{\utility(\alloc^{\beta, S, z})} \geq \inner{\beta^*}{\utility(\alloc)}$. Hence, $\inner{\beta^*}{\utility(\alloc^{\beta, S, z})} = \inner{\beta^*}{\utility(\alloc)}$. Furthermore, we have that $$\sum_{i \in S}\sum_{k \in \alloc^{\beta, S, z}_i(\val, c)} \beta_i \cdot \val_{i,k} - z \cdot c\left(\alloc^{\beta, S, z}(\val, c)\right) \geq \sum_{i \in S}\sum_{k \in \alloc_i(\val, c)} \beta_i \cdot \val_{i,k} - z \cdot c\left(\alloc(\val, c)\right)$$ for all $(\val, c)$. We now notice that is the inequality is strict on a set of measure greater than 0, then $\inner{\beta^*}{\utility(\alloc^{\beta, S, z})} > \inner{\beta^*}{\utility(\alloc)}$. Hence, $\alloc_i(\val, c)$ maximizes the same objective as $\alloc^{\beta, S, z}(\val, c)$ for almost all $(\val, c)$. Since out distribution is absolutely continuous, there is a unique maximizer for almost all $(\val, c)$. Hence, we conclude that $\alloc_i(\val, c) = \alloc^{\beta, S, z}(\val, c)$ for almost all $(\val, c)$ as claimed.
\end{proof}

We now show that an allocation in the ex-ante approximate core induced an ex-post allocation that is in the approximate core with vanishing additive error:

\begin{restatable}{theorem}{ThmExAnteCoreToRealCore}
\label{thm: ex-ante core to real core}
    Consider an ex-ante allocation policy $\alloc$ that satisfies the conditions of \cref{lem: ex-ante core generalized}. Let $\val[t] \in [0, 1]^{n \times m}$ and $c[t] \in \costFunctionSet$ be a sequence on $T$ i.i.d. random variables drawn from $\valDist$. Let 
    $$\tau = \max\left\{r \in [T]: \sum_{t=1}^r c[t]\left(\alloc\left(\val[t], c[t]\right)\right) \leq \alpha T\right\}$$ for constant $C$. 
    Let $\allocSet$ be the allocation such that, for $t \leq \tau$, $\alloc[t] = \alloc\left(\val[t], c[t]\right)$. Then, $\allocSet$ is in the $\left(\gamma, O\left(n\sqrt{\frac{\log T}{T}}\right)\right)$-approximate core with high probability.
\end{restatable}

\begin{proof}
    Fix $S \subset [n]$. By assumption, there exists $\beta \in \R^n_{> 0}$ and $z \geq 0$ satisfying the conditions of \cref{lem: ex-ante core generalized}. Let $$\alloc^{\beta, S, z}[t] = \arg\max_{\alloc \subset [n] \times [m]} \left\{\sum_{i \in S}\sum_{k \in \alloc_i} \beta_i \cdot \val_{i,k}[t] - z \cdot c[t]\left(\alloc\right)\right\}$$

     We now condition on the event:
     $$\left|\sum_{t=1}^{T}c[t]\left(\alloc^{\beta, S, z}[t]\right)-  T \cdot \E_{(\val,c) \sim \valDist}\left[c[t]\left(\alloc_i^{\beta, S,z}(\val,c)\right) \right]\right| = \left|\sum_{t=1}^{T}c[t]\left(\alloc^{\beta, S, z}[t]\right)-  T \cdot \costFunc\left(\alloc_i^{\beta, S,z}\right)\right|\leq \delta$$

     Consider any sequence of allocations $\alloc'[t]$ such that $\sum_{t=1}^{T}c[t]\left(\alloc'[t]\right) \leq \alpha T \leq \costFunc\left(\alloc_i^{\beta, S,z}\right)$.
     Then, by the definition of $\alloc^{\beta, S, z}$,
     \begin{align*}
         0 & \leq \left(\sum_{t=1}^T\sum_{i \in S}\sum_{k \in \alloc^{\beta, S, z}_i[t]} \beta_i \cdot \val_{i,k}[t] - z \cdot c[t]\left(\alloc^{\beta, S, z}[t]\right)\right) - \left(\sum_{t=1}^T\sum_{i \in S}\sum_{k \in \alloc'_i[t]} \beta_i \cdot \val_{i,k}[t] - z \cdot c[t]\left(\alloc'[t]\right)\right)\\
         &\;\;\;\;\;= \sum_{t=1}^T\sum_{i \in S}\sum_{k \in \alloc^{\beta, S, z}_i[t]} \beta_i \cdot \val_{i,k}[t] - \sum_{t=1}^T\sum_{i \in S}\sum_{k \in \alloc'_i[t]} \beta_i \cdot \val_{i,k}[t] - z \cdot \left(\sum_{t=1}^T c[t]\left(\alloc^{\beta, S, z}[t]\right)- \sum_{t=1}^Tc[t]\left(\alloc'[t]\right)\right)\\
         &\;\;\;\;\;\leq \sum_{t=1}^T\sum_{i \in S}\sum_{k \in \alloc^{\beta, S, z}_i[t]} \beta_i \cdot \val_{i,k}[t] - \sum_{t=1}^T\sum_{i \in S}\sum_{k \in \alloc'_i[t]} \beta_i \cdot \val_{i,k}[t] + z \cdot \delta.
     \end{align*}
     Hence, 
     $$\sum_{t=1}^T\sum_{i \in S}\sum_{k \in \alloc^{\beta, S, z}_i[t]} \beta_i \cdot \val_{i,k}[t] \geq \sum_{t=1}^T\sum_{i \in S}\sum_{k \in \alloc'_i[t]} \beta_i \cdot \val_{i,k}[t] - z \cdot \delta.$$

     We now further condition on the events
     $$\left|\sum_{t=1}^{T}\sum_{k \in \alloc^{\beta, S, z}_i[t]} \beta_i\cdot \val_{i,k}[t] - T \cdot \E_{(\val,c) \sim \valDist}\left[\sum_{k\in \alloc^{\beta, S, z}_i(\val,c)}\beta_i \cdot \val_{i,k} \right]\right| \leq \delta$$
     
     and
     $$\left|\sum_{t=1}^{T}\sum_{k \in \alloc_i[t]} \val_{i,k}[t] - T \cdot \E_{(\val,c) \sim \valDist}\left[\sum_{k\in \alloc_i(\val,c)} \val_{i,k} \right]\right| \leq \delta$$  for all $i \in S$ and the event $\tau \geq T - \delta$. Then, for any $\allocSet'$ such that $\costFunc(\allocSet') \leq \alpha T$:
    \begin{eqnarray*}
        (1+\gamma)\sum_{i \in S}\beta_i\cdot \utility_i(\allocSet, T) &=& \frac{(1+\gamma)}{T}\sum_{t = 1}^{\tau} \sum_{i \in S}\sum_{k \in \alloc_i[t]} \beta_i\cdot V_{i,k}[t]\\
        &=& \frac{(1+\gamma)}{T}\sum_{t = 1}^{T} \sum_{i \in S}\sum_{k \in \alloc_i[t]} \beta_i\cdot V_{i,k}[t] - \left(1 - \frac{\tau}{T}\right)\sum_{i\in S}\beta_i\\
        &\geq& (1+\gamma)\sum_{i \in [n]}\beta_i\cdot \E_{(\val,c) \sim \valDist}\left[\sum_{i\in S}\sum_{k\in \alloc_i(\val,c)}\beta_i \cdot \val_{i,k} \right] - \frac{\delta(2+\gamma)}{T}\sum_{i\in S} \beta_i \\
        &\geq& \E_{(\val,c) \sim \valDist}\left[\sum_{i\in S}\sum_{k\in \alloc_i^{\beta, S,z}(\val,c)}\beta_i \cdot \val_{i,k} \right] - \frac{\delta(2+\gamma)}{T}\sum_{i\in S} \beta_i\\
        &\geq& \sum_{t=1}^T\sum_{i\in S}\sum_{k\in \alloc_i^{\beta, S,z}[t]}\beta_i \cdot \val_{i,k} - \frac{\delta(3+\gamma)}{T}\sum_{i\in S} \beta_i \\
        &\geq& \sum_{t=1}^T\sum_{i\in S}\sum_{k\in \alloc'_i[t]}\beta_i \cdot \val_{i,k} -\frac{\delta}{T}\left((3+\gamma)\sum_{i\in S} \beta_i + z\right)\\
        &=& \utility_i(\allocSet', T)-\frac{\delta}{T}\left((3+\gamma)\sum_{i\in S} \beta_i + z\right)
    \end{eqnarray*} 
    Thus, $\allocSet$ is in the $\left(\gamma, C_S\cdot \frac{\delta }{T}\right)$-approximate core for $C_S = (3+\gamma)\sum_{i\in S} \beta_i + z$. Standard LLN results tell us that for $\delta = 40n\sqrt{T\log T}$, the probability of each of the events we have conditioned on failing is at most $\frac{2^{-n}}{2T^3}$. We note that there are $2^{n+1} + 2$ events to condition on simultaneously. Hence, the probability that any of them fail is at most $\frac{1}{T^3}$.
\end{proof}

\begin{corollary} \label[corollary]{cor: ex ante core implies expost core}
\label[corollary]{cor:ex_ante_to_ex_post_core_in_mechanism}
Fix monetary mechanism $\mech$ and a strategy profile $\phi$, and let $\alloc$ be the induced ex-ante allocation. If $\alloc$ is in the $\gamma$-approximate ex-ante core, and if $\costFunc(\phi) \leq \alpha_i$ for each agent $i$, then the realized allocation under $\nonMonMech$ is in the $(\gamma, o(1))$-approximate core with probability at least $1-o(1)$.
\end{corollary}
\begin{proof}
By \cref{prop:do_not_run_out_of_budget_high_probability}, $\min_j\tau_j(\phi)$, the first time an agent runs out of budget, is at least $T - o(T)$ with probability at least $1-o(1)$. The result follows from \cref{thm: ex-ante core to real core}.
\end{proof}

%% file: Paper_Sections/Appendix/equilibrium_proof_IC.tex
\section{Proofs of Approximate Equilibrium}
\begin{definition}
     Fix a profile of strategies $\phi = (\phi_1, \cdots, \phi_n)$. Let $\tau_j(\phi)$ be the stopping time when the budget of agent $j$ goes below $\payments_{\max}$ when the agents use the strategy profile $\phi$.
\end{definition}

\begin{proposition} \label[proposition]{prop:do_not_run_out_of_budget_high_probability}
    Fix a profile of strategies $\phi = (\phi_1, \cdots, \phi_n)$ such that $\costFunc_i(\phi) \leq \alpha_i$ for all $i$. Then,
    \[
    \min_{j\in [n]}\{\tau_j(\phi)\} \geq T - 2\sqrt{T}\log T
    \]
    with probability $1 - o(1)$.
\end{proposition}
\begin{proof}
    Since payments are bounded, we apply Hoeffding's inequality to see that $$\sum_{t=1}^{T - 2\sqrt{T}\log T}\payments_i(\phi_1(\val_1[t]), \cdots, \phi_n(\val_n[t])) \leq \costFunc_i(\phi)(T - 2\sqrt{T}\log T) + \alpha_i \sqrt{T}\log T \leq  \alpha_i T - \alpha_i \sqrt{T}\log T$$ with probability at least $1-\exp\left(-\frac{2\alpha_i^2 T \log^2 T}{(T-2\sqrt T\log T)\bar p^2}\right) = 1-o(1)$. Hence, with probability at least $1-o(1)$, $\sum_{t=1}^{T - 2\sqrt{T}\log T}\payments_i(\phi_1(\val_1[t]), \cdots, \phi_n(\val_n[t])) \leq  \alpha_i T$ for all $i$. Thus, 
    \[
        \min_{j\in [n]}\{\tau_j(\phi)\} \geq T - 2\sqrt{T}\log T
    \] with probability at least $1-o(1)$.
\end{proof}

Analogous to the simpler setting the majority of this paper considered, we define an adaptive strategy $\phi$ in the general setting as a sequence of maps $\phi_i[t]: (\R^{n \times m})^{t-1} \times [0, 1]^m \times \costFunctionSet \rightarrow \R_{+}^m$ where, under the strategy $\phi_i$, if the realized value of the agent and cost function in round $t$ are $\val_i[t]$ and $c[t]$ and the reports of agents in previous rounds are $\reportedVal[1], \cdots, \reportedVal[t-1]$, agent $i$ reports $\reportedVal_i = \phi_i(\reportedVal[1], \cdots, \reportedVal[t-1], \val_i[t], c[t])$. Let $\phi^\beta$ be the value scaling strategy in which $\phi^\beta(\reportedVal[1], \cdots, \reportedVal[t-1], \val_i[t], c[t]) = \phi^\beta(\val_i[t]) = \frac{\val_i[t]}{\beta}$. We say a strategy is time-independent if it does not depend on the round or the previous reports (ie. $ \phi_i(\reportedVal[1], \cdots, \reportedVal[t-1], \val_i[t], c[t]) =  \phi_i(\val_i[t], c[t])$). For ease of notion, we will omit the dependence on the specific cost function in this section.

\bigskip
\subsection{Approximate Equilibrium for IC Mechanisms}

\begin{theorem}\label{thm: nash equilibrium}
    Fix a mechanism $\mech$ that is (IC). Consider \cref{alg:Mon2NonMon} using this mechanism. Fix a profile of time-independent strategies for agents other than $i$, $\phi_{-i}$, and $\beta \in \R_+$ such that $\costFunc_j(\phi^\beta, \phi_{-i}) \leq \alpha_j$ for all $j \neq i$ and either $\costFunc_i(\phi^\beta, \phi_{-i}) = \alpha_i$ or $\beta_i = 0$ and $\costFunc_i(\phi^\beta, \phi_{-i}) \leq \alpha_i$. Then, for any adaptive strategy $\phi'_i$, $$\utility_i(\phi^{\beta}, \phi_{-i}, T) \geq \utility_i(\phi'_i, \phi_{-i}, T) - o(1).$$
\end{theorem}

We first prove the following proposition about monotone and incentive compatible mechanisms:

\begin{proposition}
\label[proposition]{prop: monotone + IC corollary}
Fix a mechanism $\mech$ that is (IC) and (MT). For any $\reportedVal$ and $\reportedVal'$ such that $\reportedVal_{-i}' \leq \reportedVal_{-i}$,
    
    $$\sum_{k \in \alloc_i(\reportedVal)} \reportedVal_{i,k} - \payments_i(\reportedVal) \geq \sum_{k \in \alloc_i(\reportedVal')} \reportedVal_{i,k} - \payments_i(\reportedVal').$$ 
\end{proposition}
    
\begin{proof}
    We deduce this result by observing that
    $$\sum_{k \in \alloc_i(\reportedVal)} \reportedVal_{i,k} - \payments_i(\reportedVal) \geq \sum_{k \in \alloc_i(\reportedVal_{i}, \reportedVal'_{i})} \reportedVal_{i,k} - \payments_i(\reportedVal_{i}, \reportedVal'_{i}) \geq \sum_{k \in \alloc_i(\reportedVal')} \reportedVal_{i,k} - \payments_i(\reportedVal'_{i})$$ where the first inequality follows from condition 3 of (MT) and the definition of an (IC) mechanism.
    
\end{proof}

We now prove \cref{thm: nash equilibrium}:

\begin{proof}[Proof of \cref{thm: nash equilibrium}]
    We first see that if $\beta_i = 0$ and $\costFunc_i(\phi^\beta, \phi_{-i}) \leq \alpha_i$, the agent receives all their possible value. Hence, has no incentive to deviate up to when they deplete their budget (which happens no earlier than round $T - o(T)$ by concentration).

    We now assume $\costFunc_i(\phi^\beta, \phi_{-i}) = \alpha_i$. Fix an arbitrarily (time-varying) strategy $\phi'_i$. We consider $\omega$, a sample path of values. Any strategy will produce a sequence of reports in each round $\reportedValAlt_i[t]$. Let $B_i[t-1]$ be the budget remaining for agent $i$ at the start of round $t$ when other agent use the strategy profile $\phi_{-i}$ and agent $i$ reports $\reportedValAlt_i[t]$ on this sample path. For a profile of possible reported values $\reportedVal$, let 
    $$\reportedVal^t = \left(\reportedVal_{1} \cdot \ind{B_1[t-1] \geq \payments_{\max}}, \cdots, \reportedVal_{n} \cdot \ind{B_n[t-1] \geq \payments_{\max}}\right).$$ 
    Hence, $\alloc_i(\reportedVal^t)$ is the realized allocation to agent $i$ by the non-monetary mechanism in round $t$ if agents report $\reportedVal$ (under strategy profile $(\phi'_i, \phi_{-i})$). (We drop the dependence on the cost function for notational convenience.) We define $\reportedVal^t_{-i}$ analogously.

    Let $\tau_j(\phi)$ be the stopping time for the round in which agent $j$ depletes their budget under the strategy profile $\phi$. Assume our sample path is such that the following hold:
    \begin{enumerate}
        \item $\tau^* = \min_{j\in [n]}\{\tau_j(\phi^\beta, \phi_{-i})\} \geq T - o(T)$
        \item Agent $i$ has spent at least $\alpha_i T - o(T)$
    \end{enumerate}
  \vspace{1em}
    
    For $t \in [T]$, let $\val[t]$ and $c[t]$ be the realized value profile and cost function in that round. Let $\reportedVal_{-i}[t]$ be reported values by the agents other than agent $i$ in rounds $t$ under the strategy $\phi_{-i}$.  Let $\reportedVal_{i}[t] = \frac{\val[t]}{\beta}$ be the reported value of agent $i$ under the strategy $\phi^\beta$ and $\reportedValAlt_{i}[t]$ be the reported value of agent $i$ under the (time-varying) strategy $\phi'$.

    We apply the fact that $\mech$ is monotone and (IC) to use \cref{prop: monotone + IC corollary} to deduce that:
    $$\sum_{k \in \alloc_i(\reportedVal_{i}[t], \reportedVal_{-i}[t])} \reportedVal_{i,k}[t] - \payments_i(\reportedVal_{i}[t], \reportedVal_{-i}[t]) \;\geq \sum_{k \in \alloc_i(\reportedValAlt_{i}[t], \reportedVal^t_{-i}[t])} \reportedVal_{i,k}[t] - \payments_i(\reportedValAlt_{i}[t], \reportedVal^t_{-i}[t]).$$ This holds since  $\reportedVal_{-i}[t] \geq \reportedVal^t_{-i}[t]$. We now observe that:
\begin{eqnarray*}
    \sum_{t =1}^{\tau^*}\sum_{k \in \alloc_i(\reportedVal_{i}[t], \reportedVal_{-i}[t])} \frac{\val_{i,k}[t]}{\beta} - \payments_i(\reportedVal_{i}[t], \reportedVal_{-i}[t]) &\geq&   \sum_{t =1}^{\tau^*}\sum_{k \in \alloc_i(\reportedValAlt_{i}[t], \reportedVal^t_{-i}[t])} \frac{\val_{i,k}[t]}{\beta} - \payments_i(\reportedValAlt_{i}[t], \reportedVal^t_{-i}[t])\\
    &\geq&  \sum_{t =1}^{T}\sum_{k \in \alloc_i(\reportedValAlt_{i}[t], \reportedVal^t_{-i}[t])} \frac{\val_{i,k}[t]}{\beta} - \payments_i(\reportedValAlt_{i}[t], \reportedVal^t_{-i}[t]) - \frac{T - \tau^*}{\beta}
\end{eqnarray*}

Since in rounds $t \leq \tau^*$ no agent has depleted their budget, we deduce that $\alloc_i(\reportedVal_{i}[t], \reportedVal_{-i}[t])$ is the allocation chosen when agent use the strategy profile $(\phi^\beta, \phi_{-i})$ and $\payments_i(\reportedVal_{i}[t], \reportedVal_{-i}[t])$ is the corresponding payment. Hence, $\frac{1}{T}\sum_{t =1}^{\tau^*}\sum_{k \in \alloc_i(\reportedVal_{i}[t], \reportedVal_{-i}[t])} \val_{i,k}[t]$ is lower bound on the per round value achieved by agent $i$ under the strategy profile $(\phi^\beta, \phi_{-i})$ on this sample path, denoted $\utility_i(\phi^{\beta}, \phi_{-i}, T|\omega)$. Furthermore, $\sum_{t =1}^{T}\sum_{k \in \alloc_i(\reportedValAlt_{i}[t], \reportedVal^t_{-i}[t])} \val_{i,k}[t]$ is exactly the value achieved by agent $i$ under the strategy profile $(\phi', \phi_{-i})$ on this sample path, denoted $\utility_i(\phi', \phi_{-i}, T|\omega)$.
Then,
\begin{eqnarray*}
    \utility_i(\phi^\beta, \phi_{-i}, T|\omega) - \utility_i(\phi', \phi_{-i}, T|\omega) &\geq&  \frac{\beta}{T}\sum_{t =1}^{\tau^*} \payments_i(\reportedVal_{i}[t], \reportedVal_{-i}[t]) - \frac{\beta}{T}\sum_{t =1}^{T} \payments_i(\reportedValAlt_{i}[t], \reportedVal^t_{-i}[t]) - \frac{T - \tau^*}{T}\\
    &\geq&  \frac{\beta}{T}\sum_{t =1}^{T} \payments_i(\reportedVal_{i}[t], \reportedVal_{-i}[t]) - \frac{\beta}{T}\sum_{t =1}^{T} \payments_i(\reportedValAlt_{i}[t], \reportedVal^t_{-i}[t]) -  o(1) - \frac{T - \tau^*}{T}\\
    &\geq& \beta\cdot \alpha_i - o(1) - \beta\cdot\alpha_i = -o(1)
\end{eqnarray*} where we use the assumptions about the sample path.  Due to \cref{prop:do_not_run_out_of_budget_high_probability} and standard concentration, the probability of a sample path satisfying assumption (1) and (2) is at least $1 - o(1)$. Hence, by conditioning appropriately and taking an expectation, we see that $\utility_i(\phi^{\beta}, \phi_{-i}, T) \geq \utility_i(\phi'_i, \phi_{-i}, T) - o(1) $ as claimed.
\end{proof}

%% file: Paper_Sections/Appendix/equilibrium_proof_Non_IC.tex
\bigskip
\subsection{Pacing Equilibrium for Non-IC Mechanisms}
\label{app: proof of pace eq}

\begin{lemma} \label{lem:value_scaling_equilibrium}
    
Fix a profile of time-independent strategies for agents other than $i$, $\phi_{-i}$, and $\beta \in \R_+$ such that $\costFunc_j(\phi^\beta, \phi_{-i}) \leq \alpha_j$ for all $j \neq i$ and either $\costFunc_i(\phi^\beta, \phi_{-i}) = \alpha_i$ or $\beta_i = 0$ and $\costFunc_i(\phi^\beta, \phi_{-i}) \leq \alpha_i$. Then, for any $\beta\in\R_+$, $$\utility_i(\phi^{\beta}, \phi_{-i}, T) \geq \utility_i(\phi_i^{\beta'}, \phi_{-i}, T) - o(1).$$
\end{lemma}

\begin{proof}
    We first see that if $\beta_i = 0$ and $\costFunc_i(\phi^\beta, \phi_{-i}) \leq \alpha_i$, the agent receives all their possible value. Hence, has no incentive to deviate up to when they deplete their budget (which happens no earlier than round $T - o(T)$ by concentration).

    Fix a mechanism $\mech$. Consider \cref{alg:Mon2NonMon} using this mechanism. Fix $\beta' \in\R$. Consider $\omega$, any sample path of values in which $\tau^* = \min_{j\in [n]}\{\tau_j(\phi^\beta, \phi_{-i})\} \geq T - o(T)$ holds and agent $i$ has spent at least $\alpha_i T - o(T)$ of their budget under the strategy profile $(\phi^\beta, \phi_{-i})$. (This occurs with probability at least $1-o(1)$ by \cref{prop:do_not_run_out_of_budget_high_probability}.) 
    
    On any round $t \leq \tau^*$, let $\reportedVal[t]$ be the values reported to the monetary mechanism in round $t$ under the strategy profile $(\phi^\beta, \phi_{-i})$. Hence, $\reportedVal_j[t] = \phi_j(\val_j[t]) \cdot \ind{t \leq \tau_j\left(\phi^{\beta}, \phi_{-i}\right)} = \phi_j(\val_j[t])$ for $j \neq i$ and $\reportedVal_i[t] = \frac{\val_i[t]}{\beta} \cdot \ind{t \leq \tau_i\left(\phi^{\beta}, \phi_{-i}\right)} = \frac{\val_i[t]}{\beta}$. 
    Similarly, let $\reportedVal'[t]$ be such that value reported to the monetary mechanism in round $t$ under the strategy profile $(\phi^{\beta'}, \phi_{-i})$. Hence, $\reportedVal'_j[t] = \phi_j(\val_j[t]) \cdot \ind{t \leq \tau_i\left(\phi^{\beta'}, \phi_{-i}\right)}$ for $j \neq i$ and $\reportedVal'_i[t] = \frac{\val_i[t]}{\beta'} \cdot \ind{t \leq \tau_i\left(\phi^{\beta'}, \phi_{-i}\right)}$.

    We first consider $\beta < \beta'$. Then $\phi^\beta(\val_i) > \phi^{\beta'}(\val_i)$ for all $\val_i \in [0, 1]^m$. Then,  by inspection, it should be clear that $\reportedVal_j[t] \geq \reportedVal'_j[t]$ for all $j \in [n]$ for $t \leq \tau^*$. Hence, by (MT), $\alloc^*\left(\reportedVal'_j[t]\right) \subset \alloc^*\left(\reportedVal_j[t]\right)$. Hence, we deduce:
    \begin{align*}
        \utility_i(\phi^\beta,\phi_{-i}|\omega) &\geq \frac{1}{T}\sum_{t=1}^{\tau^*}\sum_{k \in \alloc_i^*\left(\reportedVal[t]\right)} \val_{i,k}[t] \geq \frac{1}{T}\sum_{t=1}^{\tau^*}\sum_{k \in \alloc_i^*\left(\reportedVal'[t]\right)} \val_{i,k}[t]\\
        &\geq \frac{1}{T}\sum_{t=1}^{T} \sum_{k \in \alloc_i^*\left(\reportedVal'[t]\right)} \val_{i,k}[t] - \frac{T - \tau}{T}\\
        &= \utility_i(\phi^{\beta'},\phi_{-i}|\omega) - \frac{T - \tau}{T} = \utility_i(\phi^{\beta'},\phi_{-i}|\omega) - o(1)
    \end{align*} 
    where $\utility_i(\phi^\beta,\phi_{-i}|\omega)$ is the per round utility achieved by agent $i$ under the sample path $\omega$ when agents use strategy profile $(\phi^\beta,\phi_{-i})$. Since sample paths with the property occur with probability $1 -o(1)$, we conclude the claim for this case after taking an expectation.

 We now consider $\beta > \beta'$. Then, $\reportedVal_i[t] \leq \phi_i\left(\val'_i[t]\right) = \frac{\val'_i[t]}{\beta'} \cdot \ind{t \leq \tau_i\left(\phi^{\beta'}, \phi_{-i}\right)}$. By (PS), for $t$ such that $t \leq \tau^*$ and $t \leq \tau_i\left(\phi^{\beta'}, \phi_{-i}\right)$,
 \begin{align*}
     \sum_{k\in \alloc_i(\reportedVal[t])}\reportedVal_{i,k}[t] - \payments_i(\reportedVal[t]) \geq \sum_{k\in \alloc_i(\reportedVal_i'[t], \reportedVal_{-i}[t])}\reportedVal_{i,k}[t] - \payments_i(\reportedVal_i'[t], \reportedVal_{-i}[t]).
 \end{align*} Moreover, by (MT), noticing that  $(\reportedVal_i'[t], \reportedVal_{-i}[t]) \geq \reportedVal_i'[t]$,
 \begin{align*}
     \sum_{k\in \alloc_i(\reportedVal[t])}\reportedVal_{i,k}[t] - \payments_i(\reportedVal[t]) \geq \sum_{k\in \alloc_i(\reportedVal'[t])}\reportedVal_{i,k}[t] - \payments_i(\reportedVal'[t]).
 \end{align*}
 Furthermore, for $t$ such that $\tau_i\left(\phi^{\beta'}, \phi_{-i}\right) < t  \leq \tau^* $, by (IR),
 
 $$\sum_{k\in \alloc_i(\reportedVal[t])}\reportedVal_{i,k}[t] - \payments_i(\reportedVal[t]) \geq 0 = \sum_{k\in \alloc_i(0, \reportedVal_{-i}[t])}\reportedVal_{i,k}[t] - \payments_i(0, \reportedVal'_{-i}[t]) = \sum_{k\in \alloc_i(\reportedVal'[t])}\reportedVal_{i,k}[t] - \payments_i(\reportedVal'[t]).$$

 Hence we deduce that 
  \begin{align*}
     \sum_{t=1}^{T}\sum_{k\in \alloc_i(\reportedVal[t])}\reportedVal_{i,k}[t] -  \sum_{t=1}^{T}\payments_i(\reportedVal[t])  &\geq \sum_{t=1}^{\tau^*}\sum_{k\in \alloc_i(\reportedVal[t])}\reportedVal_{i,k}[t] - \payments_i(\reportedVal[t])\\
     &\geq \sum_{t=1}^{\tau^*}\sum_{k\in \alloc_i(\reportedVal'[t])}\reportedVal_{i,k}[t] - \payments_i(\reportedVal'[t])\\
     &\geq \sum_{t=1}^{T}\sum_{k\in \alloc_i(\reportedVal'[t])}\reportedVal_{i,k}[t] - \sum_{t=1}^{T}\payments_i(\reportedVal'[t]) - \frac{T- \tau^*}{\beta'}\\
     &\geq \sum_{t=1}^{T}\sum_{k\in \alloc_i(\reportedVal'[t])}\reportedVal_{i,k}[t] - \alpha_i T - \frac{T- \tau^*}{\beta'}.
 \end{align*}
 Hence, we deduce that 
$$ \utility_i(\phi^\beta,\phi_{-i}|\omega) = \frac{1}{T}\sum_{t=1}^{T}\sum_{k\in \alloc_i(\reportedVal[t])}\val_{i,k}[t]  \geq \frac{1}{T}\sum_{t=1}^{T}\sum_{k\in \alloc_i(\reportedVal_i'[t], \reportedVal_{-i}[t])}\val_{i,k}[t]  - o(T) = \utility_i(\phi^{\beta'},\phi_{-i}|\omega) - o(1).$$ 
Since sample paths with the property occur with probability $1 -o(1)$, we again conclude the claim for this case after taking an expectation.
\end{proof}

In the main body of the paper, we used the single-item $0$-$1$ cost scenario and claimed the following, which is a special case of the above Lemma.
\ThmBestResponseInVS*

%% file: Paper_Sections/Appendix/multi-item-dwl-and-core.tex
\section{\dwl in the General Setting and the \dwl to Core Reduction}
\subsection{Definition of Dead-Weight Loss in the General Setting}
\label{ssec: dwl in general setting}
In the single-item $0$-$1$ cost setting, we gave the following definition for \dwl.
\defDwl*
Then, in the general setting, we gave the following definition.
\defDwlMultiItem*

Below, we prove that the two notions of \dwl are the same in the single-item $0$-$1$ cost scenario.
\begin{proposition}
    When $m=1$ and $c(\alloc) = \ind{\alloc\neq\emptyset}$, then the two definitions of $\dwl$ are the same.
\end{proposition}
\begin{proof}
When $m=1$, agent $i$'s reported value for the single item is $\reportedVal_{i,1}$. We first consider the case where $\sum_{i\in [n]}\reportedVal_{i,1} < c$. In this case, by (CC), $\alloc^*(\reportedVal) = \emptyset$ and by (IR), $\sum_{i\in[n]}\payments_i(\reportedVal) = 0$, so the \dwl according to \cref{def:dwl} is $0$. Any allocation $\alloc'$ has $\sum_{i\in\alloc'}\sum_{k\in \alloc_i'}\reportedVal_{i,k} - c(\alloc') < 0$, so the \dwl according to \cref{def:dwl_multiitem} is also $0$.

Now consider the case where $\sum_{i\in[n]}\reportedVal_{i1}\geq c$. Then, one can easily see that in this single-item fixed-cost scenario, the maximizing $\alloc'$ in \cref{def:dwl_multiitem} is $\alloc'=[n]$, so the dead-weight loss according to \cref{def:dwl_multiitem} is
\begin{equation*}
\begin{split}
    \dwl(\mathcal M,\reportedVal, c) & = \left(\sum_{i\in [n]}\reportedVal_{i,1}-1\right)^+ - \left(\sum_{i\in\alloc^*(\reportedVal)}\reportedVal_{i,1} - \sum_{i\in[n]}\payments_i(\reportedVal)\right),
\end{split}
\end{equation*}
which is precisely the \dwl according to \cref{def:dwl}.
\end{proof}

\subsection{\dwl of the Potential Mechanism}
\label{ssec: mechanisms for multiple items}

Using the results of \cite{dobzinski2017combinatorial}, we can prove that $\mechPot$ (introduced in \cref{sec:extensions} has \dwl of at most $\harm_n-1$.
\begin{proposition}
    The Potential mechanism $\mechPot$ has $\dwl(\mechPot,\reportedVal,c)\leq \harm_n - 1$ for all $(\reportedVal, c)$.
\end{proposition}
\begin{proof}
By Lemma 4.4 (the main lemma for budget balance) in \cite{dobzinski2017combinatorial},
\begin{equation*}
    \sum_{i\in[n]}\payments_i(\reportedVal, c)\leq P_c(\alloc^*(V,c)).
\end{equation*}
Then, for any allocation $\alloc'\neq\emptyset$,
\begin{equation*}
\begin{split}
    & \frac{\left(\sum_{i\in[n]}\sum_{k\in\alloc'_i}\reportedVal_{i,k} - c(\alloc')\right) - \left(\sum_{i\in[n]}\sum_{k\in\alloc^*_i(\reportedVal, c)}\reportedVal_{i,k} - \sum_{i\in[n]}\payments_i(\reportedVal,c)\right)}{c(\alloc')}\\
    & \quad\leq \frac{\left(\sum_{i\in[n]}\sum_{k\in\alloc'_i}\reportedVal_{i,k} - c(\alloc')\right) - \left(\sum_{i\in[n]}\sum_{k\in\alloc^*_i(\reportedVal, c)}\reportedVal_{i,k} - P_c(\alloc^*(\reportedVal,c))\right)}{c(\alloc')}\\
    & \quad = \frac{\left(\sum_{i\in[n]}\sum_{k\in\alloc'_i}\reportedVal_{i,k} - P_c(\alloc')\right) - \left(\sum_{i\in[n]}\sum_{k\in\alloc^*_i(\reportedVal, c)}\reportedVal_{i,k} - P_c(\alloc^*(\reportedVal,c))\right) + P_c(\alloc') - c(\alloc')}{c(\alloc')}.
\end{split}
\end{equation*}
By definition of $\mechPot$, $\left(\sum_{i\in[n]}\sum_{k\in\alloc'_i}\reportedVal_{i,k} - P_c(\alloc')\right) - \left(\sum_{i\in[n]}\sum_{k\in\alloc^*_i(\reportedVal, c)}\reportedVal_{i,k} - P_c(\alloc^*(\reportedVal,c))\right)\leq 0$. Also, $P_c(\alloc') \leq \harm_n\cdot c(\alloc')$, as proved by \cite{dobzinski2017combinatorial}. Therefore, the above display is at most
\begin{equation*}
    \frac{P_c(\alloc') - c(\alloc')}{c(\alloc')}\leq \frac{\harm_n\cdot c(\alloc') - c(\alloc')}{c(\alloc')} = \harm_n-1,
\end{equation*}
implying $\dwl(\mechPot,\reportedVal,c)\leq \harm_n-1$. 
\end{proof}

\subsection{\dwl to Core Reduction} \label{ssec:dwl_to_core}
We start with the following lemma, which converts the bound on \dwl to a similar quantity when restricting to a subset $S$ of agents using (MT).
\begin{lemma} \label{lem:ndwl_to_subset}
Fix a monetary mechanism $\mech$, and set $\gamma = \sup_{(\reportedVal,c)}\dwl(\mech,\reportedVal,c)$. For any $S\subseteq[n]$, and allocation $\alloc'$ where $\alloc_i'=\emptyset$ for all $i\notin S$,
\begin{equation*}
    \left(\sum_{i\in S}\sum_{k\in\alloc_i'}\reportedVal_{i,k} - c(\alloc_i')\right) - \left(\sum_{i\in S}\sum_{k\in\alloc_i^*(\reportedVal, c)}\reportedVal_{i,k} - \sum_{i\in S}\payments_i(\reportedVal, c)\right) \leq \gamma\cdot c(\alloc').
\end{equation*}
\end{lemma}
\begin{proof}
Let $\reportedVal_S$ be the reported value profile $\reportedVal$ where the reports of agents $i\notin S$ are set to $0$. By (MT), 
\begin{equation*}
    \sum_{i\in S}\sum_{k\in\alloc_i^*(\reportedVal, c)}\reportedVal_{i,k} - \sum_{i\in S}\payments_i(\reportedVal, c) \geq \sum_{i\in S}\sum_{k\in\alloc_i^*(\reportedVal_S, c)}\reportedVal_{i,k} - \sum_{i\in S}\payments_i(\reportedVal_S, c).
\end{equation*}
By the assumption that $\alloc_i'=\emptyset$ for all $i\notin S$ and the definition of $\dwl$,
\begin{equation*}
    \left(\sum_{i\in S}\sum_{k\in\alloc_i'}\reportedVal_{i,k} - c(\alloc_i')\right) - \left(\sum_{i\in S}\sum_{k\in\alloc_i^*\left(\reportedVal_S, c\right)}\reportedVal_{i,k} - \sum_{i\in S}\payments_i\left(\reportedVal_S, c\right)\right) \leq \dwl(\mechPot, V_S, c)\cdot c(\alloc') \leq \gamma\cdot c(\alloc').
\end{equation*}
Combining the above two displays, we obtain the lemma statement.
\end{proof}

The following result is the generalization of \cref{lem: social cost to core approx} in the general setting.
\begin{theorem}[Normalized Dead-Weight Loss to Core Approximation]
\label{thm: social cost to core approx multi item}
    Fix a monetary mechanism $\mech$ and the induced non-monetary mechanism $\nonMonMech$. Set $\gamma = \sup_{(\reportedVal,c)}\dwl(\mech, \reportedVal)$. If agents use strategy profile $(\phi^{\beta_1},\dots,\phi^{\beta_n})$ such that, for all $i\in[n]$, $\mathcal C_i(\phi^{\beta_1},\dots,\phi^{\beta_n})\leq\alpha_i$ and either $\mathcal C_i(\phi^{\beta_1},\dots,\phi^{\beta_n})=\alpha_i$ or $\beta_i=0$, the ex-ante allocation policy induced by this strategy profile is in the ex-ante $\gamma$-approximate core.
\end{theorem}
\begin{proof}
    Let $\alloc$ be the ex-ante allocation policy induced by $\nonMonMech$ when agents use strategy profile $(\phi^{\beta_1}, \cdots, \phi^{\beta_n})$ (ie. for any realized value and cost $(\val, c)$, $\alloc\left(\val, c\right) = \alloc^*\left(\reportedVal, c\right)$, the allocation chosen by $\nonMonMech$ where $\reportedVal_{i,k} = \frac{\val_{i,k}}{\beta_i}$). Define $\payments_i(V, c)$ similarly.

    We now show that $\alloc$ satisfies the condition of \cref{def: ex-ante Core extension}. Fix $S \subset [n]$ and an allocation policy $\alloc'$ such that, for all $V, c$, $\alloc'_i(V, c) = \emptyset$ for $i \not\in S$ and $\costFunc(\alloc') \leq \sum_{i \in S}\alpha_i$.
    
     We first assume that $\beta_i = 0$ for some $i \in S$. Then $$\E_{(\val,c) \sim \valDist}\left[\sum_{k\in \alloc_i(\val,c)}\val_{i,k}\right] \geq \E_{(\val,c) \sim \valDist}\left[\sum_{k\in \alloc'_i(\val,c)}\val_{i,k}\right]$$ since agent $i$ receives access to every item. Hence, $$(1+\gamma) \cdot \E_{(\val,c) \sim \valDist}\left[\sum_{k\in \alloc_i(\val,c)}\val_{i,k}\right] \geq \E_{(\val,c) \sim \valDist}\left[\sum_{k\in \alloc'_i(\val,c)}\val_{i,k}\right].$$ 

     We, hence, assume henceforth that  $\costFunc_i(\phi^{\beta_1}, \cdots, \phi^{\beta_n}) = \alpha_i$ and $\beta_i > 0$ for all $i \in S$. Observe that
\begin{equation*}
\begin{split}
    \E_{(\val,c) \sim \valDist}\left[\sum_{i\in S}\sum_{k\in \alloc_i'(\val,c)} \frac{\val_{i,k}}{\beta_i}\right] & - \E_{(\val,c) \sim \valDist}\left[\sum_{i\in S}\sum_{k\in \alloc_i(\val,c)}\frac{\val_{i,k}}{\beta_i}\right]\\
    & = \E_{(\val,c) \sim \valDist}\left[\sum_{i\in S}\sum_{k\in \alloc_i'(\val,c)}\frac{\val_{i,k}}{\beta_i} - c(\alloc'(\val ,c))\right]-  \E_{(\val,c) \sim \valDist}\left[\sum_{i\in S}\sum_{k\in \alloc_i(\val,c)}\frac{\val_{i,k}}{\beta_i} - \sum_{i\in S}\payments_i(\val,c)\right]\\
    & \leq \gamma \cdot \E_{(V,c)\sim D}[c(\alloc'(V,c))] = \gamma\sum_{i\in S}\alpha_i,
\end{split}
\end{equation*}
where the second line follows from the fact that both $\E_{(V,c)\sim\mathcal D}[c(\alloc'(V,c))] = \E_{(V,c)\sim\mathcal D}[\payments_i(V,c)]$ since they are both equal to $\sum_{i\in S}\alpha_i$ by assumption, and the third line follows from \cref{lem:ndwl_to_subset}. Therefore,
\begin{align*}
(1 + \gamma )\cdot \E_{(\val,c) \sim \valDist}&\left[\sum_{i\in S}\sum_{k\in \alloc_i(\val,c)}\frac{\val_{i,k}}{\beta_i}\right] -  \E_{(\val,c) \sim \valDist}\left[\sum_{i\in S}\sum_{k\in \alloc'_i(\val,c)}\frac{\val_{i,k}}{\beta_i}\right]\\ 
& = \gamma\cdot \E_{(\val,c) \sim \valDist}\left[\sum_{i\in S}\sum_{k\in \alloc_i(\val,c)}\frac{\val_{i,k}}{\beta_i}\right] +\E_{(\val,c) \sim \valDist}\left[\sum_{i\in S}\sum_{k\in \alloc_i(\val,c)}\frac{\val_{i,k}}{\beta_i}\right] -  \E_{(\val,c) \sim \valDist}\left[\sum_{i\in S}\sum_{k\in \alloc'_i(\val,c)}\frac{\val_{i,k}}{\beta_i}\right]\\
&\geq \gamma\cdot \E_{(\val,c) \sim \valDist}\left[\sum_{i\in S}\sum_{k\in \alloc_i(\val,c)}\frac{\val_{i,k}}{\beta_i}\right] -\gamma\cdot  \sum_{i\in S}\alpha_i\\
&\geq 0,
\end{align*}
as claimed, where the first inequality follows from the previous display and the second inequality follows from the fact that the expected scaled values $V_{i,k}/\beta_i$ are at least the expected payments $\alpha_i$ by individual rationality of $\mech$. Using \cref{lem: ex-ante core generalized}, it follows from the above display that $\alloc$ is in the ex-ante $\gamma$-approximate core.

\end{proof}

Then, \cref{thm:mainresult_multiitem} and \cref{thm:mainresult} as a special case follow from \cref{thm: equilibrium exist,thm: ex-ante core to real core,thm: nash equilibrium,lem:value_scaling_equilibrium,thm: social cost to core approx multi item}

%% file: Paper_Sections/Appendix/lower_bounds_proofs.tex
\section{Proofs from \texorpdfstring{\cref{sec:lower_bounds}}{Lower Bounds Section}}\label{app:lower_bounds_proofs}
In this section, we prove the lower bound claims in \cref{sec:lower_bounds}. In this section, we take the single item, $0$-$1$ cost setting.

We start with the below lemma, which relates the \newObjective to sum payments and of excluded values.
\begin{lemma} \label{lem:relate_dwl_to_consumer_cost}
\begin{equation}
\label{eq:relate_dwl_to_consumer_cost}
\sup_{\reportedVal}\dwl(\mech, \reportedVal) + 1= \sup_{\reportedVal}\left(\sum_{i\in[n]} \payments_i(\reportedVal) + \sum_{i\notin \alloc(\reportedVal)}\reportedVal_i\right).
\end{equation}
\end{lemma}
\begin{proof}
If $\sum_{i\in[n]}\reportedVal_i \geq 1$, then
\[
     \dwl(\mech, \reportedVal) + 1 = \left(\sum_{i\notin\alloc(\reportedVal)}\reportedVal_i + \sum_{i\in[n]}\payments_i(\reportedVal) - 1\right)^+ + 1 = \sum_{i\in[n]} \payments_i(\reportedVal) + \sum_{i\notin \alloc(\reportedVal)}\reportedVal_i.
\]
If $\sum_{i\in[n]}\reportedVal_i < 1$, then by (IR) and (CC), it must be that $\alloc(\reportedVal) = \emptyset$. In this case,
\begin{equation*}
    \sum_{i\in[n]}\payments_i(\reportedVal) + \sum_{i\notin \alloc(\reportedVal)}\reportedVal_i = \sum_{i\in[n]}\reportedVal_i < 1.
\end{equation*}
and
\begin{equation*}
    \dwl(\mech, \reportedVal) = \left(\sum_{i\notin\alloc(\reportedVal)}\reportedVal_i + \sum_{i\in[n]}\payments_i(\reportedVal) - 1\right)^+ = \left(\sum_{i\in[n]}\reportedVal_i - 1\right)^+ = 0.
\end{equation*}
Notice that $\payments_i(\reportedVal) + \sum_{i\notin \alloc(\reportedVal)}\reportedVal_i$ is higher when $\sum_{i\in[n]}\reportedVal_i \geq 1$ in which case it is equal to $\dwl(\mech,\reportedVal)+1$, so we conclude \eqref{eq:relate_dwl_to_consumer_cost}.

\end{proof}

\ThmGeneralLowerBound*
\begin{proof}
Set
    \[
        \eta = \sup_{\reportedVal}\left(\sum_{i\in[n]}\payments_i(\reportedVal) + \sum_{i\notin\alloc(\reportedVal)}\reportedVal_i\right).
    \]
By \cref{lem:relate_dwl_to_consumer_cost}, $\eta = \gamma + 1$, so we must show there is a $(\eta - 1 - \epsilon)$-blocking coalition.
Note that $\eta < \infty$ by (CS) and (BP).

Find a reported value profile $\reportedVal^\star$ such that
\begin{align*}
\sum_{i = 1}^n \payments_i(\reportedVal^\star) + \sum_{i \not\in \alloc(\reportedVal^\star)}\reportedVal_i^\star > \eta - \epsilon.
\end{align*}
Define the value profile $\reportedVal'$ by
\begin{align*}
    \reportedVal_i' = \begin{cases}\payments_i(\reportedVal^\star) & \text{if $i\in\alloc(\reportedVal^\star)$}\\\reportedVal_i^\star & \text{if $i\notin\alloc(\reportedVal^\star)$}\end{cases}.
\end{align*}
By (IC), under a reported value profile $\reportedVal$, there is a threshold $p_i(\reportedVal_{-i})$ such that agent $i$ is included if and only if $\reportedVal_i \geq p_i(\reportedVal_{-i})$. By (MT), $p_i(\reportedVal_{-i})$ is nondecreasing in the other agents' reports $\reportedVal_{-i}$. We constructed $\reportedVal'$ such that every agent is either not included or paying exactly $p_i(\reportedVal_{-i}^\star)$. It follows that under any reported valuation profile $\reportedVal''$ such that $\reportedVal_i'' < \reportedVal_i'$ for all $i$, no agent gets allocated the good. By (SA), $\reportedVal_i'\leq 1$ for each $i$.

Let $\sigma$ be a uniformly random permutation of $[n]$, let $\alpha' < 1/2$, and let
\begin{equation}
\label{eq:bad_measure_ic_definition}
    \mu = (1-\epsilon)\cdot\begin{cases}(\reportedVal'_{\sigma(1)}, \dots, \reportedVal'_{\sigma(n)}) - \text{Uniform}([0,\epsilon]^n) & \text{with probability $\alpha'$}\\\text{Uniform}\left(\left[1, \frac{1}{1-\epsilon}\right]\right)e_i + \sum_{j\neq i}\text{Uniform}([0,\epsilon])e_j & \text{with probability $\frac{\alpha'}{n(1-n\epsilon)}$}\\\text{Uniform}([0,\epsilon]^n) & \text{otherwise}\end{cases}.
\end{equation}
It is readily checked that $\mu$ is a well-defined absolutely continuous measure on $[0,1]^n$ for $\epsilon$ sufficiently small (recalling that $\reportedVal_i' \leq 1$). Suppose each player $i$ uses value scaling factor $\beta = 1-\epsilon$. Set the shares as $\alpha_i = \mathcal C_i(\phi^{\beta}, \dots, \phi^{\beta})$ so that each agent spends their budget exactly in expectation. Let $\alpha = \sum_i\alpha_i$ be the total budget.

Consider a round in which no agent has run out of budget, and let $x\sim\mu$ be the values drawn in that round. Let $E$ be the $\alpha'$ probability event in the top case of \eqref{eq:bad_measure_ic_definition}, and let $E_i$ be the $\frac{\alpha'}{n(1-n\epsilon)}$ probability event in the second case of \eqref{eq:bad_measure_ic_definition}. Suppose event $E$ happens. Then, $\frac{x_i}{\beta} = \frac{x_i}{1-\epsilon} < \reportedVal_i'$ for all $i$, so no agent gets allocated the good, so every agent's payment is $0$ by individual rationality. For any agent $j$, if neither event $E$ nor event $E_j$ happens, $\frac{x_j}{\beta} = \frac{x_j}{1-\epsilon} \leq \epsilon$, so agent $j$ pays at most $\epsilon$ by individual rationality. Now fix an agent $i$. If event $E_i$ occurs, then $\frac{x_i}{\beta} = \frac{x_i}{1-\epsilon} \geq 1$. By (SA), agent $i$ is allocated the good. Since agents $j\neq i$ pay at most $\epsilon$, by cost covering, agent $i$ pays at least $1-n\epsilon$. Thus, $\alpha_i = \mathcal C_i(\phi^{\beta}, \dots, \phi^{\beta}) \geq (1-n\epsilon)\Pr(E_i) = \alpha'/n$. This implies $\alpha\geq \alpha'$.

Let $\{U_i\}_{i\in[n]}$ be the induced ex-ante allocation; that is, $U_i = \{x\in[0,1]^n:i\in\alloc(x)\}$. When event $E_i$ occurs, $\frac{x_i}{\beta} = \frac{x_i}{1-\epsilon} \leq \frac{1}{1-\epsilon} + \epsilon$. Thus,
\begin{equation}
\begin{split}
\label{eq:upper_bound_bad_utility}
    \E\left[\frac{x_i}{\beta}\ind{x\in U_i}\right] = \E\left[\frac{x_i}{\beta}\mathbbm1_E\ind{x\in U_i}\right] + \E\left[\frac{x_i}{\beta}\mathbbm 1_{E_i}\ind{x\in U_i}\right] + \E\left[\frac{x_i}{\beta}\mathbbm 1_{(E\cup E_i)^c}\ind{x\in U_i}\right]\\
     \leq 0 + \left(\frac{1}{1-\epsilon} + \epsilon\right)\Pr(E_i) + \epsilon\Pr((E\cup E_i)^c)\leq \frac{\alpha'}{n(1-\epsilon)} + 2\epsilon.
\end{split}
\end{equation}

Now consider $A = \left\{x\in[0,1]^n:\frac{x_i}{\beta} \geq \epsilon\text{ for every $i$}\right\}$. Notice that the event that $x\in A$ is a subset of the event that $E$ happens, so $\mu(A) \leq \Pr(E) = \alpha'\leq \alpha$. We have
\begin{equation}
\begin{split}
\label{eq:lower_bound_good_utility}
\E\left[\sum_{i\in[n]}\frac{x_i}{\beta}\ind{x\in A}\right] \geq \sum_{i\in[n]}\frac{x_i}{\beta}\mathbbm1_E\ind{\frac{x_i}{\beta}\geq \epsilon\;\forall i} = \alpha' \E\left[\sum_{i\in[n]}(\reportedVal_i' - \epsilon)\ind{\reportedVal_i' - \epsilon\geq \epsilon\;\forall i}\right]\\
\geq \alpha' \E\left[\sum_{i\in[n]}\reportedVal_i'\ind{\sum_{i\in[n]}\reportedVal_i' \geq 2n\epsilon}\right] - n\epsilon = \alpha'\eta  - \epsilon - n\epsilon
\end{split}
\end{equation}
where the last equality holds for $\epsilon$ sufficiently small, noting that $\sum_{i\in[n]}\reportedVal_i' > \eta - \epsilon$. Notice that by symmetry,
\[
    \E\left[\frac{x_j}{\beta}\ind{x\in A}\right] = \frac1n\cdot\E\left[\sum_{i\in[n]}\frac{x_i}{\beta}\ind{x\in A}\right] \geq \frac1n\cdot\left(\frac{\alpha'}{n}\cdot\eta-\epsilon-n\epsilon\right) = \frac{\alpha'}{n}\cdot\eta - \frac\epsilon n - \epsilon.
\]
for any particular agent $j$. It follows from the above, \eqref{eq:upper_bound_bad_utility} and \eqref{eq:lower_bound_good_utility} that
\[
    \frac{\E\left[\frac{x_j}{\beta}\ind{x\in A}\right]}{\mathbb E\left[\frac{x_j}{\beta}\ind{x\in X_j}\right]} \geq \frac{\frac{\alpha'}{n}\cdot\eta - \epsilon/n-\epsilon}{\alpha'/(n(1-\epsilon))+2\epsilon}.
\]
Thus, there is an ex-ante $\zeta$-blocking coalition for
\[
    \zeta = \frac{\frac{\alpha'}{n}\cdot\eta - \epsilon/n-\epsilon}{\alpha'/(n(1-\epsilon))+2\epsilon} - 1\to \eta -1 = \gamma
\]
as $\epsilon\to 0$.

\end{proof}

To combine the above theorem with previous lower bounds, we prove the following lemma, which implies \cref{cor: lower bound result for ic mechanisms} using the results of \cite{dobzinski2018shapley}.
\begin{lemma} \label{lem:high_social_cost_implies_high_dwl}
Suppose $\mech$ does not provide better than an $\eta$ multiplicative approximation to the social cost; that is, denoting the allocation of $\mech$ by $\alloc(\reportedVal)$,
\begin{align*}
    \sup_{\reportedVal}\frac{c(\alloc(\reportedVal)) + \sum_{i\notin\alloc(\reportedVal)}\reportedVal_i}{\min_{\alloc'}\left(c(\alloc') + \sum_{i\notin\alloc'}\reportedVal_i\right)} \geq \eta.
\end{align*}
Then,
\begin{align*}
    \sup_{\reportedVal}\dwl(\mech, \reportedVal) \geq \eta - 1.
\end{align*}
\end{lemma}
\begin{proof}
    If $\eta\leq 1$, there is nothing to prove, so assume $\eta > 1$. Let $0 < \epsilon < 1$. Find $\reportedVal$ such that
\begin{equation}
\label{eq:high_social_cost_instantiation}
\frac{c(\alloc(\reportedVal)) + \sum_{i\notin\alloc(\reportedVal)}\reportedVal_i}{\min_{\alloc'}\left(c(\alloc') + \sum_{i\notin\alloc'}\reportedVal_i\right)} > \eta-\epsilon.
\end{equation}
We must have $\sum_{i\in[n]}\reportedVal_i \geq 1$: if $\sum_{i\in[n]}\reportedVal_i < 1$, then by (CC), $\alloc(\reportedVal)=\emptyset$, so both the numerator and denominator of the left-hand side of \eqref{eq:high_social_cost_instantiation} are $\sum_{i\in[n]}\reportedVal_i$, so it cannot be greater than $\eta-\epsilon > 1$. When $\sum_{i\in[n]}\reportedVal_i \geq 1$, the denominator of the left-hand side of \eqref{eq:high_social_cost_instantiation} is $1$, so
\begin{align*}
    \dwl(\mech,\reportedVal) = \left(\sum_{i\notin\alloc(\reportedVal)}\reportedVal_i + \sum_{i\in[n]}\payments_i(\reportedVal) - 1\right)^+ \geq \sum_{i\notin\alloc(\reportedVal)}\reportedVal_i + c(\alloc(\reportedVal)) - 1 \geq \eta - \epsilon - 1 \to \eta-1.
\end{align*}
\end{proof}

%% file: Paper_Sections/Appendix/proof_of_regularity.tex
\section{Proof of Regularity} \label{sec:regularity}

In the single item, $0$-$1$ cost setting, we show that the proportional mechanism $\mechProp$ satisfies all the regularity conditions as in \cref{app:assumptions_on_mech}.

\begin{lemma}
The proportional mechanism $\mechProp$ satisfies (IR), (CC), (MT), (CS), (PS), (BP), and (ME).  
\end{lemma}
\begin{proof}
(ME) and (BP) are immediate from the definition. (CC) follows since any agent that reports a value of at least $1$ causes the good to be allocated regardless of other agents' reports. (IR) follows from the fact that an agent is only charged a nonzero amount if $\repTotalVal \geq 1$, in which case $\payments_i = \reportedVal_i / \repTotalVal \leq \reportedVal_i$. (CC) follows from the fact that if the good is allocated, then $\sum_i\payments_i(\reportedVal) = \reportedVal_i / \repTotalVal = 1$, exactly the cost of the good. 

Let us prove (MT). Let $\reportedVal\geq \reportedVal'$. The conditions 1., 2., 4., and 5. are immediate. For condition 3., if $\alloc_i(\reportedVal') = \emptyset$, then the condition is obviously satisfied. Otherwise, $\alloc_i(\reportedVal') \neq\emptyset$, so $\alloc_i(\reportedVal) \neq\emptyset$ by condition 1., and then
\[
    \reportedVal_i - \payments_i(\reportedVal) = \reportedVal_i - \frac{\reportedVal_i}{\sum_{j\in[n]}\reportedVal_j} \geq \reportedVal_i' - \frac{\reportedVal_i'}{\reportedVal_i' + \sum_{j\in[n]\setminus\{i\}}\reportedVal_j} \geq \reportedVal_i' - \frac{\reportedVal_i'}{\reportedVal_i' + \sum_{j\in[n]\setminus\{i\}}\reportedVal_j'} = \reportedVal_i' - \payments_i(\reportedVal'),
\]
using the monotonicity of the function $x\mapsto x - \frac{x}{x+y}$ over $x\geq 0$ for $y\geq 0$ for the first inequality.

To prove (PS), fix $\reportedVal_{-i}$ and consider any $\reportedVal_i \leq \reportedVal_i'$, and we must show
\[
    \reportedVal_i\ind{\alloc^*(\reportedVal_i, \reportedVal_{-i})\neq\emptyset} - \payments_i(\reportedVal_i, \reportedVal_{-i}) \geq \reportedVal_i\ind{\alloc^*(\reportedVal_i', \reportedVal_{-i})\neq\emptyset} - \payments_i(\reportedVal_i', \reportedVal_{-i}).
\]
If $\alloc^*(\reportedVal_i, \reportedVal_{-i}) = \emptyset$, then $\alloc^*(\reportedVal_i, \reportedVal_{-i}) = \emptyset$ by (MT), and $\payments_i(\reportedVal_i, \reportedVal_{-i}) = \payments_i(\reportedVal_i', \reportedVal_{-i}) = 0$ by (IR), so the inequality holds. Otherwise, if $\alloc^*(\reportedVal_i, \reportedVal_{-i})$, then \[\payments_i(\reportedVal_i, \reportedVal_{-i})  = \frac{\reportedVal_i}{\sum_{j\in[n]}\reportedVal_j} \leq \frac{\reportedVal_i'}{\reportedVal_i' + \sum_{j\in[n]\setminus\{i\}}\reportedVal_j} = \payments_i(\reportedVal_i', \reportedVal_{-i}),\] establishing the inequality.

\end{proof}

In the single item, $0$-$1$ cost setting, we now show that the Moulin mechanism $\mechMou$ is regular and (IC). It should be immediate from the definition of the mechanism this is is (IC), (IR), (CC), (CS), (PS), (BP) and (ME). We now prove that this mechanism is (MT):
\begin{lemma}
    The Moulin Mechanism is Monotone (MT).
\end{lemma}
\begin{proof}
    We prove the sub-property as follows:
    \begin{enumerate}
        \item[Prop 1:] Notice that by definition, for all $i \in \alloc^*\left(\reportedVal'\right)$, $\reportedVal_i \geq \reportedVal'_i \geq \frac{1}{\left|\alloc^*\left(\reportedVal'\right)\right|} \geq \frac{1}{\left|\alloc_i^*\left(\reportedVal'\right) \cup \alloc_i^*\left(\reportedVal\right)\right|}$. Similarly, for all $i \in \alloc^*\left(\reportedVal\right)$, $\reportedVal_i \geq \frac{1}{\left|\alloc^*\left(\reportedVal\right)\right|} \geq \frac{1}{\left|\alloc_i^*\left(\reportedVal'\right) \cup \alloc_i^*\left(\reportedVal\right)\right|}$. Hence, we conclude that there exists at least $\left|\alloc_i^*\left(\reportedVal'\right) \cup \alloc_i^*\left(\reportedVal\right)\right|$ agents such that $\reportedVal_i \geq \frac{1}{\left|\alloc_i^*\left(\reportedVal'\right) \cup \alloc_i^*\left(\reportedVal\right)\right|}$. Thus, $\left|\alloc^*\left(\reportedVal\right)\right| = \left|\alloc_i^*\left(\reportedVal'\right) \cup \alloc_i^*\left(\reportedVal\right)\right|$. Hence, $\alloc^*\left(\reportedVal'\right) \subset \alloc^*\left(\reportedVal\right)$.\\

        \item[Prop 2:] Observe that if $i \not\in \alloc^*\left(\reportedVal'_i, \reportedVal_{-i}\right)$, $\payments_i\left(\reportedVal'_i, \reportedVal_{-i}\right) = 0$ and the result is obvious. For $i \not\in \alloc^*\left(\reportedVal'_i, \reportedVal_{-i}\right)$, it should be clear that agent $i$ increasing their report cannot cause any other agents to receive the good. Hence, $\payments_i\left(\reportedVal\right) = \payments_i\left(\reportedVal'_i, \reportedVal_{-i}\right)$. \\

        \item[Prop 3:] Notice that if $i \not\in \alloc^*\left(\reportedVal'\right)$, then $ \reportedVal'_{i,k}\cdot \ind{i \in \alloc^*\left(\reportedVal'\right)} - \payments_i\left(\reportedVal'\right) = 0$; hence, the result is obvious.  Otherwise, by Prop 1, $i \in \alloc^*\left(\reportedVal\right)$. Hence, $\payments_i\left(\reportedVal'\right) = \frac{1}{|\alloc^*\left(\reportedVal'\right)|} \geq \frac{1}{|\alloc^*\left(\reportedVal\right)|} = \payments_i\left(\reportedVal\right)$. Thus, the result follows. \\
        
        \item[Prop 4:] Notice that $\payments_i\left(\reportedVal\right) = \payments_i\left(\reportedVal'_i, \reportedVal_{-i}\right)$ implies that $\left|\alloc^*_i\left(\reportedVal\right)\right| = \left|\alloc_i^*\left(\reportedVal'_i, \reportedVal_{-i}\right) \right|$. Hence, the result follows by Prop 1.\\

        \item[Prop 5:] This follows from the fact than an agent reporting a higher value cannot cause more agents to be included in the allocation if the original agent was already included.
        \end{enumerate}
\end{proof}

We now show that the Moulin mechanism achieves a $(\harm_n-1)-$approximation of deadweight-loss. This follows from the bound on ``social cost'' from \cite{roughgarden2009quantifying} and \cref{lem:relate_dwl_to_consumer_cost}, but for completeness, we give the full proof below.
\begin{lemma}
    $\dwl(\mechMou,\reportedVal) \leq \harm_n-1$
\end{lemma}
\begin{proof}
    By \cref{lem:relate_dwl_to_consumer_cost}, 
$\sup_{\reportedVal}\dwl\left(\mechMou, \reportedVal\right)= \sup_{\reportedVal}\left(\sum_{i\in[n]} \payments_i\left(\reportedVal\right) + \sum_{i\notin \alloc^*\left(\reportedVal\right)}\reportedVal_i\right) - 1$. We first consider $\reportedVal$ such that the good is allocated. Then $\payments_i\left(\reportedVal\right) = 1$. We now show that $\sum_{i\notin \alloc^*\left(\reportedVal\right)}\reportedVal_i \leq \harm_n-1$. Let $r = \left|[n]-\alloc^*\left(\reportedVal\right)\right| < n$ and assume WLOG that $[n]-\alloc^*\left(\reportedVal\right) = [r]$ and $\reportedVal_i \leq \reportedVal_{j}$ for $i < j$. We notice that $\reportedVal_1 < \frac{1}{n}$, otherwise $\reportedVal_i \geq \frac{1}{n}$ for all $i \geq 1$. Thus, the mechanism would allocate to all agents. By identical argument, we see that  $\reportedVal_2 < \frac{1}{n-1}$ and, more generally, $\reportedVal_i < \frac{1}{n-i+1}$ for all $i \leq r$. Hence, $\sum_{i\notin \alloc^*\left(\reportedVal\right)}\reportedVal_i < \sum_{i=1}^r \frac{1}{n-i+1} \leq \sum_{i=1}^{n-1} \frac{1}{n-i+1} = \harm_n-1$.

On the other hand, when the good is not allocated, the same argument show that $\sum_{i \in [n]}\reportedVal_i < \harm_n$. In both cases, $\sup_{\reportedVal}\dwl\left(\mechMou, \reportedVal\right) \leq \harm_n-1$.

\end{proof}

We now consider the general setting. Due to \cite{dobzinski2017combinatorial}, we know that the Potential mechanism is (IC), (IR) and (CC). From the definition of the mechanism, it should also be clear that it is (CS), (BP) and (ME). We now prove that this mechanism is (MT):

\begin{theorem}\label[lemma]{lem: potential mech is monotone on agent}
The Potential Mechanism with a submodular cost is monotone (MT).
\end{theorem}

We prove each condition separately as follows.

\begin{lemma}
\label[lemma]{prop: monotone subset lemma}
    Consider the Potential Mechanism $\mechPot$ with a submodular cost function. For any $\reportedVal$ and $\reportedVal'$ such that $\reportedVal \geq \reportedVal'$,
    $\alloc^*\left(\reportedVal'\right) \subset \alloc^*\left(\reportedVal\right)$.
\end{lemma}
\begin{proof}
     We consider the allocation $\hat{\alloc}$ such that $\hat{\alloc}_j = \alloc^*_j\left(\reportedVal\right) \cup \alloc^*_j\left(\reportedVal'\right)$ for $j \in [n]$. We see that
     
    \begin{eqnarray*}
        \sum_{j = 1}^n \sum_{k \in \hat{\alloc}_j} \reportedVal_{j,k} - P_c(\alloc^*) &=&  \sum_{j = 1}^n \sum_{k \in \alloc^*_j\left(\reportedVal\right)} \reportedVal_{j,k} + \sum_{j = 1}^n \sum_{k \in \alloc^*_j\left(\reportedVal'\right)/\alloc^*_j\left(\reportedVal\right)} \reportedVal_{j,k} -P_c\left(\alloc^*\left(\reportedVal'\right) \cup \alloc^*\left(\reportedVal\right)\right)\\
        &\geq&  \sum_{j = 1}^n \sum_{k \in \alloc^*_j\left(\reportedVal\right)} \reportedVal_{j,k} + \sum_{j = 1}^n \sum_{k \in \alloc^*_j\left(\reportedVal'\right)/\alloc^*_j\left(\reportedVal\right)} \reportedVal'_{j,k}\\
        &&\;\;\;\;\;-\; P_c\left(\alloc^*\left(\reportedVal'\right)\right) - P_c\left(\alloc^*\left(\reportedVal\right)\right) + P_c\left(\alloc^*\left(\reportedVal'\right) \cap \alloc^*\left(\reportedVal\right)\right)\\
        &\geq&  \left[\sum_{j = 1}^n \sum_{k \in \alloc^*_j\left(\reportedVal\right)} \reportedVal_{j,k}  - P_c\left(\alloc^*\left(\reportedVal\right)\right)\right] + \left[\sum_{j = 1}^n \sum_{k \in \alloc_j(\reportedVal')} \reportedVal'_{j,k}- P_c\left(\alloc^*\left(\reportedVal'\right)\right)\right]\\
       &&\;\;\;\;\;-\;  \left[\sum_{j = 1}^n \sum_{k \in \alloc_j(\reportedVal') \cap \alloc^*_j\left(\reportedVal\right)} \reportedVal'_{j,k} - P_c\left(\alloc^*\left(\reportedVal'\right) \cap \alloc^*\left(\reportedVal\right)\right)\right]\\
       &\geq&  \left[\sum_{j = 1}^n \sum_{k \in \alloc^*_j\left(\reportedVal\right)} \reportedVal_{j,k}  - P_c\left(\alloc^*\left(\reportedVal\right)\right)\right]
    \end{eqnarray*} 
    where we use the fact that $c$ submodular implies $P_c$ submodular (from \cite{dobzinski2017combinatorial}) and $\reportedVal \geq \reportedVal'$ to achieve the first inequality and the fact that $\alloc^*\left(\reportedVal'\right)$ maximizes $\sum_{j = 1}^n \sum_{k \in \alloc_j} \reportedVal'_{j,k}- P_c\left(\alloc\right)$. Note that $\left|\hat{\alloc}\right| > \left|\alloc^*\left(\reportedVal\right)\right|$ if $\alloc^*\left(\reportedVal'\right) \not\subset \alloc^*\left(\reportedVal\right)$. Hence, we deduce that $\alloc^*\left(\reportedVal'\right) \subset \alloc^*\left(\reportedVal\right)$ by the maximality of $\alloc^*\left(\reportedVal\right)$ with respect to size and $\sum_{j = 1}^n \sum_{k \in \alloc_j} \reportedVal_{j,k}- P_c\left(\alloc\right)$.
\end{proof}

\begin{lemma}
\label[lemma]{prop: monotone payment lemma}
    Consider the Potential Mechanism $\mechPot$ with a submodular cost function. For any $\reportedVal$ and $\reportedVal'$ such that $\reportedVal \geq \reportedVal'$,
    $\payments_i\left(\reportedVal\right) \geq \payments_i\left(\reportedVal'_i, \reportedVal_{-i}\right)$.
\end{lemma}
\begin{proof}
    By the definition of the VCG payment rule, $$\payments_i\left(\reportedVal\right) = \left[\sum_{j\neq i}\sum_{k \in \alloc^*_j\left(0, \reportedVal_{-i}\right)} \reportedVal_{j,k} - P_c\left(\alloc^*\left(0, \reportedVal_{-i}\right)\right)\right] - \left[\sum_{j\neq i}\sum_{k \in \alloc^*_j\left(\reportedVal\right)} \reportedVal_{j,k} - P_c\left(\alloc^*\left(\reportedVal\right)\right)\right]$$ 
    where we consider a maximizer in which $\alloc_i^*\left(0, \reportedVal_{-i}\right) = \emptyset$. Hence, our result reduces to showing that 

    $$\sum_{j\neq i}\sum_{k \in \alloc^*_j\left(\reportedVal'_i, \reportedVal_{-i}\right)} \reportedVal_{j,k} - P_c\left(\alloc^*\left(\reportedVal'_i, \reportedVal_{-i}\right)\right) \geq \sum_{j\neq i}\sum_{k \in \alloc^*_j\left(\reportedVal\right)} \reportedVal_{j,k} - P_c\left(\alloc^*\left(\reportedVal\right)\right).$$

    We observe that, by the definition of $\alloc^*\left(\reportedVal'_i, \reportedVal_{-i}\right)$,
    \begin{eqnarray*}
        \sum_{k \in \alloc^*_i\left(\reportedVal'_i, \reportedVal_{-i}\right)}\reportedVal'_{i,k} + \sum_{j\neq i}\sum_{k \in \alloc^*_j\left(\reportedVal'_i, \reportedVal_{-i}\right)} \reportedVal_{j,k} - P_c\left(\alloc^*\left(\reportedVal'_i, \reportedVal_{-i}\right)\right) &\geq& \sum_{k \in \alloc^*_i\left(\reportedVal\right)}\reportedVal'_{i,k} + \sum_{j\neq i}\sum_{k \in \alloc^*_j\left(\reportedVal\right)} \reportedVal_{j,k} - P_c\left(\alloc^*\left(\reportedVal\right)\right)\\
        &\geq& \sum_{k \in \alloc^*_i\left(\reportedVal'_i, \reportedVal_{-i}\right)}\reportedVal'_{i,k} + \sum_{j\neq i}\sum_{k \in \alloc^*_j\left(\reportedVal\right)} \reportedVal_{j,k} - P_c\left(\alloc^*\left(\reportedVal\right)\right)
    \end{eqnarray*}
    where we use the result that  $\alloc^*_i\left(\reportedVal'_i, \reportedVal_{-i}\right) \subset \alloc^*_i\left(\reportedVal\right)$. The result then follows by subtracting $\sum_{k \in \alloc^*_i\left(\reportedVal'_i, \reportedVal_{-i}\right)}\reportedVal'_{i,k}$ from both sides of the inequality.
    
\end{proof}

\begin{lemma}
\label{lem: monotone mechanism extension}
    Consider the Potential Mechanism $\mechPot$ with a submodular cost function. For any $\reportedVal$ and $\reportedVal'$ such that $\reportedVal \geq \reportedVal'$, 
    $$\sum_{k \in \alloc^*_i\left(\reportedVal\right)} \reportedVal_{i,k} - \payments_i\left(\reportedVal\right) \geq \sum_{k \in \alloc^*_i\left(\reportedVal'\right)} \reportedVal'_{i,k} - \payments_i\left(\reportedVal'\right).$$ 
\end{lemma}
\begin{proof}
    We begin by showing the intermediate result,
    $$\sum_{k \in \alloc^*_i\left(\reportedVal\right)} \reportedVal_{i,k} - \payments_i\left(\reportedVal\right) \geq \sum_{k \in \alloc^*_i\left(\reportedVal_i, \reportedVal'_{-i}\right)} \reportedVal_{i,k} - \payments_i\left(\reportedVal_i, \reportedVal'_{-i}\right).$$

    We first note that $\alloc^*_i\left(0, \reportedVal'_{-i}\right) \subset \alloc^*_i\left(0, \reportedVal_{-i}\right) \cap \alloc^*_i\left(\reportedVal'\right)$ holds immediately from \cref{prop: monotone subset lemma} since $\left(0, \reportedVal'_{-i}\right) \leq \left(0, \reportedVal_{-i}\right)$ and $\left(0, \reportedVal'_{-i}\right) \leq \reportedVal'$. By the definition of the VCG payment rule, 
    $$\payments_i\left(\reportedVal\right) = \sum_{j\neq i}\sum_{k \in \alloc^*_j\left(0, \reportedVal_{-i}\right)} \reportedVal_{j,k} - P_c\left(\alloc^*\left(0, \reportedVal_{-i}\right)\right) - \sum_{j\neq i}\sum_{k \in \alloc^*_j\left(\reportedVal\right)} \reportedVal_{j,k} + P_c\left(\alloc^*\left(\reportedVal\right)\right)$$ where we consider a maximizer in which $\alloc^*_i\left(0, \reportedVal_{-i}\right) = \emptyset$. 
    We now observe that

    \begin{eqnarray*}
        \sum_{k \in \alloc^*\left(\reportedVal\right)} \reportedVal_{i,k} - \payments_i\left(\reportedVal\right) &=& \sum_{k \in \alloc^*_i\left(\reportedVal\right)} \reportedVal_{i,k} - \left[\sum_{j\neq i}\sum_{k \in \alloc^*_j\left(0, \reportedVal_{-i}\right)} \reportedVal_{j,k} - P_c\left(\alloc^*\left(0, \reportedVal_{-i}\right)\right) - \sum_{j\neq i}\sum_{k \in \alloc^*_j\left(\reportedVal\right)} \reportedVal_{j,k} + P_c\left(\alloc^*\left(\reportedVal\right)\right)\right]\\
        &=& \left[\sum_{k \in \alloc^*_i\left(\reportedVal\right)} \reportedVal_{i,k} + \sum_{j\neq i}\sum_{k \in \alloc^*_j\left(\reportedVal\right)/\alloc^*_j\left(0, \reportedVal_{-i}\right)} \reportedVal_{j,k}\right] - \left[P_c\left(\alloc^*\left(\reportedVal\right)\right) - P_c\left(\alloc^*\left(0, \reportedVal_{-i}\right)\right) \right]\\
        &=& \left[\sum_{j=1}^n\sum_{k \in \alloc^*_j\left(\reportedVal\right)/\alloc^*_j\left(0, \reportedVal_{-i}\right)} \reportedVal_{j,k}\right] - \left[P_c\left(\alloc^*\left(\reportedVal\right)\right) - P_c\left(\alloc^*\left(0, \reportedVal_{-i}\right)\right) \right]
    \end{eqnarray*}
    Then,
    \begin{eqnarray*}
        \left[\sum_{k \in \alloc^*_i\left(\reportedVal\right)} \reportedVal_{i,k} - \payments_i\left(\reportedVal\right)\right]  &=& \left[\sum_{k \in \alloc^*_i\left(\reportedVal'\right)} \reportedVal'_{i,k} - \payments_i\left(\reportedVal'\right)\right] + \left[\sum_{j=1}^n\sum_{k \in \alloc^*_j\left(\reportedVal\right)/\alloc^*_j\left(0, \reportedVal_{-i}\right)} \reportedVal_{j,k}\right] - \left[P_c\left(\alloc^*\left(\reportedVal\right)\right) - P_c\left(\alloc^*\left(0, \reportedVal_{-i}\right)\right) \right]\\
        &&\;\;\;\;\;-\; \left[\sum_{j=1}^n\sum_{k \in \alloc_j^*\left(\reportedVal'\right)/\alloc^*_j\left(0, \reportedVal'_{-i}\right)} \reportedVal'_{j,k}\right] +\left[P_c\left(\alloc^*\left(\reportedVal'\right)\right) - P_c\left(\alloc^*\left(0, \reportedVal'_{-i}\right)\right) \right].
    \end{eqnarray*}
    Rearranging, we achieve
    \begin{eqnarray*}
        \left[\sum_{k \in \alloc^*_i\left(\reportedVal\right)} \reportedVal_{i,k} - \payments_i\left(\reportedVal\right)\right]  &=& \left[\sum_{k \in \alloc_i^*\left(\reportedVal'\right)} \reportedVal'_{i,k} - \payments_i\left(\reportedVal'\right)\right] + \left[\sum_{j=1}^n\sum_{k \in \alloc_j^*\left(\reportedVal\right)} \reportedVal_{j,k} -P_c\left(\alloc^*\left(\reportedVal\right)\right) \right] - \left[\sum_{j=1}^n\sum_{k \in \alloc^*_j\left(0, \reportedVal_{-i}\right)} \reportedVal_{j,k}\right] \\
        &&\;\;\;\;\;-\; \left[\sum_{j=1}^n\sum_{k \in \alloc^*_j\left(\reportedVal'\right) \cap \alloc^*_j\left(0,\reportedVal_{-i}\right)} \reportedVal'_{j,k}\right] +\left[\sum_{j=1}^n\sum_{k \in \alloc_j^*\left(0, \reportedVal'_{-i}\right)} \reportedVal'_{j,k}\right] - \left[\sum_{j=1}^n\sum_{k \in \alloc^*_j\left(\reportedVal'\right) /\alloc^*_j\left(0,\reportedVal_{-i}\right)} \reportedVal'_{j,k}\right]\\
        &&\;\;\;\;\;+\; \left[P_c\left(\alloc^*\left(\reportedVal'\right)\right)  + P_c\left(\alloc^*\left(0, \reportedVal_{-i}\right)\right) - P_c\left(\alloc^*\left(0, \reportedVal'_{-i}\right)\right) \right] .
    \end{eqnarray*}\\

    where we use the fact that $\alloc^*\left(0, \reportedVal'_{-i}\right) \subset \alloc^*\left(\reportedVal'\right)$. Since $P_c$ is submodular, $$P_c\left(\alloc^*\left(\reportedVal'\right)\right) + P_c\left(\alloc^*\left(0, \reportedVal_{-i}\right)\right) \geq P_c\left(\alloc^*\left(\reportedVal'\right) \cup \alloc^*\left(0, \reportedVal_{-i}\right)\right) + P_c\left(\alloc^*\left(\reportedVal'\right) \cap \alloc^*\left(0, \reportedVal_{-i}\right)\right).$$

    Furthermore, since $\reportedVal' \leq \reportedVal$, $$\sum_{j=1}^n\sum_{k \in \alloc_j^*\left(\reportedVal'\right) / \alloc_j^*\left(0,\reportedVal_{-i}\right)} \reportedVal'_{j,k} \leq \sum_{j=1}^n\sum_{k \in \alloc_j^*\left(\reportedVal'\right) / \alloc_j^*\left(0,\reportedVal_{-i}\right)} \reportedVal_{j,k}.$$

    Applying these inequalities and combining terms, we achieve 
    \begin{align*}
        &\left[\sum_{k \in \alloc^*_i\left(\reportedVal\right)} \reportedVal_{i,k} - \payments_i\left(\reportedVal\right)\right]  \geq \left[\sum_{k \in \alloc^*_i\left(\reportedVal'\right)} \reportedVal'_{i,k} - \payments_i\left(\reportedVal'\right)\right] \\
        &\;\;\;\;\;\;\;\;\;\;+\;  \left[\sum_{j=1}^n\sum_{k \in \alloc^*_j\left(\reportedVal\right)} \reportedVal_{j,k} -P_c\left(\alloc^*\left(\reportedVal\right)\right) \right]   - \left[\sum_{j=1}^n\sum_{k \in \alloc^*_j\left(0, \reportedVal_{-i}\right)\cup \alloc^*_j\left(\reportedVal'\right)} \reportedVal_{j,k}- P_c\left(\alloc^*\left(\reportedVal'\right) \cup \alloc^*\left(0,\reportedVal_{-i}\right)\right)\right]\\
        &\;\;\;\;\;\;\;\;\;\;+\; \left[\sum_{j=1}^n\sum_{k \in \alloc^*_j\left(0, \reportedVal'_{-i}\right)} \reportedVal'_{j,k}- P_c\left(\alloc^*\left(0, \reportedVal'_{-i}\right)\right)\right] - \left[\sum_{j=1}^n\sum_{k \in \alloc^*_j\left(\reportedVal'\right) \cap \alloc^*_j\left(0,\reportedVal_{-i}\right)} \reportedVal'_{j,k} - P_c\left(\alloc^*\left(\reportedVal'\right) \cap \alloc^*\left(0,\reportedVal_{-i}\right)\right)\right]
    \end{align*}

We now note that $$\left[\sum_{j=1}^n\sum_{k \in \alloc^*\left(\reportedVal\right)} \reportedVal_{j,k} -P_c\left(\alloc^*\left(\reportedVal\right)\right) \right]   \geq \left[\sum_{j=1}^n\sum_{k \in \alloc^*\left(0, \reportedVal_{-i}\right)\cup \alloc^*\left(\reportedVal'\right)} \reportedVal_{j,k}- P_c(\alloc^*\left(\reportedVal'\right) \cup \alloc(0,\reportedVal_{-i}))\right]$$ by the maximility of $\alloc^*\left(\reportedVal\right)$. Furthermore, we see that $\alloc^*_i\left(0, \reportedVal'_{-i}\right) = \emptyset$ and $\alloc^*_i\left(\reportedVal'\right) \cap \alloc^*_i\left(0,\reportedVal_{-i}\right)= \emptyset$. Since $\alloc^*_i\left(0, \reportedVal'_{-i}\right)$ maximizes  
$\sum_{j\neq i}\sum_{k \in \alloc_j} \reportedVal_{j,k} -P_c(\alloc)$, 
$$\left[\sum_{j=1}^n\sum_{k \in \alloc^*_j\left(0, \reportedVal'_{-i}\right)} \reportedVal'_{j,k}- P_c\left(\alloc^*\left(0, \reportedVal'_{-i}\right)\right)\right] \geq \left[\sum_{j=1}^n\sum_{k \in \alloc^*_j\left(\reportedVal'\right) \cap \alloc^*_j\left(0,\reportedVal_{-i}\right)} \reportedVal'_{j,k} - P_c\left(\alloc^*\left(\reportedVal'\right) \cap \alloc^*\left(0,\reportedVal_{-i}\right)\right)\right].$$
Hence, we conclude the intermediate result. To prove the main result, we observe that 
$$\sum_{k \in \alloc^*_i\left(\reportedVal\right)} \reportedVal_{i,k} - \payments_i\left(\reportedVal\right) = \left[\sum_{j = 1}^n\sum_{k \in \alloc^*_j\left(\reportedVal\right)} \reportedVal_{j,k} - P_c\left(\alloc^*\left(\reportedVal\right)\right)\right] - \left[\sum_{j\neq i}\sum_{k \in \alloc^*_j\left(0, \reportedVal_{-i}\right)} \reportedVal_{j,k} - P_c\left(\alloc^*\left(0, \reportedVal_{-i}\right)\right)\right].$$ Let $\reportedVal'' = \left(\reportedVal_i, \reportedVal'_{-i}\right)$. Hence,
\begin{align*}
    \left[\sum_{k \in \alloc^*_i\left(\reportedVal''\right)} \reportedVal''_{i,k} - \payments_i\left(\reportedVal''\right)\right]& - \left[\sum_{k \in \alloc^*_i\left(\reportedVal'\right)} \reportedVal'_{i,k} - \payments_i\left(\reportedVal'\right)\right]\\
    &= \left[\sum_{j = 1}^n\sum_{k \in \alloc^*_j\left(\reportedVal''\right)} \reportedVal_{j,k} - P_c\left(\alloc^*\left(\reportedVal''\right)\right)\right] - \left[\sum_{j = 1}^n\sum_{k \in \alloc^*_j\left(\reportedVal'\right)} \reportedVal_{j,k} - P_c\left(\alloc^*\left(\reportedVal'\right)\right)\right]\\
    &\geq \left[\sum_{j = 1}^n\sum_{k \in \alloc^*_j\left(\reportedVal'\right)} \reportedVal''_{j,k} - P_c\left(\alloc^*\left(\reportedVal'\right)\right)\right] - \left[\sum_{j = 1}^n\sum_{k \in \alloc^*_j\left(\reportedVal'\right)} \reportedVal_{j,k} - P_c\left(\alloc^*\left(\reportedVal'\right)\right)\right] \geq 0,
\end{align*} proving the main result in combination with the intermediate result.
\end{proof}

\begin{lemma}
\label{lem: monotone mechanism prop 4}
    Consider the Potential Mechanism $\mechPot$ with a submodular cost function. For any $\reportedVal$ and $\reportedVal'$ such that $\reportedVal \geq \reportedVal'$,
    $\payments_i\left(\reportedVal\right) = \payments_i\left(\reportedVal'_i, \reportedVal_{-i}\right)$ implies that $\alloc^*\left(\reportedVal\right) = \alloc^*\left(\reportedVal'_i, \reportedVal_{-i}\right) $
\end{lemma}
\begin{proof}
    We assume WLOG that $\reportedVal_{-i} = \reportedVal'_{-i}$. Then $\reportedVal' =  \left(\reportedVal'_i, \reportedVal_{-i}\right)$. Since $\payments_i\left(\reportedVal\right) = \payments_i\left(\reportedVal'\right)$, we deduce that 
    $$\sum_{j \neq i}\sum_{k \in \alloc^*_j\left(\reportedVal\right)} \reportedVal'_{j,k} - P_c\left(\alloc^*\left(\reportedVal\right)\right)  = \sum_{j \neq i}\sum_{k \in \alloc^*_j\left(\reportedVal\right)} \reportedVal_{j,k} - P_c\left(\alloc^*\left(\reportedVal\right)\right)  = \sum_{j \neq i}\sum_{k \in \alloc^*_j\left(\reportedVal'\right)} \reportedVal'_{j,k} - P_c\left(\alloc^*\left(\reportedVal'\right)\right).$$ Hence,

    $$\sum_{k \in \alloc^*_i\left(\reportedVal\right)} \reportedVal'_{j,k} +\sum_{j \neq i}\sum_{k \in \alloc^*_j\left(\reportedVal\right)} \reportedVal'_{j,k} - P_c\left(\alloc^*\left(\reportedVal\right)\right)  \geq \sum_{k \in \alloc^*_i\left(\reportedVal'\right)} \reportedVal'_{j,k} +  \sum_{j \neq i}\sum_{k \in \alloc^*_j\left(\reportedVal'\right)} \reportedVal'_{j,k} - P_c\left(\alloc^*\left(\reportedVal'\right)\right)$$ where we use the fact that $\alloc^*\left(\reportedVal'\right) \subset \alloc^*\left(\reportedVal\right)$. We see that by the maximality of $\alloc^*\left(\reportedVal'\right)$ in both value and cardinality, $\left|\alloc^*\left(\reportedVal\right)\right| = \left|\alloc^*\left(\reportedVal'\right)\right|$. Hence,  $\alloc^*\left(\reportedVal\right) = \alloc^*\left(\reportedVal'\right)$, proving the result.
\end{proof}

\begin{lemma}
\label{lem: monotone mechanism prop 5}
    Consider the Potential Mechanism $\mechPot$ with a submodular cost function. For any $\reportedVal$ and $\reportedVal'$ such that $\reportedVal \geq \reportedVal'$, $\alloc_i^*\left(\reportedVal\right) = \alloc_i^*\left(\reportedVal'_i, \reportedVal_{-i}\right)$ implies $\alloc^*\left(\reportedVal\right) = \alloc^*\left(\reportedVal'_i, \reportedVal_{-i}\right)$.
\end{lemma}
\begin{proof} Since $\mechPot$ is (IC), $\payments_i\left(\reportedVal\right) > \payments_i\left(\reportedVal'_i, \reportedVal_{-i}\right)$ cannot hold since when the agent has value $\reportedVal_i$ they would be incentivized to report $\reportedVal'_i$, which cannot be the case. Hence, $\payments_i\left(\reportedVal\right) \leq \payments_i\left(\reportedVal'_i, \reportedVal_{-i}\right)$. By similar reasoning $\payments_i\left(\reportedVal\right) \geq \payments_i\left(\reportedVal'_i, \reportedVal_{-i}\right)$. Hence, $\payments_i\left(\reportedVal\right) = \payments_i\left(\reportedVal'_i, \reportedVal_{-i}\right)$. Thus, the result follows by \cref{lem: monotone mechanism prop 4}.
\end{proof}